\documentclass[bibyear]{aa} % if the references are not structured 
\usepackage{graphicx}
\usepackage{txfonts}
\def\f#1   {Fig.~\ref{#1}}
\def\s#1   {Sect.~\ref{#1}}
\def\tab#1   {Table~\ref{#1}}
\def\eq#1   {Eq.~\ref{#1}}
\def\t#1   {Table~\ref{#1}}

\def\comm#1   {{\tt (COMMENT: #1) }}

\def\kms{~km~s$^{\mathrm{-1}}$}

\def\i                {{\em i}}

\def\msol              {$\mathrm{M}_{\odot}$}

\def\kms{~km~s$^{\mathrm{-1}}$}
\def\smo	{Smol\v{c}i\'{c}}
\def\ntot {28}

\begin{document}

   \title{
  (Sub)millimetre interferometric imaging of a sample of COSMOS/AzTEC submillimetre galaxies III. Environments 
  }

   \author{
   V.~Smol\v{c}i\'{c}\inst{1}, 
   O.~Miettinen\inst{1}, 
   N.~Tomi\v{c}i\'{c}\inst{1},  
   G.~Zamorani\inst{2}, 
   A.~Finoguenov\inst{3}, 
   B.~C.~Lemaux\inst{4},  
   M.~Aravena\inst{5}, 
      P.~Capak\inst{6}, 
   Y-K.~Chiang\inst{7}, 
   F.~Civano\inst{8},
   I.~Delvecchio\inst{1},
   O.~Ilbert\inst{4}, 
   N.~Jurlin\inst{1}, 
   A.~Karim\inst{9}, 
   C.~Laigle\inst{10,11}, 
   O.~Le F{\`e}vre\inst{4}, 
   S.~Marchesi\inst{8},
   H.~J..~McCracken\inst{11}, 
   D.~A.~Riechers\inst{12}, 
   M.~Salvato\inst{13}, 
   E.~Schinnerer\inst{14}, 
   L.~Tasca\inst{4},
   S.~Toft\inst{15}
   }

   \institute{
   Department of Physics, University of Zagreb, Bijeni\v{c}ka cesta 32, HR-10000 Zagreb, Croatia \\ \email{vs@phy.hr} 
\and 
   INAF - Osservatorio Astronomico di Bologna, via Ranzani, 1, I-40127, Bologna, Italy 
\and 
   Department of Physics, University of Helsinki, PO Box 64, 00014 Helsinki, Finland 
\and 
   Aix Marseille Universit\'e, CNRS, LAM (Laboratoire d'Astrophysique de Marseille) UMR 7326, 13388, Marseille, France 
\and 
   N\'ucleo de Astronom\'{\i}a, Facultad de Ingenier\'{\i}a, Universidad Diego Portales, Av. Ej\'ercito 441, Santiago, Chile 
\and 
   Department of Astronomy, California Institute of Technology, MC 249-17, 1200 East California Blvd, Pasadena, CA 91125, USA 
\and 
   Department of Astronomy, University of Texas at Austin, 1 University Station C1400, Austin, TX 78712, USA
\and 
   Harvard-Smithsonian Center for Astrophysics, 60 Garden Street, Cambridge, MA 02138, USA 
\and 
   Argelander Institute for Astronomy, Auf dem H\"{u}gel 71, Bonn, 53121, Germany 
\and
Sorbonne Universit\'{e}s, UPMC Univ Paris 06, UMR 7095, Institut d'Astrophysique de Paris, F-75005, Paris, France 
\and 
  Institut d'Astrophysique de Paris, UMR 7095 CNRS, Universit\'{e} Pierre et Marie Curie, 98 bis Boulevard Arago, F-75014 Paris, France 
\and 
   Department of Astronomy, Cornell University, 220 Space Sciences Building, Ithaca, NY 14853, USA
\and  
   Max-Planck-Institut f\"{u}r Extraterrestrische Physik, Postfach 1312, D-85741, Garching bei M\"{u}nchen, Germany
\and 
   Max Planck Institut f\"{u}r Astronomie, K\"{o}nigstuhl 17, 69117 Heidelberg, Germany
\and
   Dark Cosmology center, Niels Bohr Institute, University of Copenhagen, Juliane Mariesvej 30, DK-2100 Copenhagen, Denmark
}

   \date{Received ; accepted}

% \abstract{}{}{}{}{} 
% 5 {} token are mandatory
\authorrunning{Smol\v{c}i\'{c} et al.}
\titlerunning{Environments of COSMOS SMGs}

  \abstract{
We investigate the environment of 23 submillimetre galaxies (SMGs) drawn from a signal-to-noise (S/N)-limited sample of SMGs originally discovered in the James Clerk Maxwell Telescope (JCMT)/AzTEC 1.1~mm continuum survey of a Cosmic Evolution Survey (COSMOS) subfield and then followed up with the Submillimetre Array and Plateau de Bure Interferometer at 890~$\mu$m and 1.3~mm, respectively. These SMGs already have well defined multiwavelength counterparts and redshifts. We additionally analyse the environment of four COSMOS SMGs spectroscopically confirmed to lie at redshifts $z_{\rm spec}>4.5$, and one at $z_\mathrm{spec}=2.49$ resulting in a total SMG sample size of \ntot . We search for overdensities using the COSMOS photometric redshifts based on over 30 UV-NIR photometric measurements including the new UltraVISTA Data Release 2 and \textit{Spitzer}/SPLASH data, and reaching an accuracy of $\sigma_{\Delta z/(1+z)}=0.0067~(0.0155)$ at $z<3.5~(>3.5)$. To identify overdensities we apply the Voronoi tessellation analysis, and estimate the  redshift-space overdensity estimator $\delta_{\rm g}$ as a function of distance from the SMG and/or overdensity center. We test and validate our approach via simulations,  X-ray detected groups/clusters, and spectroscopic verifications using VUDS and zCOSMOS catalogues  which show that even with photometric redshifts in the COSMOS field we can efficiently retrieve overdensities out to $z\approx5$. 
Our results yield that  11/23 (48)\% JCMT/AzTEC 1.1~mm SMGs occupy overdense environments. Considering the entire JCMT/AzTEC 1.1~mm S/N$\geq4$ sample, and taking the expected fraction of spurious detections into account,  yields that 35-61\% of the SMGs in the S/N-limited sample occupy overdense environments. 
We perform an X-ray stacking analysis in the 0.5-2~keV band using a $32\arcsec$ aperture and our SMG positions, and find statistically significant detections. For our $z<2$ subsample we find  an average flux of $(4.0\pm0.8) \times 10^{-16}$ erg s$^{-1}$ cm$^{-2}$ and  a corresponding total mass of M$_{200} = 2.8\times 10^{13}$~\msol . 
The $z>2$ subsample yields an average flux of $(1.3\pm0.5) \times 10^{-16}$ erg s$^{-1}$ cm$^{-2}$ and  a  corresponding total  mass of M$_{200}=2\times 10^{13}$~\msol . Our results suggest a higher occurrence of SMGs occupying overdense environments at $z\geq3$, than at $z<3$. This may be understood if highly star forming galaxies  can only be formed in the highest peaks of the density field tracing the most massive dark matter haloes at early cosmic epochs, while at later times cosmic structure may have matured sufficiently that more modest overdensities correspond to sufficiently massive haloes to form SMGs.
}


   \maketitle
%
%________________________________________________________________

\section{Introduction}

Massive clusters of galaxies -- the largest gravitationally
bound objects in the Universe -- are common in the local  Universe, and they have been found up to $z=2.00$ (Gobat et al.\ 2011, 2013). 
In a $\Lambda$CDM Universe it is expected that such systems
descend from proto-clusters -- early overdensities of massive galaxies that merge hierarchically and have the potential to form a virialised galaxy
cluster  by the present-day ($z=0$; \cite{miley04}; \cite{hatch2011}; \cite{chiang2013}). 
Given the hierarchical growth of the proto-clusters, an enhanced galaxy interaction and merger rate within the systems is expected. 
This can trigger luminous active galactic nuclei (AGN) and/or starburst phenomena 
in the overdensity/proto-cluster member galaxies (e.g. \cite{lehmer2009}; \cite{daddi2009}; \cite{capak2011}; \cite{walter2012}). Indeed, to-date a few proto-clusters at $z>4$ 
have been found hosting dusty luminous starburst galaxies, the so-called submillimetre galaxies or SMGs (\cite{daddi2009}; \cite{capak2011}; \cite{walter2012}),  per-definition detected at wavelengths beyond 850~$\mu$m and up to a few millimeters  (e.g.,  \cite{blain2002}).

With star formation rates (SFRs) of $\sim100-1\,000$~M$_{\sun}$~yr$^{-1}$, SMGs are acknowledged to be the most powerful starbursts in the  Universe 
(see \cite{blain2002} and \cite{casey2014} for reviews). They are promising candidates for the progenitors of the most massive elliptical galaxies seen in the present-day  Universe (e.g. \cite{lilly1999}; \cite{fu2013}; \cite{ivison2013}; \cite{toft2014}; 
\cite{simpson2014}). Because early-type galaxies 
are predominantly found in clusters, the question arises whether SMGs are preferentially found to reside in galaxy overdensities, or are more likely to represent 
a field galaxy population -- an issue that has not yet been systematically studied. Nonetheless, there is a number of examples of SMGs found to-date that are associated with galaxy overdensities, as summarised in \t{table:literature} . 

A number of studies in the past analysed the spatial clustering of SMGs (e.g. Blain et al.\ 2004; Cowley et al.\ 2015; Hickox et al.\ 2012; see also Almeida et al.\ 2011). For example, 
Hickox et al. (2012) derived a spatial comoving correlation length of 
$r_0=7.7_{-2.3}^{+1.8}h^{-1}\simeq10.8_{-3.2}^{+2.5}$~ comoving Mpc (we here adopt $h=0.71$; see below) for $z=1-3$ SMGs, drawn from the $870~\mu$m LABOCA ECDFS 
(Extended \textit{Chandra} Deep Field South) Submillimetre Survey (LESS; \cite{weiss2009}). This suggests that the probability for an SMG to have 
another SMG within $r\lesssim10.8_{-3.2}^{+2.5}$~comoving Mpc is significantly higher than for a random spatial distribution. Their analysis further suggests that 
the $z = 0$ descendants of SMGs are typically massive  ($\sim$2--3$~L_{\star}$, where $L_{\star}$ is a characteristic galaxy luminosity) elliptical galaxies residing in moderate- to high-mass groups [$\log(M_\mathrm{halo} [h^{-1} \mathrm{M_\odot}]) = 13.3^{+0.3}_{-0.5}$], in agreement with the general view of 
this evolutionary sequence. 
Miller et al. (2015) recently studied whether SMG overdensities are tracers of the most massive structures of dark matter. Their 
simulation results suggest that SMG associations might be underdense in dark matter mass, while dark matter overdensities might be devoid of SMGs. 
The authors concluded that such complex bias of SMGs make them 
a poor tracer of dark matter overdensities.

A systematic study of overdensities around SMGs was performed by Aravena et al.\ (2010) who studied the environment of MAMBO (Max-Planck Millimetre Bolometer) 1.2~mm-detected SMGs in the COSMOS field (\cite{bertoldi2007}) searching for overdensities in $BzK$-selected galaxies. They identified three statistically significant compact overdensities at $z=1.4-2.5$, and concluded that only $\sim30\%$ of radio-identified bright SMGs in that redshift range form in galaxy number density peaks. As already noted by Aravena et al. (2010), studies purely based on the angular two-point correlation function measure the average clustering properties of SMGs, and these might be dominated by only a few significantly clustered systems. A systematic  study of  SMGs thus carries the advantage of identifying individual overdensities and studying the statistical clustering properties of the full sample, as well as the physical connection between the galaxies and their environment (e.g. \cite{capak2011}; \cite{riechers2014}). To-date, however, no environmental analysis of individual SMGs in a well-selected sample of SMGs with secure counterparts has been performed. 
Here we  perform such an analysis based on a sample of 23 SMGs drawn from the COSMOS JCMT/AzTEC 1.1~mm SMG sample, and associated with secure multiwavelength counterparts and (spectroscopic or photometric) redshifts (\cite{scott2008}; \cite{younger2007}, 2009; \cite{smolcic2012}; \cite{miettinen2015}).  Given the sparsity of well-selected SMG samples with secure counterparts and redshifts as well as exquisite data for large-scale structure studies this SMG sample is supplemented with four COSMOS 
SMGs that lie at spectroscopic redshifts of $z_{\rm spec}>4.5$, and one that lies at $z_{\rm spec} = 2.49$ and was part of the SMG sample of Aravena et al. (2010). Hence, a total of 28 SMG environments are analysed in the present work.

Our SMG sample and observational data are described in Sect.~2. The overdensity analysis is 
presented in Sect.~3.  The results are presented in Sect.~4, the spectroscopic verification of the identified overdensities in Sect.~5. The results are discussed in Sect.~6, and summarised in Sect.~7.
We adopt a concordance $\Lambda$CDM cosmology, with the Hubble constant $H_0 =71$~km~s$^{-1}$~Mpc$^{-1}$ 
[$h \equiv H_0/(100\,{\rm km~s^{-1}~Mpc^{-1}})=0.71$], total (dark+luminous baryonic) matter density $\Omega_{\rm m}=0.27$, 
and dark energy density $\Omega_{\Lambda}=0.73$ (\cite{spergel2007}; \cite{larson2011}). 
Magnitudes in the present paper refer to the AB magnitude system (\cite{oke1974}). 

\begin{table}
\renewcommand{\footnoterule}{}
\caption{Galaxy overdensities hosting SMGs in order of increasing redshift.}
{\scriptsize
\begin{minipage}{1\columnwidth}
\centering
\label{table:literature}
\begin{tabular}{c c c}
\hline\hline 
Field or Source ID & $z$ & Reference \\ 
\hline 
Cosbo-16\tablefootmark{a} & 1.4 & \cite{aravena2010} \\
XCS J2215.9-1738 & 1.46 & \cite{ma2015} \\
Cosbo-6 & 1.6 & \cite{aravena2010} \\
GOODS-N & 1.99 & \cite{chapman2009} \\
MRC1138-262 & 2.16 & \cite{dannerbauer2014} \\
Cosbo-3\tablefootmark{b} & 2.3 & \cite{aravena2010}; Casey et al.\ 2015 \\
53W002 & 2.39 & \cite{smail2003} \\ 
SSA22 & 3.09 & \cite{chapman2001}; \cite{tamura2009} \\
4C+41.17 & 3.8 & \cite{ivison2000}; \cite{stevens2003} \\
GN20/20.2 & 4.05 & \cite{daddi2009}\\
HDF850.1 & 5.2 & \cite{walter2012} \\ 
AzTEC3\tablefootmark{b} & 5.3 & \cite{capak2011}\\
\hline 
\end{tabular} 
\tablefoot{\tablefoottext{a}{The association of the target SMG (Cosbo-16) with the $z\sim1.4$ overdensity is uncertain because of the 
uncertainty in the photometric redshift of the SMG.}\tablefoottext{b}{The Cosbo-3 and AzTEC3 systems are analysed in the present study. 
We note that we use the revised spectroscopic redshift of $z_{\rm spec}=2.49$ for Cosbo-3 (Riechers et al., in prep.; see also \smo \ et al.\ 2012b).}}
\end{minipage} 
}
\end{table}

\section{Data}
\label{sec:data}

\subsection{SMG sample}

\subsubsection{Main-SMG sample}

The main sample of 23 SMGs studied here ({\em main-SMG sample} hereafter) is taken from the JCMT/AzTEC 1.1~mm COSMOS survey S/N-limited sample identifying sources at $18\arcsec$ angular resolution within 0.15~deg$^2$ in the COSMOS field (Scott et al. 2008). 
Younger et al.\ (2007, 2009) followed-up sources AzTEC1--15 with the Submillimetre Array (SMA) at 890~$\mu$m wavelength and $\lesssim2\arcsec$ angular resolution, and identified 17 counterparts to the initially selected 15 SMGs. Sources AzTEC16--30 were followed-up by Miettinen et al.\ (2015) with the Plateau de Bure Interferometer (PdBI) at 1.3~mm and they identified 22 counterparts at this wavelength, yielding a total of 39 AzTEC1--30  counterparts identified at 890~$\mu$m (S/N~$\geq3.9$) and 1.3~mm (S/N~$\geq4.5$) wavelength. Using the COSMOS panchromatic data multi-wavelength counterparts were then associated to a subset of these sources by Younger et al.\ (2007, 2009), \smo \  et al.\ (2012a), Miettinen et al.\ (2015), and photometric redshifts computed  via a standard $\chi^2$ minimization SED fitting procedure and galaxy templates optimized for SMGs (\cite{smolcic2012, miettinen2015}; see also \cite{smolcic2012b} for details). A comparison between spectroscopic and photometric redshifts yields an accuracy of such computed photometric redshifts of $\sigma_{\Delta z/(1+z)}=0.09$ (see \cite{smolcic2012, smolcic2012b} for details). For the SMGs with no multi-wavelength counterparts lower-redshift limits were computed using the millimeter-to-radio-flux ratio technique (see \cite{miettinen2015}). 

An analysis of the number of spurious sources in the sample of the 39  counterparts of AzTEC1--30 detected at 890~$\mu$m (S/N~$\geq3.9$) and 1.3~mm (S/N~$\geq4.5$)  with the SMA and PdBI, respectively, has been performed by both Younger et al.\ (2009), and Miettinen et al.\ (2015). While all sources with associated multi-wavelength counterparts are estimated to be real, 8 spurious sources are expected among the 16 detections without associated multi-wavelength counterparts (i.e., 2 within the AzTEC7--15 SMA--890~$\mu$m sample, and 6 within the AzTEC16--30 PdBI--1.3~mm sample). 

In summary, within the sample of 39  counterparts of AzTEC1--30 detected at 890~$\mu$m and 1.3~mm at $\lesssim2\arcsec$ angular resolution, 7 have been associated with counterparts with spectroscopic redshifts, 16 with counterparts for which photometric redshifts ($\sigma_{\Delta z/(1+z)}=0.09$) have been determined, and 16 for which only lower-redshift ($z\gtrsim2$) limits could be estimated. Out of the last 16 sources, 8 are estimated to be spurious. We here exclude the SMGs with only lower redshift limits, and list in \t{table:sample} \ the sources analyzed here. Taking the expected spurious source fraction into account we estimate that the excluded fraction of SMGs amounts to 26\%[$=100\%\times(16-8)/(39-8)$] of the total sample. This fraction is taken into account in the interpretation of our results in later Sections.

\subsubsection{Additional-SMG sample}

Given that well-selected samples of SMGs with secure counterparts and redshifts, as well as exquisite data for large-scale structure studies are still sparse in the literature, here we also analyze the environments of five more SMGs in the COSMOS field ({\em additional-SMG sample} hereafter), four of which are spectroscopically confirmed to lie at $z_{\rm spec} > 4.5$ (see \cite{smolcic2015} for more details), and one additional SMG (Cosbo-3) at $z_{\rm spec} = 2.49$ identified by Bertoldi et al.\ (2007), and further studied by Aravena et al. (2010), and \smo \ et al.\ (2012b). 
The additional-SMG sample has a high degree of complementarity with the main-SMG sample. 
While the inclusion of this sample induces a slight deviation from a S/N-limited sample, the main purpose of a S/N-limited sample is to minimize contamination from spurious sources, which the galaxies in the additional-SMG sample are clearly not. Furthermore, any potential bias from the inclusion of these sources is more than compensated for by every galaxy in the additional-SMG sample having a secure spectroscopic redshift, which dramatically decreases the uncertainty in overdensity association.
In total, this yields \ntot \ SMGs analyzed here, the details of which are listed in Table~\ref{table:sample}.

\begin{table}
\renewcommand{\footnoterule}{}
\caption{
 Source list. The upper part lists the 23 sources of our main-SMG sample with photometric or spectroscopic redshifts, drawn from a (sub-)mm-interferometric follow-up sample of the 
S/N-limited JCMT/AzTEC 1.1 mm SMG sample (see text for details).
The lower part of the table lists 5 additional, spectroscopically confirmed SMGs at $z=2.49$ and $z>4.5$ in the COSMOS field (our additional-SMG sample).
}
{\scriptsize
\begin{minipage}{1\columnwidth}
%\begin{center}
\label{table:sample}
\begin{tabular}{c c c c c}
\hline\hline 
Source ID & $\alpha_{2000.0}$\tablefootmark{e} & $\delta_{2000.0}$\tablefootmark{e} & Redshift\tablefootmark{a} & $z$ reference\tablefootmark{a}\\
       & [h:m:s] & [$\degr$:$\arcmin$:$\arcsec$] & & \\ 
\hline
AzTEC1 &  09 59 42.86 & +02 29 38.2 & $z_{\rm spec}=4.3415$ & 1 \\
AzTEC2 &  10 00 08.05 & +02 26 12.2 & $z_{\rm spec}=1.125$ & 2 \\
AzTEC3 &  10 00 20.70 & +02 35 20.5 & $z_{\rm spec}=5.298$ & 3\\
AzTEC4 &  09 59 31.72 & +02 30 44.0 & $z_{\rm phot}=4.93_{-1.11}^{+0.43}$ & 4 \\
AzTEC5 &  10 00 19.75 & +02 32 04.4 & $z_{\rm phot}=3.05_{-0.28}^{+0.33}$ & 4 \\
AzTEC7 &  10 00 18.06 & +02 48 30.5 & $z_{\rm phot}=2.30\pm0.10$ & 4 \\
AzTEC8 &  09 59 59.34 & +02 34 41.0 & $z_{\rm spec}=3.179$ & 6  \\
AzTEC9 &  09 59 57.25 & +02 27 30.6 & $z_{\rm phot}=1.07_{-0.10}^{+0.11}$ & 4 \\
AzTEC10 &  09 59 30.76 & +02 40 33.9 & $z_{\rm phot}=2.79_{-1.29}^{+1.86}$ & 4 \\
AzTEC11-N\tablefootmark{b} & 10 00 08.91 & +02 40 09.6 & $z_{\rm spec}=1.599$ & 7  \\
AzTEC11-S\tablefootmark{b} & 10 00 08.94 & +02 40 12.3 & $z_{\rm spec}=1.599$ & 7 \\
AzTEC12 &  10 00 35.29 & +02 43 53.4 & $z_{\rm phot}=2.54_{-0.33}^{+0.13}$ & 4 \\
AzTEC14-W & 10 00 09.63 & +02 30 18.0 & $z_{\rm phot}=1.30_{-0.36}^{+0.12}$ & 4 \\
AzTEC15 & 10 00 12.89 & +02 34 35.7 & $z_{\rm phot}=3.17_{-0.37}^{+0.29}$ & 4  \\
AzTEC17a & 09 59 39.194 & +02 34 03.83 & $z_{\rm spec}=0.834$ & 7 \\
AzTEC17b & 09 59 38.904 & +02 34 04.69 & $z_{\rm phot}=4.14_{-1.73}^{+0.87}$ & 5 \\
AzTEC18 & 09 59 42.607 & +02 35 36.96 & $z_{\rm phot}=3.00_{-0.17}^{+0.19}$ & 5 \\
AzTEC19a & 10 00 28.735 & +02 32 03.84 & $z_{\rm phot}=3.20_{-0.45}^{+0.18}$ & 5 \\
AzTEC19b & 10 00 29.256 & +02 32 09.82 & $z_{\rm phot}=1.11\pm0.10$ & 5 \\ 
AzTEC21a & 10 00 02.558 & +02 46 41.74 & $z_{\rm phot}=2.60_{-0.17}^{+0.18}$ & 5 \\
AzTEC21b & 10 00 02.710 & +02 46 44.51 & $z_{\rm phot}=2.80_{-0.16}^{+0.14}$ & 5 \\
AzTEC23 & 09 59 31.399 & +02 36 04.61 & $z_{\rm phot}=1.60_{-0.50}^{+0.28}$ & 5 \\
AzTEC26a & 09 59 59.386 & +02 38 15.36 & $z_{\rm phot}=2.50_{-0.14}^{+0.24}$ & 5 \\
AzTEC29b & 10 00 26.561 & +02 38 05.14 & $z_{\rm phot}=1.45_{-0.38}^{+0.79}$ & 5 \\
\hline 
Cosbo-3 & 10 00 56.95 & +02 20 17.79 & $z_{\rm spec}=2.490$ & 6\\
J1000+0234 & 10 00 54.484 & +02 34 35.73 & $z_{\rm spec}=4.542$ & 8\\
AzTEC/C159 & 09 59 30.420 & +01 55 27.85 & $z_{\rm spec}=4.569$ & 9 \\
Vd-17871\tablefootmark{c} & 10 01 27.075 & +02 08 55.60 & $z_{\rm spec}=4.622$ & 10, 9 \\
AK03\tablefootmark{d} & 10 00 18.744 & +02 28 13.53 & $z_{\rm spec}=4.747$ & 9 \\
\hline 
\end{tabular} \\
%\end{center}
\tablefoottext{a}{The $z_{\rm spec}$ and $z_{\rm phot}$ values are the spectroscopic redshift and optical-NIR photometric redshift, respectively. }\\\tablefoottext{b}{AzTEC11 was resolved into two 890~$\mu$m sources (N and S) by Younger et al. (2009). The two components are probably physically related, i.e. are at the same redshift (Miettinen et al., 2015). In the present paper, we adopt AzTEC11-N as the target SMG.}\\
\tablefoottext{c}{The SMG has two components with a projected angular separation of $1\farcs5$, and they lie at the same $z_{\rm spec}$ (Karim et al., in prep.).}\\
\tablefoottext{d}{The AK03 SMG is also composed of two components (N and S) whose angular separation is $\sim0\farcs9$ in the optical. The $z_{\rm spec}=4.747$ refers to the northern component, and the southern component has a comparable $z_{\rm phot}$   of $(4.40\pm0.10)$ or $(4.65\pm0.10)$,  depending on the template set used (\cite{smolcic2015}).} \\
\tablefoottext{e}{The coordinates given in columns~(2) and (3) for AzTEC1--15 refer to the SMA 890~$\mu$m peak position (\cite{younger2007}, 2009),
while those for AzTEC17--29 are the PdBI 1.3~mm peak positions (\cite{miettinen2015}). For J1000+0234 the coordinates refer to the PdBI-detected $^{12}$CO$(4-3)$ line emission peak (\cite{schinnerer2008}), while those for AzTEC/C159, Vd-17871, and AK03 refer to the VLA 3~GHz peak position (\cite{smolcic2015}), and that for Cosbo-3 from the CARMA peak position (\cite{smolcic2012b}).\\
The redshift references in the last column are as follows: \\
1=\cite{yun2015}; \\
2=Faustino Jimenez Andrade et al., in prep.; \\
3=\cite{riechers2010} and \cite{capak2011}; \\
4=\cite{smolcic2012}; \\
5=\cite{miettinen2015}; \\
6=Riechers et al., in prep.; \\
7=Salvato et al., in prep.; \\
8=\cite{capak2008} and \cite{schinnerer2008}; \\
9=\cite{smolcic2015}; \\
10=Karim et al., in prep.
}
\end{minipage} 
}
\end{table}

\subsection{Redshift data}

\subsubsection{Catalogues}

We use the most up-to-date version of the COSMOS photometric redshift catalogue (Laigle et al.\ 2016), 
which includes $Y$, $J$, $H$,  $K_{\rm s}$ 
data from the UltraVISTA Data Release 2, 
and new SPLASH 3.6 and 4.5~$\mu$m \textit{Spitzer}/IRAC data (\cite{capak2007}, in prep.; \cite{mccracken2012}; \cite{ilbert2013}). 
The catalogue was selected using the  $zYJHK_{\rm s}$ stacked mosaic, and, thus, 
the number density of galaxies is high and  well suited for our overdensity analysis. 
We here also make use of the COSMOS spectroscopic redshift catalogue (Salvato et al., in prep.), compiling all available spectroscopic redshifts, both internal to the COSMOS collaboration and from the literature [zCOSMOS (\cite{lilly2007}, 2009), IMACS (\cite{trump2007}), MMT (\cite{prescott2006}), VIMOS Ultra Deep Survey (VUDS, \cite{lefevre2014}), Subaru/FOCAS (T.~Nagao et al., priv. comm.), and SDSS DR8 (\cite{aihara2011})]. 

\subsubsection{Photometric redshift accuracy}
\label{sec:photz}

A key issue for the analysis presented here is whether photometric redshifts can efficiently be used for galaxy overdensity identification, especially at high redshifts ($z\gtrsim3.5$) where the photometric redshift accuracy is lower compared to lower redshifts (e.g. Laigle et al.\ 2016). 
Hence, to test this, in \f{fig:specphot} \ (left panels) we first compare the photometric ($z_\mathrm{phot}$) and spectroscopic ($z_\mathrm{spec}$) redshifts for 15,527 sources with secure spectroscopic redshifts up to $z_\mathrm{phot}=3.5$ ($i^+\la25$). In the figure we also show the distribution of $\Delta z / (1+z_\mathrm{spec})$, where $\Delta z $ is the difference between the spectroscopic and photometric redshifts. Consistent with the results from Ilbert et al. (2013), we find that the standard deviation of this distribution is only $\sigma_{\Delta z/(1+z)}=0.0067$, verifying the excellent photometric redshift accuracy at these redshifts.

To test whether the COSMOS  photometric-redshift catalogue can be efficiently used to search for overdensities at $z>3.5$, in \f{fig:specphot} \ (right panels) we show a comparison between photometric ($3.5\lesssim z_{\rm phot} \lesssim 6$) and secure spectroscopic redshifts for 240 high-redshift sources in the COSMOS field.   
We find excellent agreement between the photometric and spectroscopic redshifts, with a standard deviation of $\sigma_{\Delta z/(1+z)}=0.0155$. We also note that Le F{\`e}vre et al.\ (2015) recently compared the VUDS spectroscopic redshifts with the same Ilbert et al. (2013) photometric redshift values as here and found similar results (see their Fig.~11). 
Because of the well known degeneracy in the spectral energy distribution fitting procedure between the Balmer-4\,000~$\AA$ and Ly$\alpha$-1\,215~$\AA$ breaks, 
a fraction of high redshift ($z>2$) sources might have a photometric redshift of less than unity (see, for example, Fig.~11 in Le Fevre et al.\ 2015), 
making a high redshift photometric redshift sample partially incomplete. We estimate a completeness of  $80\%$ of  our high-redshift photometric redshift selected sample based on the fraction of sources with $z_\mathrm{phot}<3.5$ within the COSMOS spectroscopic sample with secure spectroscopic redshifts of $z_\mathrm{spec}>3.5$. In summary, the photometric redshift accuracy in the COSMOS field is accurate enough for an overdensity analysis up to $z<6$, as we will show below.

\begin{figure*}
\includegraphics[bb=50 0 432 432, scale=0.7]{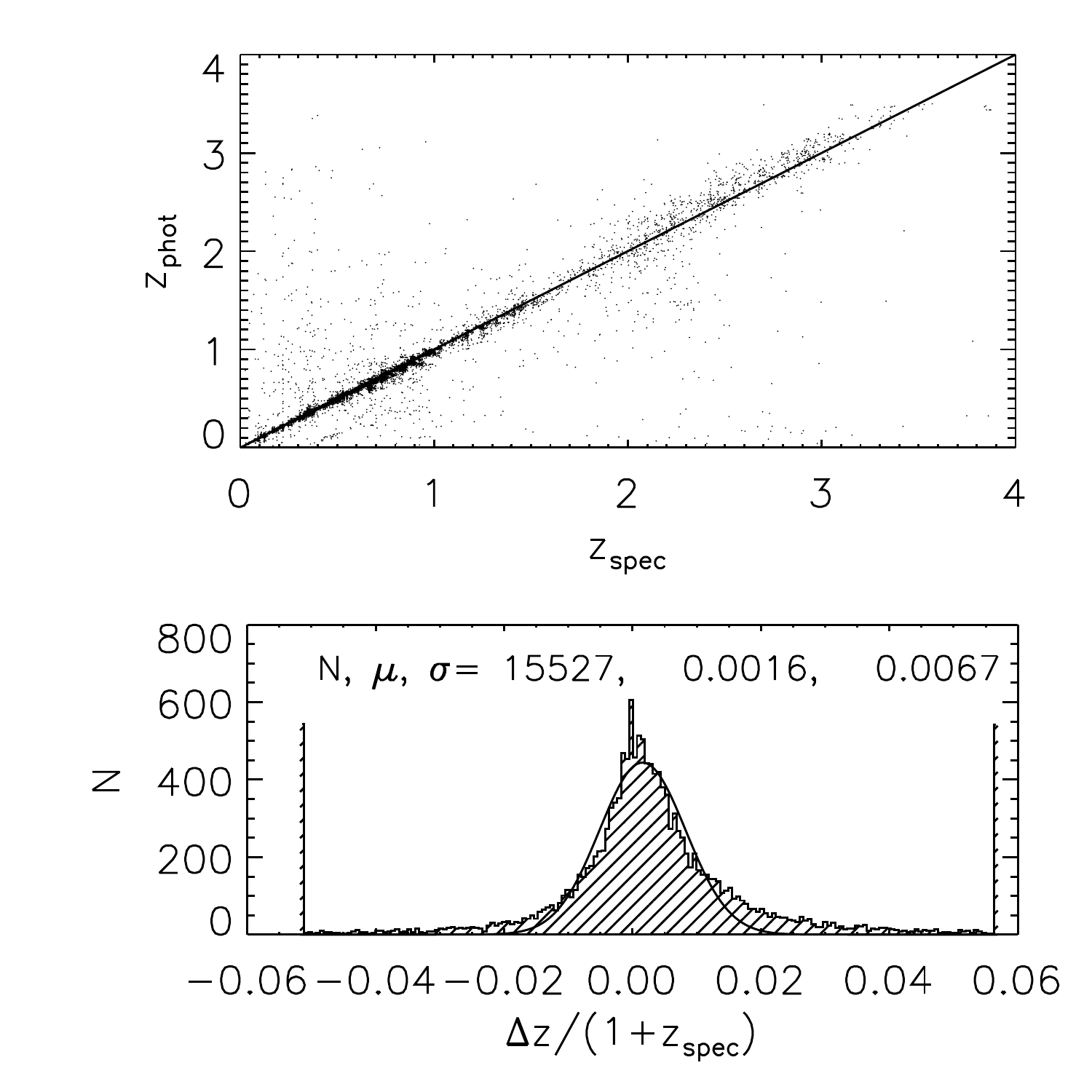}
\includegraphics[bb=70 0 432 432, scale=0.7]{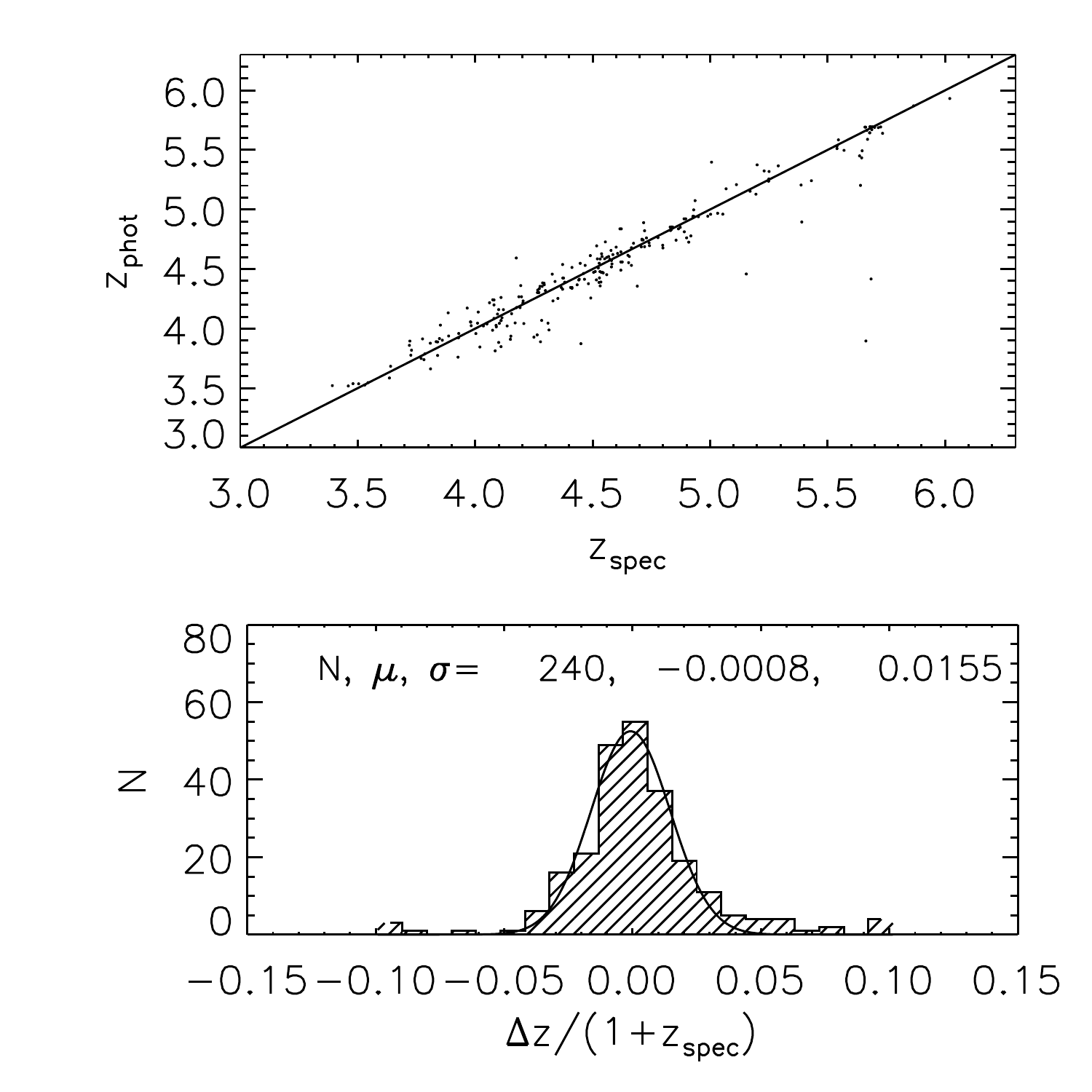}
\caption{
Comparison between photometric ($z_\mathrm{phot}$) and spectroscopic ($z_\mathrm{spec}$) redshifts in the COSMOS field for galaxies with $z_\mathrm{phot}\leq3.5$ (left panels) and $z_\mathrm{phot}>3.5$ (right panels). The top panels show a direct comparison (with the one-to-one line indicated). The bottom panels show the distribution of the difference between the  spectroscopic and photometric redshifts normalised to $1+z_\mathrm{spec}$. A Gaussian fit  to the distribution is shown and the number of sources, the mean, and standard deviation are indicated in the panels.
\label{fig:specphot}}
\end{figure*}

\section{Methodology}

We search for galaxy overdensities around our SMGs, and determine their surface density profiles in a multistep process, described in detail in Sect.~\ref{sec:method}. Around each SMG's position we first compute the galaxy overdensity parameter ($\delta_\mathrm{g}$) as a function of projected radius ($r$) for $r=0.5, 1, 2.5, 5$~arc minutes to assess potential small scale overdensities centered at the SMG. We then apply the Voronoi tessellation analysis (VTA; e.g.\ Ramella et al.\ 2001; \smo \ et al.\ 2007) at a $25\arcmin\times25\arcmin$ area surrounding the SMG to display the identified small scale overdensity and/or to identify further potential overdensities on larger scales close to the given SMG.  As SMGs may not necessarily be located in the very centers of their host overdensities, using the VTA-overdense cells we recompute the center of the overdensity. Using this newly defined overdensity center we then re-compute the galaxy overdensity parameter ($\delta_\mathrm{g}$) of the structure as a function of projected radius in small steps ($dr=0\farcm1$) out to $r = 12\farcm5$.  We describe the quantification of the significance of the identified overdensities and false detection probabilities in \s{sec:simul} , and we present further tests and verifications of the method in Appendix~\ref{sec:mc}.

\subsection{Searching for overdensities using photometric redshifts}
\label{sec:method}

To search for overdensities around our SMGs, we use the position and redshift of the target SMG. To analyze the environment of each SMG we use galaxies drawn from the COSMOS photometric redshift catalogue with $i^+\leq25.5$ assuring sample completeness and the most accurate photometric redshifts (see previous Section and Laigle et al. 2016).
To select large enough redshift bins to account for a variety of magnitude dependent photometric redshift uncertainties (see Ilbert et al.\  2013; Laigle et al.\ 2016) we use all the galaxies in the photometric redshift catalog with
 $z_\mathrm{phot}=z_\mathrm{SMG}\pm 3\sigma_{\Delta z/(1+z)} (1+z_\mathrm{SMG})$, where  $z_\mathrm{SMG}$ is the redshift of the SMG [we take $\sigma_{\Delta z/(1+z)}=0.007~(0.016)$ for $z_\mathrm{SMG}\leq3.5~(>3.5)$; see Fig.~\ref{fig:specphot} and Sect.~\ref{sec:photz} \ for details]. In Appendix~\ref{sec:1.5sig} we present the results using all galaxies within a factor of 2 narrower bin [i.e., with $z_\mathrm{phot}=z_\mathrm{SMG}\pm 1.5\sigma_{\Delta z/(1+z)} (1+z_\mathrm{SMG})$].

\subsubsection{Small-scale overdensity search with central SMGs}
\label{sec:sss}

 We first  search for small scale overdensities around our SMGs by computing the galaxy overdensity parameter in the vicinity of each SMG, i.e.\ within a distance of  $r = 0\farcm5,\, 1\arcmin,\, 2\farcm5,\, 5\arcmin$ from the given SMG. 
The galaxy overdensity parameter, i.e. contrast above the background field, as a function of radius is defined as
\begin{equation}
\label{eq:delta}
\delta_{\rm g}(r)\equiv \frac{\Sigma_{\rm r}(r)-\Sigma_{\rm bg}}{ \Sigma_{\rm bg}}=\frac{\Sigma_{\rm r}(r)}{\Sigma_{\rm bg}}-1\, ,
\end{equation}
where $\Sigma_{\rm r}$ and $\Sigma_{\rm bg}$ are  
the local galaxy, and the background galaxy surface density, respectively. The numerator in the first equality, $\Sigma_{\rm r}-\Sigma_{\rm bg}\equiv \Delta \Sigma$, 
is the deviation of the local galaxy number density from that in the  background field. The value of $\Sigma_{\rm r}$ was calculated 
as $\Sigma_{\rm r}=N_{\rm r}/A_{\rm r}$, where $N_{\rm r}$ is the number of galaxies within the given redshift bin around the center inside the search window of area 
$A_{\rm r}=\pi \times r^2$ with $r$ the search radius. Similarly, $\Sigma_{\rm bg}=N_{\rm bg}/A_{\rm bg}$, where $N_{\rm bg}$ is the number of galaxies satisfying
the photometric redshift criterion within the large background area $A_{\rm bg}$. This was taken to be the effective area of the full COSMOS field for our small-scale SMG overdensity assessment. 
 In the computation masked areas (due to e.g. saturated stars or corrupted data) in the COSMOS field have properly been taken into account.

We, furthermore, compute for each radius the Poisson probability of observing $\geq N_\mathrm{r}$ objects when the expected number is $n_\mathrm{r}=\Sigma_\mathrm{bg} \times A_\mathrm{r}$, 
$p(\geq N_\mathrm{r},n_\mathrm{r}) = 1 - \Sigma_{i=0}^{N_\mathrm{r}} (e^{-n_\mathrm{r} } n_\mathrm{r}^i / i!)$, and consider $p\leq0.05$ to be robust overdensity values.

According to the above definition, 
$\delta_{\rm g}=0 \Leftrightarrow \Delta \Sigma=0$ means there is no observed overdensity, while $\delta_{\rm g}<0$ indicates an underdensity. 
As a general rule, the higher the value of $\delta_{\rm g}$, 
the higher the probability that the identified overdensity structure is a genuine galaxy  (proto-)cluster/group (e.g., \cite{chiang2013}).

\subsubsection{Voronoi tessellation analysis (VTA) and overdensity center relocation}
\label{sec:vta}

We next apply the VTA onto galaxies surrounding our SMGs. Voronoi tessellation is an efficient method to identify and visualise overdensities, and it has already been applied for the search of galaxy clusters 
by e.g. Ramella et al. (2001), Kim et al. (2002), Lopes et al. (2004), Smol{\v c}i{\'c} et al. (2007), and Oklop{\v c}i{\'c} et al. (2010). 
A major advantage of the VTA is that overdensities are identified irrrespective of their shape or galaxy properties.  In general, a Voronoi tessellation -- in the two-dimensional case as in the present work -- is a method to partition a plane into convex 
polygons called Voronoi cells (\cite{dirichlet1850}; \cite{voronoi1908}). When a 2D distribution of distinct points, generally called 
nuclei (that in our case are galaxies), is decomposed through the Voronoi tessellation algorithm, each resulting cell encloses exactly 
one seed point, and every position within a given cell is closer to the cell's nucleus than to any other nucleus in the plane (\cite{icke1987}). 
The higher the surface density of nuclei is, the smaller is the area of the cells; an ``overdensity'' region can be uncovered this way.

We apply the VTA on galaxies drawn from the COSMOS photometric redshift catalogue [$i^+\leq25.5$,  $z_\mathrm{phot}=z_\mathrm{SMG}\pm 3\sigma_{\Delta z/(1+z)} (1+z_\mathrm{SMG})$], as defined above, over an area larger than $25\arcmin\times25\arcmin$ surrounding the given SMG.
To identify overdensities in the analysed region we make use of the inverse of the areas of the VTA cells (which effectively correspond to the local surface densities of given galaxies; $n_\mathrm{g}$ hereafter). 
We first generate 10 Mock catalogs with the same number of galaxies as present in the inner $1$~deg$^2$ of the COSMOS field, but with randomly  distributed positions over this area (note that such a size is large enough to contain both overdensities, underdensities, and field galaxies). 
We then apply the VTA on each Mock catalog and generate a cumulative distribution of $n_\mathrm{g}$, averaged over the 10 iterations. 
Following Ramella et al.\ (2001) we define the value of $n_\mathrm{g}$ that corresponds to the 80\% quantile as the threshold above which a VTA cell is considered 'overdense'.
 
\subsubsection{Large-scale overdensity search for off-center SMGs}
\label{sec:lss0}

As the small-scale $\delta_\mathrm{g}$ computation (\s{sec:sss} ), and the VTA (\s{sec:vta} ) are centered at the position of the given SMG, 
we next redefine the center of the potential overdensity by computing the median right ascension and median declination of VTA-identified overdense cells within $r \leq x \arcmin$ from the SMG, where $x$ is taken between $1\arcmin$ and $10\arcmin$ and determined in such a way that the surface density  of the overdense VTA cells (i.e., number of overdense VTA cells over $x^2\cdot\pi$) is maximized within this circular area. 

With such defined overdensity centers we then re-compute the galaxy overdensity parameter, $\delta_\mathrm{g}$ 
[Eq.~(1)] out to $r = 12\farcm5$, in steps of $dr = 0\farcm1$. To assess an average value of the background surface density and quantify its fluctuation for this analysis we  use nine differently positioned, circular, and not overlapping $A_{\rm bg}=706.86$~arcmin$^2$ areas to compute the median value of the background surface density, $\tilde{ \Sigma }_{\rm bg}$, and its standard deviation, $\sigma_{\Sigma_{\rm bg}}$. This allows us 
to quantify the significance of $\delta_{\rm g}(r)>0$ values by assigning errors on $\delta_{\rm g}(r)=0$, i.e. $\sigma_{\rm \delta_g=0}(r)$. The errors will reflect statistical fluctuations for $ \Delta \Sigma=0$, and we compute them by taking $\Sigma_{\rm r}$=$\Sigma_{\rm bg}$ and statistically propagating the errors of  $\Sigma_{\rm r}$ and $\Sigma_{\rm bg}$ ($\equiv\sigma_{\Sigma_{\rm r}}$ and $\sigma_{\Sigma_{\rm bg}}$, where the latter was defined above). This yields
\begin{equation}
\label{eq:deltaerr}
\sigma_{\delta_{\rm g}=0}(r) =  \frac{1}{\tilde{\Sigma}_{\rm bg}}   \left[   \sigma_{\Sigma_{\rm r}(r)}^2 +  \sigma_{\Sigma_{\rm bg}}^2 \right]^{1/2}\,.
\end{equation}
As $\Sigma_{\rm r}(r)$ is derived from rather small areas ($A_\mathrm{r}$), $\sigma_{\Sigma_{\rm r}}$ will be dominated by  Poisson errors (of the number of expected galaxies based on $\Sigma_{\rm bg}$). Hence, we take $\sigma_{\Sigma_{\rm r}}(r)=  \sqrt{\Sigma_\mathrm{bg} \times A_\mathrm{r}}/ A_\mathrm{r}= \sqrt{\Sigma_\mathrm{bg}/A_\mathrm{r}}$.  Finally, to consider the overdensity centered at the newly defined center as significant, at any given radius we require $\delta_{\rm g}\geq t \cdot \sigma_{\rm \delta_g=0}$, where $t$ is determined based on our simulations, detailed in the next subsection, in such a way that it reassures $\leq20\%$ of chance detections.

\subsection{Assessing significance and false detection probabilities}
\label{sec:simul}

To quantify false detection probabilities we make use of  10 Mock catalogs generated  for each SMG containing the same number of galaxies with $z=z_\mathrm{SMG}\pm 3\sigma_{\Delta_z/(1+z)}(1+z)$ present in the inner $1$~deg$^2$ of the COSMOS field, but with randomly  distributed positions over this area (see \s{sec:vta} ). For 1,000 random SMG positions we then search within each Mock catalog for the number of occurrences with  $N_\mathrm{r}\ge 1, 2, 3$ within $r=0.5\arcmin,1\arcmin$. This corresponds to 10,000 Mock realizations per SMG. For each SMG in our sample this then sets the false detection probability within our small-scale overdensity search  (see \s{sec:sss} ) out to $r=1\arcmin$. 

To  quantify the false detection probability within our large-scale overdensity search with off-center SMGs (see \s{sec:lss0} ), we apply the same procedure as for the real SMGs to 100 random SMG positions. We first apply the VTA on the Mock catalogs (including the random-SMG position per realization), then redefine the potential overdensity center for each random-SMG, and finally derive $\delta_\mathrm{g}$ as a function of $r$, centered at the newly defined overdensity center. We lastly determine the number of $\delta_\mathrm{g}/\sigma_{\delta_\mathrm{g}}\ge 3,4,...10$ detections with the requirement that the distance between the random-SMG position and the newly defined overdensity center is less or equal to the same distance, as determined for the real SMG. We set a  $\delta_\mathrm{g}/\sigma_{\delta_\mathrm{g}}$ threshold value such that the fraction of such occurrences is  $\leq0.2$. This process then yields, for each SMG in our sample, a unique $\delta_\mathrm{g}/\sigma_{\delta_\mathrm{g}}$ threshold to assess the significance of the overdensity analyzed around the newly defined center.

\section{Analysis and results}
\label{sec:results}

The overdensity search results are shown in Fig.~\ref{figure:voronoi}, where for each SMG in our sample we show 
i) the integral $\delta_\mathrm{g}$ as a function of projected circular radius, centered on the SMGs position and out to $r=5\arcmin$,  indicating the number of sources within a given radius ($N_\mathrm{r}$ which also includes the SMG of interest),
ii) the Voronoi diagram over a $25\arcmin \times 25\arcmin$ area centered on the SMG's position with overdense VTA cells, and the redefined overdensity center indicated, and 
 iii) $\delta_\mathrm{g}$ as a function of radius out to $r = 12\farcm5$ away from the newly defined overdensity center, and its significance. In \s{sec:smallscale} \ we first investigate the small-scale ($r\leq1\arcmin$) overdensities centered on the SMG positions, and in \s{sec:general} \ we investigate overdensities surrounding SMGs on larger scales and not necessarily with a centrally positioned SMG.

\subsection{Evidence for small ($r\leq1\arcmin$) scale overdensities around central SMGs}
\label{sec:smallscale}

In \t{table:smallscale} \ we list the number of sources (if larger than 1) within $r=0.5\arcmin$ or $r=1\arcmin$ for the 28 SMGs in our sample. For each SMG we also list the Poisson probability of finding $\geq N_\mathrm{r}$ sources within this radius, as well as the false detection probability (P$_\mathrm{FD}$) to find $\geq N_\mathrm{r}$ within $r$ based on the results of our simulations using random SMG positions and Mock galaxy catalogs (see \s{sec:simul} \ for details). If we set the false detection probability to P$_\mathrm{FD}\leq5\%$, we find 5 out of 23 (22\%) systems  in our main-SMG sample, and 3 out of 5 (60\%) systems in our additional-SMG sample with $N_\mathrm{r}\geq2$ values at $r\leq1\arcmin$ away from the SMG. This amounts to a  total of 8/28 (29\%) systems with significant small-scale overdensities. Note that with such a false detection probability we expect $\leq1.15$ false detections in the main-SMG sample, and $\leq0.25$ in the additional-SMG sample. If we restrict our main-SMG sample to only the SMGs with spectroscopic redshifts, we find 2/6 (=33\%) SMGs with small-scale overdensities. % 

As discussed in \s{sec:data} \ our main-SMG sample was drawn from a S/N-limited sample of AzTEC/JCMT sources, and in addition to the 23 SMGs in our main-SMG sample 8 further SMGs  have lower-redshift limit estimates only and were, thus, not analyzed here. If these 8 SMGs were to exhibit the same small-scale environment properties as the 74\% (=23/(23+8)) of the sample analyzed here, the fraction of SMGs in the flux limited sample with  small-scale overdensities would remain the same (22\%). In the two extreme cases where all 8 systems would either be or not be associated with small-scale overdensities, this fraction would range from 35\% to 61\%. In Appendix~\ref{sec:1.5sig} we compute these fractions for a narrower redshift width [$\Delta z = z_\mathrm{SMG}\pm 1.5\sigma_{\Delta z/(1+z)} (1+z_\mathrm{SMG})$], and find statistically consistent results.

The scales ($r\leq1\arcmin$) tested here correspond to physical scales of less than 510, 470, 425, 380 kpc at $z=2,3,4,5$, respectively. These are the typical sizes of (proto-)groups (e.g. Diener et al.\ 2013 used a  projected physical diameter of 500 kpc to search for proto-groups at $1.8<z<3$). We thus further test whether any of our SMG  systems has been detected via extended X-ray emission arising from thermal bremsstrahlung radiation of the intra-group/-cluster medium, heated during the gravitational collapse of the primordial density peaks.

The COSMOS field has been a target of several large programmes with both \textit{Chandra} and {\it XMM-Newton}  (\cite{hasinger2007}; \cite{finoguenov2007}; \cite{elvis2009}; Civano et al., subm.), allowing for a detection of emission from the hot gas of galaxy groups. The published catalogues combine X-ray and optical/NIR data to assign redshifts to the groups, and are limited to $z\simeq 1$ (Finoguenov et al. 2007; George et al. 2011), while most of our SMGs are at $z>1$.

The expected diameter of a galaxy group at $z\gtrsim2$ at the sensitivity limit of the X-ray survey is about $0\farcm5$ and brighter groups and clusters are expected to have even larger sizes. We find that only one source in our sample, Cosbo-3, is directly detected as an extended X-ray source in the 0.5-2~keV band. This is described in more detail in Wang et al.\ (in prep.) who estimate an X-ray luminosity of $L_\mathrm{X}=7\times10^{43}$~erg~s$^{-1}$ in the rest-frame 0.1-2.4 keV band, and a total mass within $r_\mathrm{200}=38\arcsec$ of $M_\mathrm{200} = 4\times10^{13}$~\msol .
We have further verified that none of the extended X-ray sources in the COSMOS field, unidentified in the optical/NIR data, matches the position of any of the remaining SMGs in our sample.  We note that the sensitivity of combined X-ray data only reaches roughly $\sim4\times 10^{13}$~\msol \ at the redshifts of our SMG sample.
To gain a deeper insight into the halo masses of our SMGs  we have performed a stacking analysis for the SMG systems (excluding Cosbo-3) outside the area confused with emission from other groups. We combine the data from \textit{Chandra} and \textit{XMM-Newton} in the 0.5--2 keV band, after subtracting the background and removing the emission on spatial scales of 16$^{\prime\prime}$, which removes point sources, X-ray jets as well as cores of groups and clusters (for details see e.g. Finoguenov et al. 2015). We have separately searched for SMG halo detection on scales of $4^{\prime\prime}-16^{\prime\prime}$ using \textit{Chandra} data  only, finding none.

 We extract the stacked X-ray flux using a 32$^{\prime\prime}$  diameter aperture centered on the SMG, removing source duplications. 
The selected aperture matches well the expected size of the X-ray emission from groups slightly below the X-ray detection threshold.  The choice of aperture is determined by the source confusion on the flux level of interest, as discussed in Finoguenov et al. (2015). We perform a correction for missing flux when reporting the final values below. Given the redshift dependence of conversion factors, we have split the sample into two, $z<2$ (8 systems) and $z>2$ subsamples (19). To infer the total mass, we calculate the total luminosity, using the K-corrections from the $L-T$ (luminosity -- temperature) relation and apply the $L-M$ (luminosity -- mass) relation of Leauthaud et al. (2010), using the iterative procedure described in Finoguenov et al. (2007). The scaling relation connecting the X-ray luminosity and the total mass at those fluxes (albeit at $z<1.6$) has been verified through the stacked weak lensing analysis as well as clustering in Finoguenov et al. (2015). Furthermore, an agreement between the clustering analysis and X-ray stacking has been presented at $z\sim2$ by Bethermin et al. (2014).

The $z<2$ subsample yields a $5\sigma$ detection with an average flux of $(4.0\pm0.8) \times 10^{-16}$ erg s$^{-1}$ cm$^{-2}$ and  a corresponding   total  mass of M$_{200} = 2.8\times 10^{13}$~\msol \  assuming $z=1.5$. We can rule out a major contribution to the flux from point sources, based on the depth of \textit{Chandra} observations.

The $z>2$ subsample yields a marginally significant signal with average flux of $(1.3\pm0.5) \times 10^{-16}$ erg s$^{-1}$ cm$^{-2}$ and  a corresponding   total mass of M$_{200}=2.5\times 10^{13}$~\msol  \ for redshifts between 2 and 3. Assuming $z=4$ we estimate a total mass of M$_{200}=2.0 \times 10^{13}$~\msol . At those fluxes we cannot rule out the contribution of point sources.

\subsection{Evidence for overdensities around SMGs}
\label{sec:general}

While in the previous section we have analysed the environments of our SMGs at $r\leq1\arcmin$ scales, here we investigate the general environment of the SMGs in our sample, on small and large scales, without requiring that the SMG necessarily resides in the center of the overdensity.  As described in \s{sec:method} \ for each SMG the potential overdensity center was automatically relocated by taking the median RA and Dec of overdense VTA cells surrounding the SMG. This was done within a radius that maximizes the number density of such cells within the encompassing circular area. The new center positions and the radii maximizing the VTA-overdense number densities are indicated in the middle panels in \f{figure:voronoi} , and the distances between the target SMGs and the newly defined overdensity centers are tabulated  in \t{table:lss}  . In  Fig.~\ref{figure:voronoi}  \ we also show 
 $\delta_\mathrm{g}$ as a function of radius out to $r = 12\farcm5$ away from the overdensity center, as well as its significance ($\delta_\mathrm{g}/\sigma_{\delta_\mathrm{g}=0}$). For each SMG the   $\delta_\mathrm{g}/\sigma_{\delta_\mathrm{g}=0}$ threshold above which our simulations yield $\leq20\%$ false detection probability is indicated in the panels. 
 We find 9/23 (39\%) systems within our main-SMG sample, and  4/5 (80\%) in our additional-SMG sample with $\delta_\mathrm{g}/\sigma_{\delta_\mathrm{g}=0}$ higher than the corresponding threshold. Out of the 6 SMGs in the main-SMG sample with  spectroscopic redshifts we find that 3 (50\%) of the systems show evidence for overdensities. 
Combining this with the results of the small-scale overdensity search with central SMGs presented in the previous subsection we find a total of 11/23 (48\%), 3/6 (50\%) and 4/5 (80\%) overdense systems in our main-SMG sample, main-SMG subsample with spectroscopic redshifts assigned to the SMGs, and additional-SMG sample, respectively. The combined results are tabulated in \t{table:all} . The analysis for a narrower redshift width [$\Delta z =z_\mathrm{SMG}\pm 1.5\sigma_{\Delta z/(1+z)} (1+z_\mathrm{SMG})$] yields systematically lower fractions, yet consistent within the Poisson errors (see Appendix~\ref{sec:1.5sig} for details).

\section{Spectroscopic verification of overdensity candidates}
\label{sec:specverif}
  
As our analysis is based on photometric redshifts, spectroscopic redshifts are required in order to confirm each identified 
overdensity. In order to fully assess the entire sample of photometric redshift overdensities reported here 
requires a dedicated spectroscopic campaign of the area surrounding each SMG (as in, e.g. Riechers et al.\ 2014). Though such
a comparison is not yet possible, a large number of spectroscopic redshifts at $z>2$ in the COSMOS field exist mainly from the VUDS (Le F\`evre et al.\ 2015) and the zCOSMOS-deep surveys (Lilly et al., in prep.) from which we can make a preliminary
assessment of the reliability of our photometric redshift measurements in recovering overdense regions. The VUDS spectroscopic 
redshift survey targeted $\sim8,000$ sources ($i^+\la25$) at $2<z\lesssim6$ covering 1~deg$^2$ in three fields (COSMOS, ECDFS, VVDS-02h), 
of which approximately half are in the COSMOS field. The zCOSMOS survey assembled over 20,000 spectra in the COSMOS field in two phases, 
the magnitude limited zCOSMOS-bright survey ($I_\mathrm{AB}<22.5$, Lilly et al.\ 2007, 2009) and the zCOSMOS-deep, with targets selected 
by a variety of colour and magnitude cuts (limited to $i^\prime\la24.5$), from which we rely almost exclusively from the latter. 
The combination of the two surveys puts reliable\footnote{Here we consider flag=X2,X3,X4 to be reliable for both surveys where X=0,1,2, 
see Le F\`evre et al.\ (2015) for further details on the flagging system} spectroscopic redshifts of nearly 4,000 unique galaxies 
at $z>2$ at our disposal across the full COSMOS field. Based on the VUDS and zCOSMOS coverage of the COSMOS field, out of the 28 SMGs analysed 
here, 11\footnote{AzTEC1, AzTEC3, AzTEC5, AzTEC15, AzTEC17b, AzTEC18, Vd-17871, AK03, J1000+0234, AzTEC/C159, Cosbo-3.} are 
in the redshift range where both surveys are sensitive ($z>2$) and their projected surroundings ($R<5\arcmin$) are sampled densely 
enough to \emph{potentially} identify an overdensity.  
 In the previous sections we identified overdensities associated with 8 out of these 11 systems\footnote{AzTEC1, AzTEC3, AzTEC5, AzTEC15, Vd-17871, AK03, J1000+0234, Cosbo-3}. 

For each overdensity candidate at $z>2$ we searched for associated spectroscopic members using a redshift window of 
$z_\mathrm{SMG}\pm 3\sigma_{\Delta_z/(1+z)}(1+z)$ and a $5\arcmin$ radius for the spatial filter centered on the overdensity 
candidate center. This size roughly matches that of proto-clusters found in VUDS (Cucciati et al.\ 2014; Lemaux et al.\ 2014) and
in simulations (e.g. Chiang et al.\ 2013). For known overdensities in VUDS (hereafter ``proto-structures"), the size of the filter 
was shrunk to the bounds of the proto-structure as determined in a method similar to that of Lemaux et al.\ (2014).
The number of spectroscopic members of each overdensity candidate\footnote{This number only included the SMG if its redshift was confirmed 
independently through VUDS/zCOSMOS.} was used to determine the overdensity ($\delta_\mathrm{g}$) by contrasting with an identical measure in 
1,000 proto-structure-sized volumes (i.e. the filter size) in random locations in the field where similar spectroscopic sampling 
existed. For overdensity candidates at $2 < z < 4$, the central redshift of each of the 1,000 realisations was set to a random value between 
$2.2 < z < 4.2$ to determine the density properties of the field at these epochs. For candidates at $z>4$, a random redshift between $4<z<5$ 
was used for each realisation as the number density of VUDS sources begins to decline rapidly above $z\gtrsim4$\footnote{The number density of 
zCOSMOS-faint galaxies begins to decline rapidly at $z\ga2.5$, and therefore this sample is sub-dominant to VUDS at these redshifts.}. The magnitude 
of the overdensity, $\delta_\mathrm{g}$, along with its associated uncertainty was estimated for each overdensity candidate by fitting either  a truncated Gaussian 
or a Poissonian function, the latter  being generally reserved for the higher redshift overdensities where the number of galaxies in the filter is extremely small,  to the distribution of the number of galaxies recovered 
in the 1,000 observations of random spatial and redshift positions in the ranges given above (see Lemaux et al.\ 2014 for more details). 
 
Out of the 11 candidates with the requisite coverage, we find three SMGs (AzTEC5,  J1000+0234, Cosbo-3) in environments 
which are, spectroscopically, significantly overdense systems with respect to the general VUDS+zCOSMOS field (i.e. $\ga2.5\sigma$, $\delta_\mathrm{g}\ga4$, with 
an average overdensity of $\langle \delta_\mathrm{g} \rangle \sim 5$), and three SMGs (AzTEC18, AK03, Vd-17871) in environments which are potentially overdense (i.e. $\ga 2\sigma$, $\delta_\mathrm{g}\ga1.5$). This amounts to 3/8 (38\%), and potentially 5/8 (62\%) of the overdensities found to be statistically significant in the previous sections using our photometric redshift analysis. Note that the spectroscopic analysis  retrieves one additional potential overdensity (AzTEC18), not identified within out photometric-redshift-based approach. However, this overdensity is identified when using a narrower redshift bin (see Appendix~\ref{sec:1.5sig}, and \s{sec:salt} ).

We note that while the set of VUDS/zCOSMOS proto-structures in this 
field comprises a pure sample of genuine overdensities, the sample is not complete even in those areas which we have denoted as 
having the requisite spectroscopic coverage. The selection criteria of both the VUDS and zCOSMOS surveys is such that galaxies with 
stellar population older than $\sim$ 1 Gyr are essentially absent, a population which may be more prevalent in high-density environments 
even at these redshifts (e.g. Kodama et al.\ 2007). In addition, the spatial sampling of the VUDS and zCOSMOS surveys in the area comprising 
the overdensity candidates is non-uniform in nearly all cases (see \f{fig:spec} \ for an example) and is also a strong function of 
redshift, further diminishing our ability to spectroscopically confirm some candidates and limiting the conclusions that can be drawn from 
this analysis. 
Thus, a dedicated spectroscopic study of the remaining overdensity candidates is 
necessary to either confirm or deny their genuineness. 

To-date such spectroscopic studies have  confirmed for example the proto-cluster nature of the structure associated with AzTEC3 (Capak et al.\ 2010; Riechers et al.\ 2014). Furthermore, Karim et al.\ (in prep.) show that the SMG Vd-17871 has a spectroscopically identified close companion, in accordance with the above estimated elevated galaxy density in its surroundings.
Lastly, the structure associated with the SMG Cosbo-3, initially identified by Aravena et al.\ (2010), here spectroscopically identified has been independently confirmed, and studied in more detail based on optical and mm- spectroscopic, and X-ray data in Casey et al.\ (2015) and Wang et al.\ (in prep.). In summary, we conclude that 4 (AzTEC5, J1000+0234 and Cosbo-3 based on VUDS and zCOSMOS data and the structure associated with AzTEC3 via dedicated follow-up) out of 8 overdensity candidates tested here (i.e. $\sim 50\%$) have been independently spectroscopically identified. Adding to this the two spectroscopically identified ambiguous cases (AK03 and Vd-17871) this amounts to 6/8, i.e.\ 75\% of the tested systems.

\section{Discussion}

\subsection{Fraction of SMGs in overdense environments}
\label{sec:nature}

Within our main-SMG sample we have identified 5/23 overdensities showing evidence of small-scale overdensities, and six additional overdensities showing statistically significant overdensities with the SMG off-center. This yields a fraction of 11/23 (48\%) SMGs in our main-SMG sample located in overdense environments. In \s{sec:data} \ we estimated that in addition to these 23 SMGs drawn from a S/N-limited sample, 8 further sources with only lower-redshift limits are present (i.e., 16 in total out of which 8 are expected to be spurious). If these 8 sources with lower redshift limits on average have the same properties as the sources analyzed, then the estimated fraction of SMGs occupying overdensities does not change. In the two extreme cases, however, where all 8 systems would  either occupy or not occupy  overdensities this fraction would then range from 35\% to 61\%. 
Within our additional-SMG sample, containing Cosbo-3 at $z_\mathrm{spec}=2.49$, as well as AzTEC/C159, Vd-17871, J1000+1234, and AK03 at $z_\mathrm{spec}>4.5$ we find  4/5 (80\%) systems associated with overdensitites.

Only the spectroscopically verified Cosbo-3 proto-cluster ($z=2.49$) has been directly detected in the X-rays, and has an estimated X-ray mass of $M_\mathrm{200}=4\times10^{13}$~\msol \ (Wang et al., in prep.). Stacking the X-ray data to search for extended halos occupied by the SMGs in our sample in \s{sec:smallscale} \ we derived 
for the $z<2$ subsample a  total  mass of M$_{200} = 2.8\times 10^{13}$~\msol  \ assuming $z=1.5$. For the $z>2$ subsample we found  a   total mass of M$_{200}=2-2.5\times 10^{13}$~\msol  \ given the span in assumed redshift (from $z=2$ to 4). This yields dark-matter halo masses of the SMGs analyzed here in the range of M$_{200} = 2-3\times 10^{13}$~\msol , in reasonable agreement with  dark-matter halo masses   derived by Hickox et al.\ (2012) based on a clustering analysis of LABOCA SMGs in the ECDFS field (log($M_\mathrm{halo}[h^{-1}\mathrm{M_\odot}])=12.8^{+0.3}_{-0.5}$).

If we split our SMG sample in two redshift bins and use the errors on measurements based on a small number of events (Gehrels 1986), at $z<3$ we find $36^{+17}_{-15}$\% ($40^{+16}_{-15}$\%)   overdensities in the main- (main+additional-) SMG sample, while this fraction increases to $67^{+18}_{-22}$\% ($69^{+14}_{-18}$\%) at $z\geq3$. Although the errors on these fractions are large, this could suggest that the occurrence of SMGs in overdensities at high redshift is higher. This is consistent with
the picture that massive, highly star forming
galaxies  can only be formed in the highest peaks of the
density field tracing the most massive dark matter haloes at the
earliest epochs ($z\geq3$; e.g., Kaufman et al.\ 1999; Springel et al.\
2005; see also discussion by Riechers et al.\
2013). At later times, cosmic structure may have matured
sufficiently that more modest overdensities correspond to sufficiently
massive haloes to form SMGs (e.g., Hopkins et al.\ 2008;
Gonzalez et al.\ 2010; Amblard et al.\
2013). However, a more detailed theoretical exploration of this
aspect in the future is desirable in cosmological
simulations, as well as more constraining observational data, as discussed below.

\subsection{A grain of salt}
\label{sec:salt}

We have performed a statistical analysis of the environments of two samples of SMGs,  i) a well-defined  sample obtained via interferometric (SMA and PdBI) follow-up of a S/N-limited sample identified at 1.1~mm with AzTEC/JCMT over a 0.15~deg$^2$ COSMOS subfield with photometric or spectroscopic redshifts assigned to the SMGs, ii) five additional SMGs with secure spectroscopic
redshifts, one at $z_{\rm spec} = 2.49$, and four at  $z_{\rm spec} > 4.5$. The analysis was performed using uniform selection and overdensity identification criteria for all SMGs in our sample, thus rendering a robust statistical result on the fraction of SMGs occupying overdense environments under the assumptions made. We here discuss a few caveats that may have an effect on the results of individual  overdensities hosting our SMGs.

The analysis performed here is based on a galaxy sample limited to $i^+\leq25.5$, which assures the most accurate photometric redshifts in the COSMOS field (see Sect.~\ref{sec:photz}). It, thus, does not take into account fainter galaxies that may be present in vicinity of, and be gravitationally bound to the SMGs analyzed here. 
Furthermore, a 20\% incompleteness  in the photometric redshift catalog at $z>3.5$ (estimated in \s{sec:photz} , and due to the well known degeneracy in the spectral energy distribution fitting procedure between the Balmer-4\,000~$\AA$ and Ly$\alpha$-1\,215~$\AA$ breaks) may bias the results at high redshift low. On the other hand, catastrophic failures in photometric redshift solutions, estimated to amount  to $<15\%$ for $3<z<6$ (Laigle et al.\ 2016), could bias the results in some cases high. Furthermore, wrongly assigned redshifts to  SMGs, and multiplicity of the SMGs at scales $\lesssim2\arcsec$ may bias the results in an unpredictable manner (see \cite{koprowski2014}; Aravena et al., in prep.; Miettinen et al., in prep).

Other caveats that may bias the derived overdensity fraction low are the width and centering of the redshift bin used for the analysis. The SMG redshift may not necessarily be coincident with the host overdensity's redshift, nor does the extent of the overdensity in the z-direction need to be of certain width, but may vary from case-to-case. For example, as shown in Appendix~\ref{sec:1.5sig} using a narrower redshift bin [$\Delta z=z_\mathrm{SMG}\pm 1.5\sigma_{\Delta z/(1+z)} (1+z_\mathrm{SMG})$] we identify three additional overdensities\footnote{AzTEC17a, AzTEC18, and AzTEC/C159, out of which AzTEC18 has been verified also as a potentially spectroscopic overdensity (see \s{sec:specverif} ).}. Adding these to those identified above would then yield
 a total of 13/23 (56\%), and 5/5 (100\%) overdensities in our main-, and additional-SMG samples, respectively, and an estimated range of 42-68\% of SMGs occupying overdensities in our S/N-limited 1.1mm-selected SMG sample (accounting for the 
  expected spurious fraction in the sample).
Furthermore, as already
shown by Chiang et al.\ (2013) using photometric redshifts to search for overdensities dilutes the $\delta_\mathrm{g}$ signal (see their Fig.\ 13), and we are, thus, likely to be biased in the present analysis towards detecting the most prominent overdensities only.

With the data in-hand it is impossible to address all the above described caveats in a robust way. The optimal solution would be a dedicated spectroscopic follow-up of all galaxies surrounding SMGs drawn from a well-selected, S/N-limited SMG sample within which each SMG has been assigned a secure spectroscopic redshift. With facilities like ALMA such samples may become reality in the near future.

\section{Summary and conclusions}

We have carried out a search for galaxy overdensities around a sample of \ntot \ SMGs in the COSMOS field lying at redshifts of $z\simeq0.8-5.3$. We have searched for overdensities using the COSMOS photometric redshifts based on over 30 UV-NIR photometric measurements including the new UltraVISTA Data Release 2 and \textit{Spitzer}/SPLASH data, and reaching an accuracy of $\sigma_{\Delta z/(1+z)}=0.0067~(0.0155)$ at $z<3.5~(>3.5)$. To identify overdensities we have applied the Voronoi tessellation analysis, and derived the $\delta_{\rm g}$ redshift-space overdensity estimator. Within the search it was not required that the SMG is in the center of the potential overdensity. The approach was tested and validated via simulations,  X-ray detected groups/clusters, and spectroscopically verified overdensities which show that even with photometric redshifts in the COSMOS field we can efficiently retrieve overdensities out to $z\approx5$. Our main conclusions, based on this analysis are summarised as follows.

\begin{enumerate}

\item Out of \ntot \ SMGs in our sample 15 show evidence for associated galaxy overdensities. Separating these into our main-SMG sample of 23 SMGs drawn from a S/N-limited 1.1mm AzTEC/JCMT SMG sample, and our additional-SMG sample consisting of 5 SMGs with spectroscopic redshifts at $z=2.49$, and $z>4.5$ we find evidence for 11/23, and 4/5 systems, respectively, occupying overdense environments. Considering all sources in the S/N-limited AzTEC/JCMT sample, and taking into account the fraction of spurious sources, our results suggest that 35-61\% of the SMGs in the 1.1~mm S/N-limited sample occupy overdense environments.
As discussed in \s{sec:salt} , these fractions are likely to be lower limits, as our analysis based on photometric redshifts may have missed the less prominent overdensities.\\

\item Performing an X-ray stacking analysis using  a $32\arcsec$ aperture centered at the SMG positions for our $z<2$ subsample, we have found  an average 0.5-2~keV X-ray flux of $(4.0\pm0.8) \times 10^{-16}$ erg s$^{-1}$ cm$^{-2}$ and  a corresponding total mass of M$_{200} = 2.8\times 10^{13}$~\msol \ assuming $z=1.5$, and for the 
 $z>2$ subsample  an average flux of $(1.3\pm0.5) \times 10^{-16}$ erg s$^{-1}$ cm$^{-2}$ and  a corresponding total mass of M$_{200} =2.5\times 10^{13}$~\msol \ for redshifts between 2 and 3 (M$_{200} =2\times 10^{13}$~\msol \ assuming $z=4$). 

\end{enumerate}

Our results suggest a higher occurrence of SMGs occupying overdense environments at $z\geq3$, than at $z<3$. This may be understood if massive, highly star forming galaxies  can only be formed in the highest peaks of the density field tracing the most massive dark matter haloes at early cosmic epochs, while at later times cosmic structure may have matured sufficiently that more modest overdensities correspond to sufficiently massive haloes to form SMGs. Further theoretical and observational efforts are needed to investigate this further.

\begin{acknowledgements}

This research was funded by the European Union's Seventh Framework programme 
under grant agreement 337595 (ERC Starting Grant, 'CoSMass'). AF wishes  to acknowledge Finnish Academy award, decision 266918.
This work was supported in part by the National Science Foundation under Grant No. PHYS-1066293 and the hospitality of the Aspen Center for Physics. This work is partly based on data products from observations made with ESO Telescopes at the La Silla Paranal 
Observatory under ESO programme ID {\tt 179.A-2005} and and {\tt 185.A-0791} and on data 
products produced by TERAPIX and the Cambridge Astronomy Survey Unit on behalf of the 
UltraVISTA consortium. This research has made use of NASA's Astrophysics Data System, and the NASA/IPAC Infrared Science Archive, 
which is operated by the JPL, California Institute of Technology, under contract with the NASA. 
We greatfully acknowledge the contributions of the entire COSMOS collaboration consisting of more than 100 scientists. 
More information on the COSMOS survey is available at {\tt http://www.astro.caltech.edu/$\sim$cosmos}. We thank the VUDS team for making the data in the COSMOS field
     available prior to public release

\end{acknowledgements}

\clearpage
\begin{table}
\renewcommand{\footnoterule}{}
\caption{
Small-scale overdensity search results. The systems with statistically significant overdensities are indicated in bold-faced.
}
{\small
\begin{minipage}{1\columnwidth}
%\centering
\label{table:smallscale}
\begin{tabular}{l l c c c c c }
\hline
name & radius & $N_\mathrm{r}$ & Poisson probability &  false detection \\ 
    & [$\arcmin$] & & $p(\geq N_\mathrm{r}, n_\mathrm{r})$&  probability, P$_\mathrm{FD}$ \\ 
\hline 
   {\bf AzTEC1 } & 0.5 & 2 & 0.055 & 0.031 \\
    AzTEC2 & 0.5 & 2 & 0.177 & 0.538 \\
    {\bf AzTEC3 } & 0.5 & 2 & 0.007 & 0.000 \\
    {\bf AzTEC4 } & 1.0 & 2 & 0.070 & 0.046 \\
    {\bf AzTEC5 } & 0.5 & 3 & 0.094 & 0.019 \\
    AzTEC7 & 1.0 & 4 & 0.193 & 0.276 \\
    AzTEC8 & 0.5 & 2 & 0.091 & 0.096 \\
    AzTEC9 & 1.0 & 3 & 0.200 & 0.919 \\
   AzTEC10 & 0.5 & 2 & 0.119 & 0.159 \\
   AzTEC11 & 1.0 & 5 & 0.200 & 0.753 \\
   AzTEC12 & 1.0 & 3 & 0.190 & 0.426 \\
 AzTEC14-W & 1.0 & 7 & 0.200 & 0.729 \\
   AzTEC15 & 0.5 & 2 & 0.091 & 0.094 \\
  AzTEC17a & 0.5 & 4 & 0.184 & 0.166 \\
  AzTEC17b & 1.0 & 3 & 0.162 & 0.142 \\
   AzTEC18 & 0.5 & 2 & 0.102 & 0.116 \\
  AzTEC19a & 1.0 & 4 & 0.183 & 0.151 \\
  AzTEC19b & 1.0 & 6 & 0.200 & 0.915 \\
  AzTEC21a & 1.0 & 3 & 0.191 & 0.437 \\
  AzTEC21b & 1.0 & 2 & 0.194 & 0.758 \\
   {\bf AzTEC23 } & 0.5 & 5 & 0.157 & 0.039 \\
  AzTEC26a & 1.0 & 2 & 0.190 & 0.669 \\
  AzTEC29b & 0.5 & 2 & 0.166 & 0.437 \\
  \hline
 {\bf   Cosbo-3 } & 0.5 & 4 & 0.104 & 0.003 \\
{\bf J1000+0234 } & 0.5 & 2 & 0.046 & 0.022 \\
 AzTECC159         & -- & -- & -- & -- \\
  {\bf Vd-17871 } & 0.5 & 2 & 0.040 & 0.018 \\
      AK03         & -- & -- & -- & -- \\
  \hline 
\end{tabular} 
\end{minipage} }
\end{table}

  \begin{table}
\renewcommand{\footnoterule}{}
\caption{
Large-scale overdensity search results. The systems with statistically significant overdensities are indicated in bold-faced.
}
{\small
\begin{minipage}{1\columnwidth}
%\centering
\label{table:lss}
\begin{tabular}{l l c }
\hline
SMG & SMG  \\ 
    &  distance [$\arcmin$]  \\
\hline 
 {\bf   AzTEC1 } & 0.281 \\
    AzTEC2  & 1.673 \\
 {\bf     AzTEC3 } &  0.020 \\
    AzTEC4 & 11.525 \\
  {\bf    AzTEC5 } &  0.000 \\
    AzTEC7 &  1.588 \\
  {\bf    AzTEC8 } &  0.066 \\
    AzTEC9  & 2.695 \\
 {\bf    AzTEC10 }  & 0.958 \\
   AzTEC11 &  0.792 \\
 {\bf    AzTEC12 }  & 0.697 \\
 AzTEC14-W  & 1.952 \\
  {\bf   AzTEC15 } &  0.212 \\
  AzTEC17a  & 1.406 \\
  AzTEC17b &  1.389 \\
   AzTEC18  & 1.269 \\
  AzTEC19a  & 0.875 \\
  AzTEC19b  & 1.277 \\
 {\bf   AzTEC21a }  & 1.424 \\
  AzTEC21b  & 1.239 \\
   AzTEC23  & 1.270 \\
  AzTEC26a &  2.424 \\
 {\bf   AzTEC29b}  &  0.549 \\
  \hline
  {\bf   Cosbo-3}  &  0.066 \\
 {\bf J1000+0234 } &  0.000 \\
 AzTECC159  & 2.199 \\
 {\bf   Vd-17871}  &  0.000 \\
   {\bf     AK03 } &  1.223 \\
\hline 
\end{tabular} \\

\end{minipage} }
\end{table}

  \begin{table*}
\renewcommand{\footnoterule}{}
\caption{
Overdensity search results  based on small- and large-scale analyses. Only sources associated with statistically significant overdensities are listed. Source redshifts and assumed distances between the SMG and overdensity center are also given. Spectroscopically verified systems (as described in detail in \s{sec:specverif} ) are indicated in bold-faced.
}
{\small
\begin{minipage}{1\columnwidth}
%\centering
\label{table:all}
\begin{tabular}{l l c c c c c}
\hline
SMG & redshift & small-scale  & large-scale  & SMG & spectroscopically \\ 
    & &  overdensity & overdensity & distance [$\arcmin$]  & verified \\
\hline 
AzTEC1 &       4.3415\tablefootmark{a} &  $\surd$ & $\surd$ & 0.281 & -- \\  
AzTEC3 &       5.298\tablefootmark{a} & $\surd$ & $\surd$ & 0.027 & $\surd$\\ 
AzTEC4 &       4.93 &    $\surd$& --& 11.525 & -- \\ 
AzTEC5 &       3.05 &  $\surd$ & $\surd$ & 0.000 & $\surd$\\  
AzTEC8 &       3.179\tablefootmark{a} & & $\surd$ &  0.066 & --\\  
AzTEC10 &       2.79 &   -- & $\surd$ & 0.958 & -- \\
AzTEC12 &       2.54 &  -- &$\surd$ & 0.697 & -- \\  
AzTEC15 &       3.17 &   -- & $\surd$ & 0.212 & -- \\   
AzTEC21a &       2.6 &    -- & $\surd$ & 1.424 & --\\
AzTEC23 &       1.6 &    $\surd$ & -- & 0.000 & -- \\  
AzTEC29b &       1.45 &   -- & $\surd$ & 0.549 & --\\
\hline
Cosbo-3 &       2.49\tablefootmark{a} &    $\surd$& $\surd$ & 0.000 & $\surd$\\  
J1000+0234 &       4.542\tablefootmark{a} &   $\surd$ &$\surd$ &  0.000 & $\surd$\\  
Vd-17871 &       4.622\tablefootmark{a} & $\surd$ &$\surd$ &0.000 & $\surd$\tablefootmark{b} \\    
AK03&       4.757\tablefootmark{a} & -- & $\surd$  &   1.223 & $\surd$\tablefootmark{b} \\  
\hline 
\end{tabular} \\
\tablefoottext{a}{Spectroscopic redshift}\\
\tablefoottext{b}{Possible verification}
\end{minipage} }
\end{table*}

\begin{figure*}
\begin{center}
\includegraphics[width=0.31\textwidth]{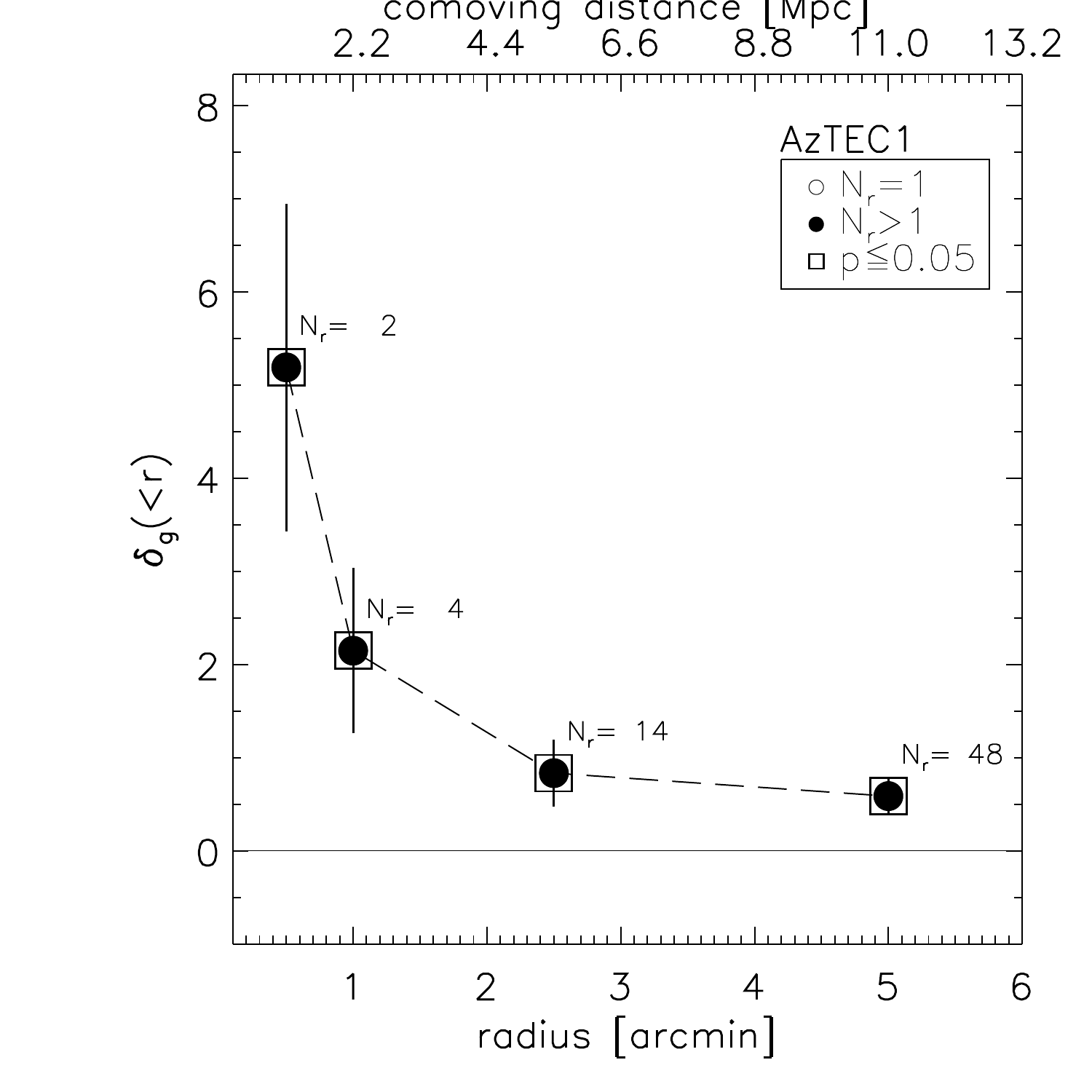}
\includegraphics[width=0.31\textwidth]{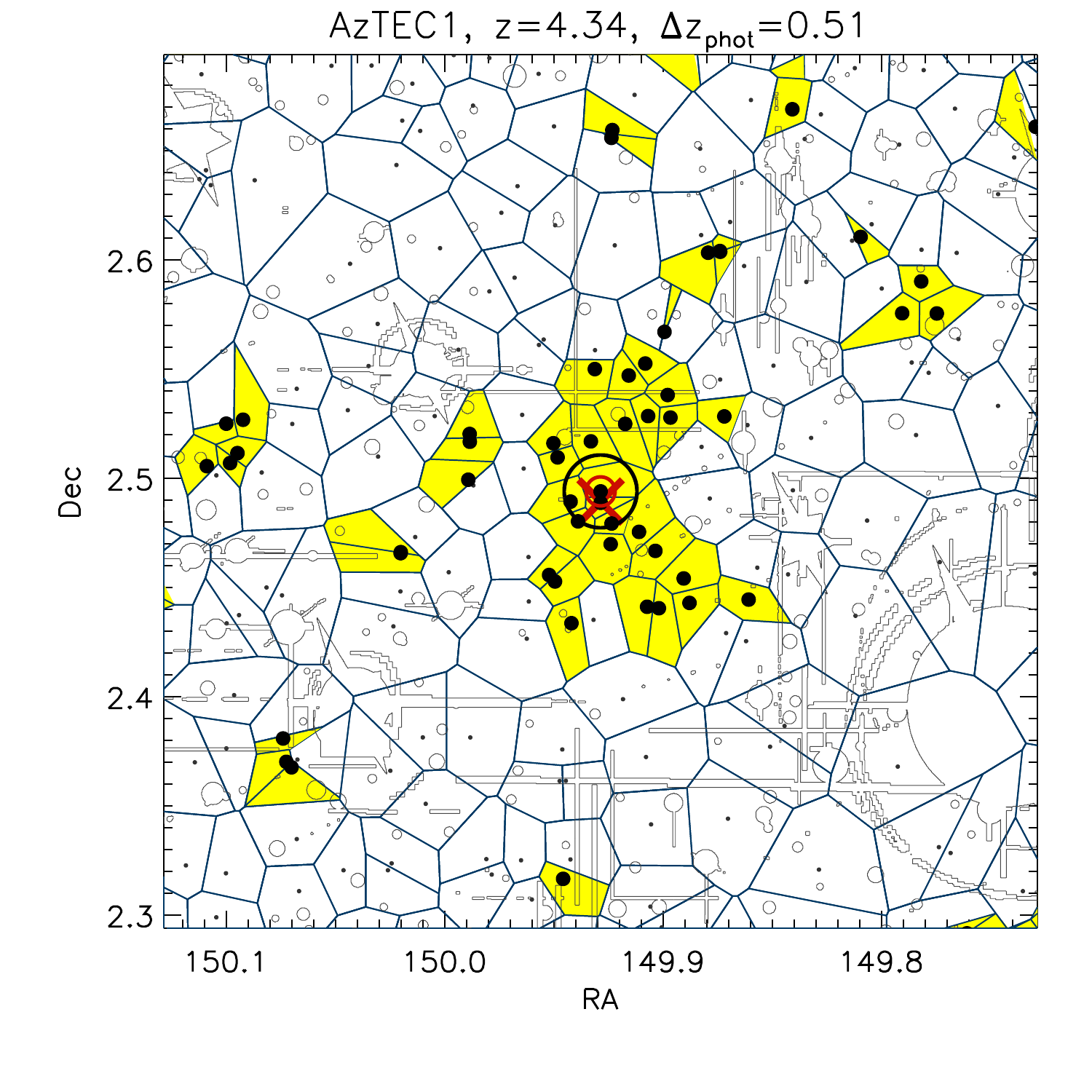}
\includegraphics[width=0.31\textwidth]{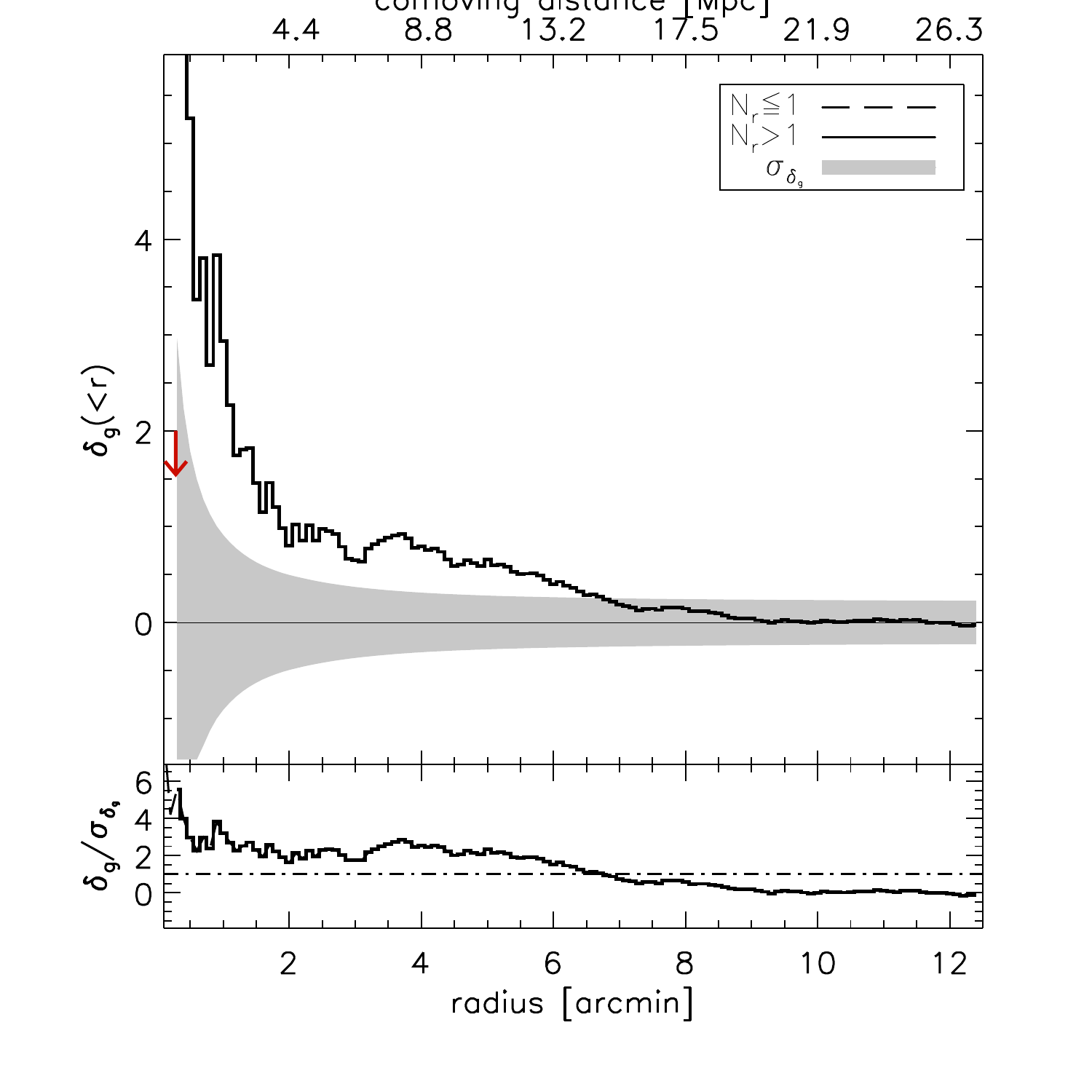}\\
\includegraphics[width=0.31\textwidth]{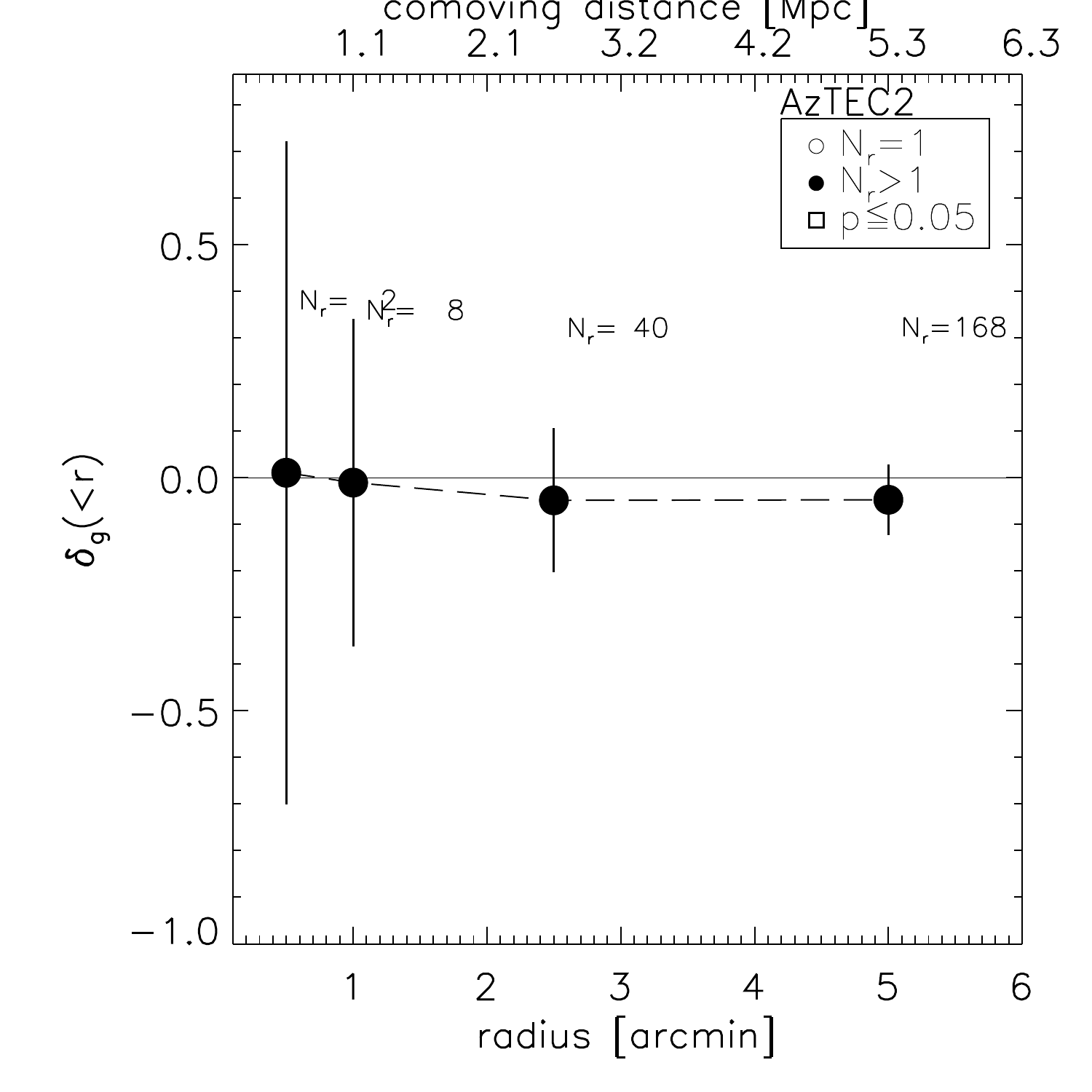}
\includegraphics[width=0.31\textwidth]{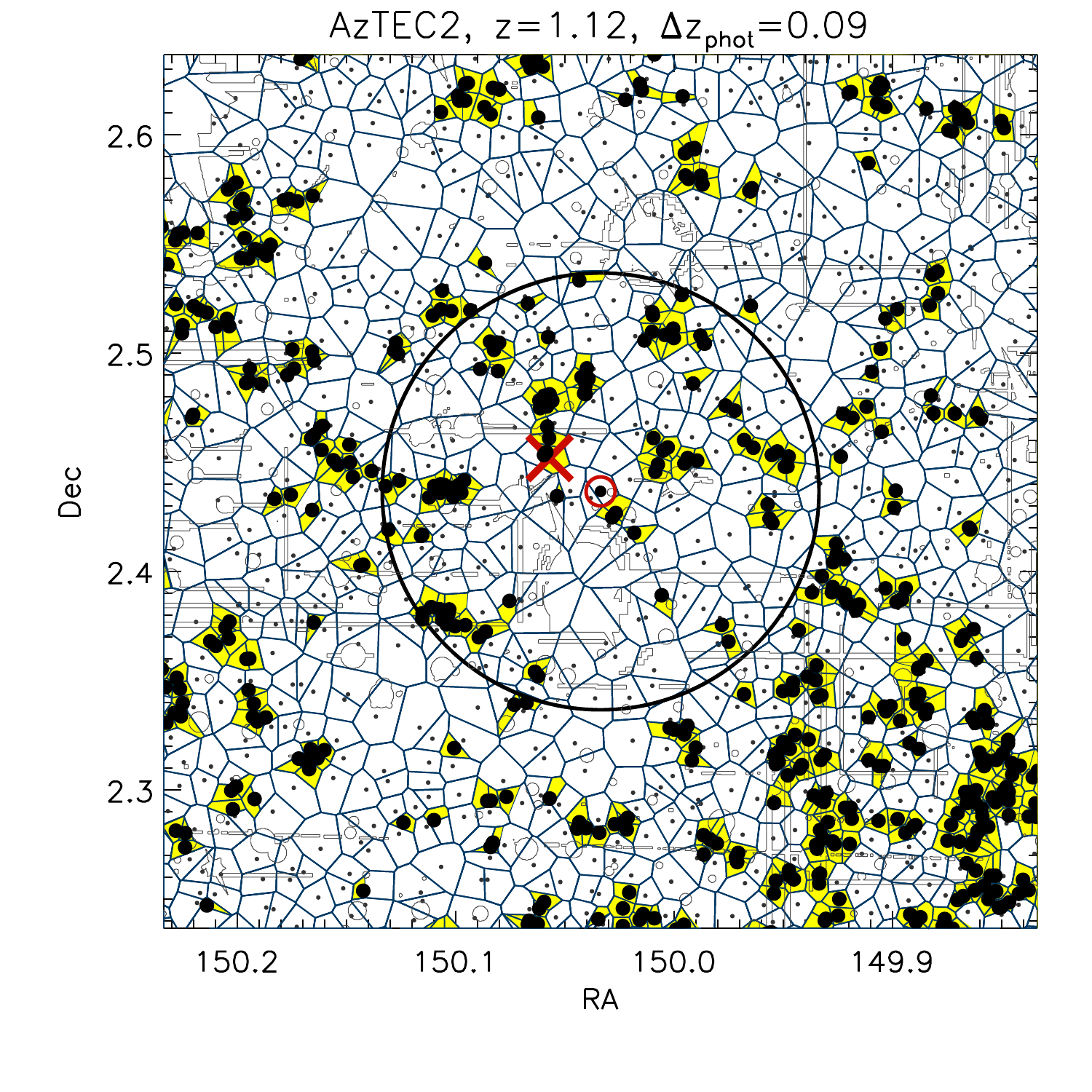}
\includegraphics[width=0.31\textwidth]{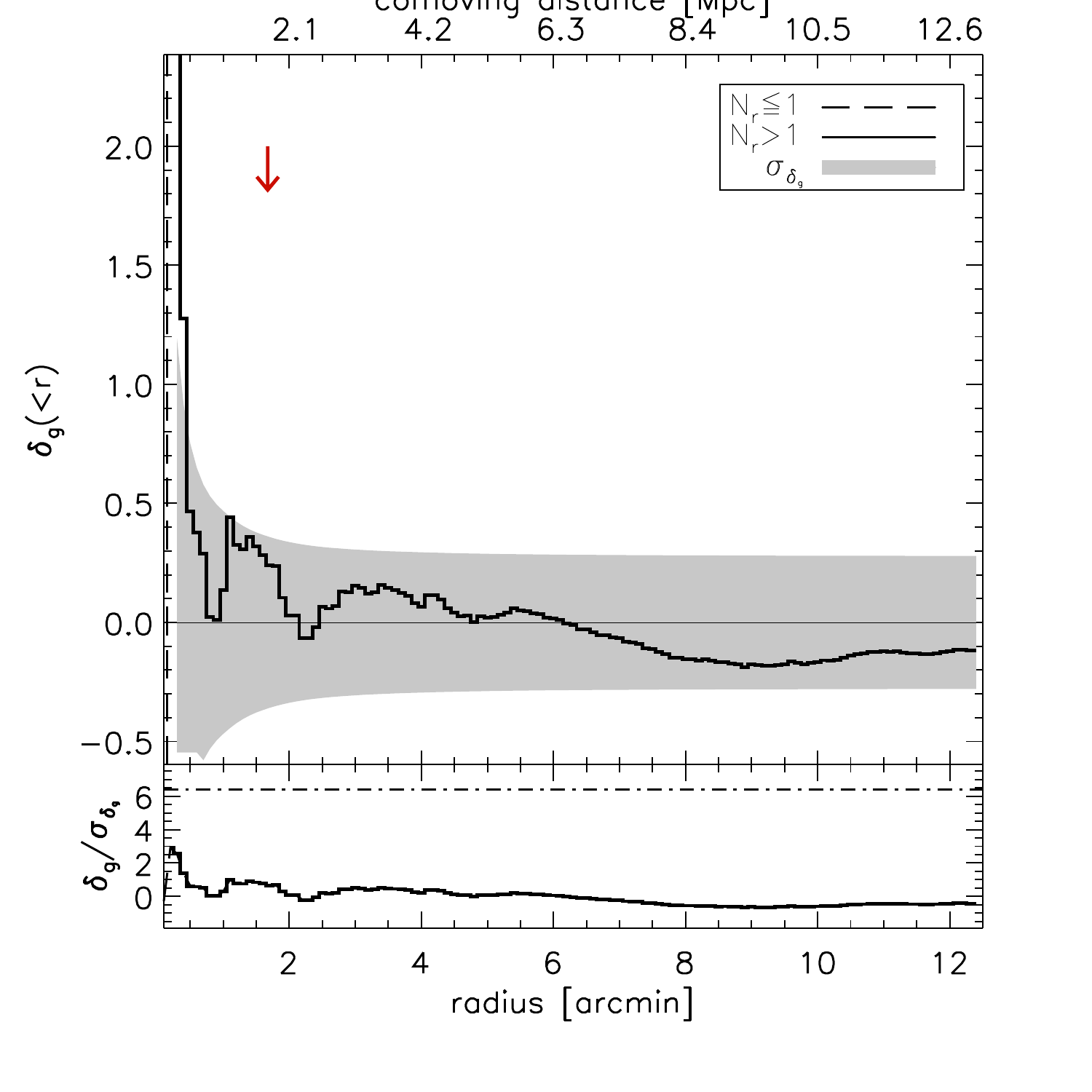}\\
\includegraphics[width=0.31\textwidth]{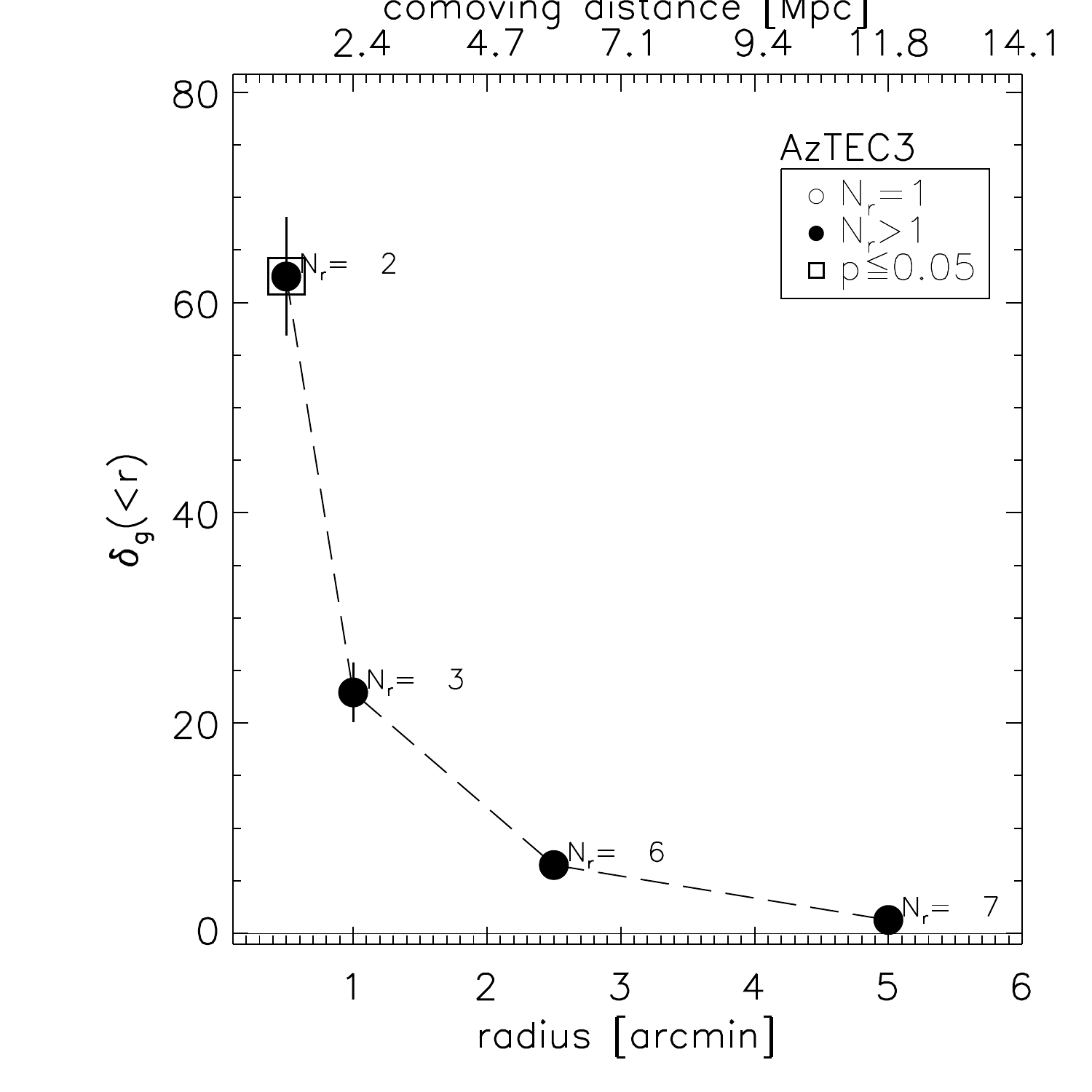}
\includegraphics[width=0.31\textwidth]{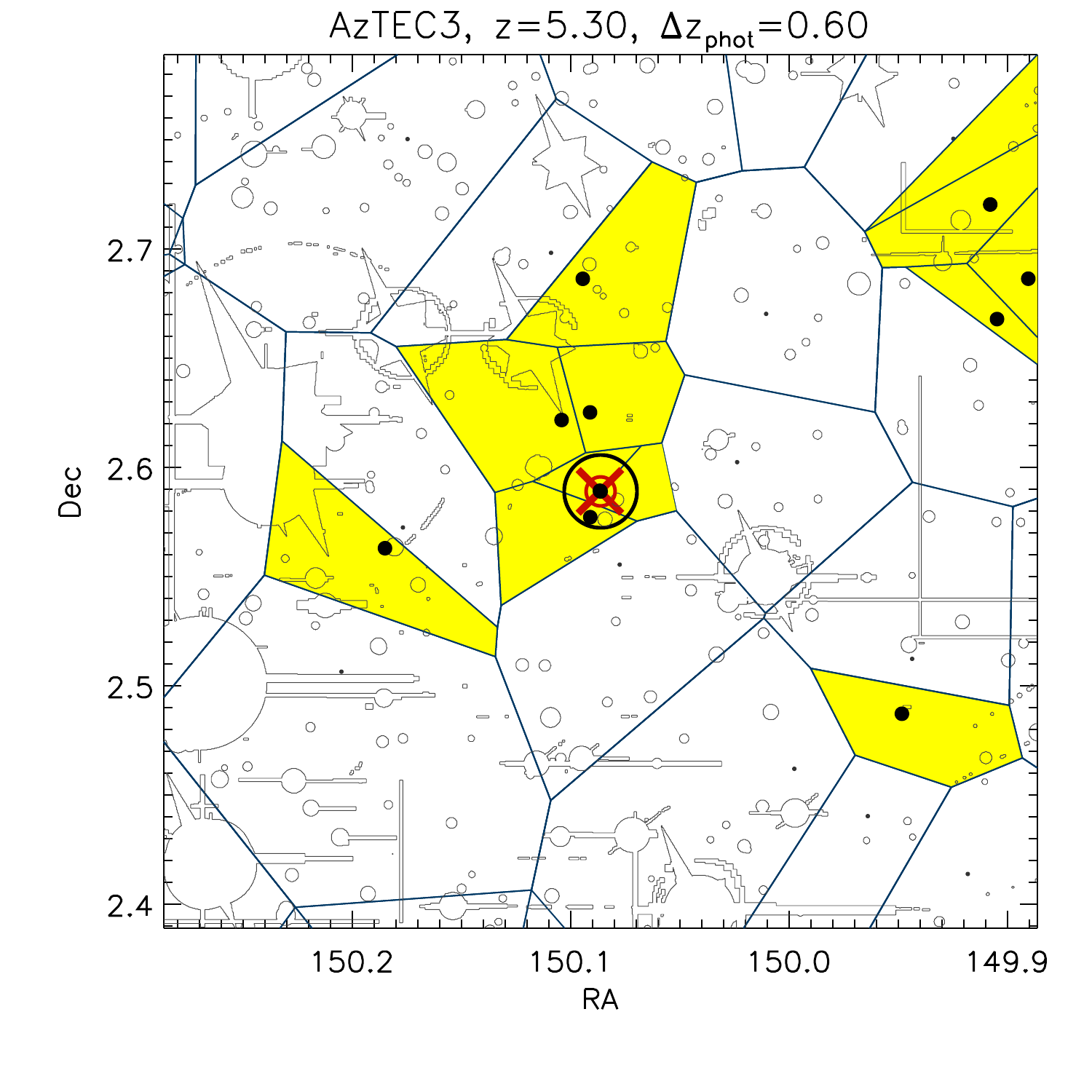}
\includegraphics[width=0.31\textwidth]{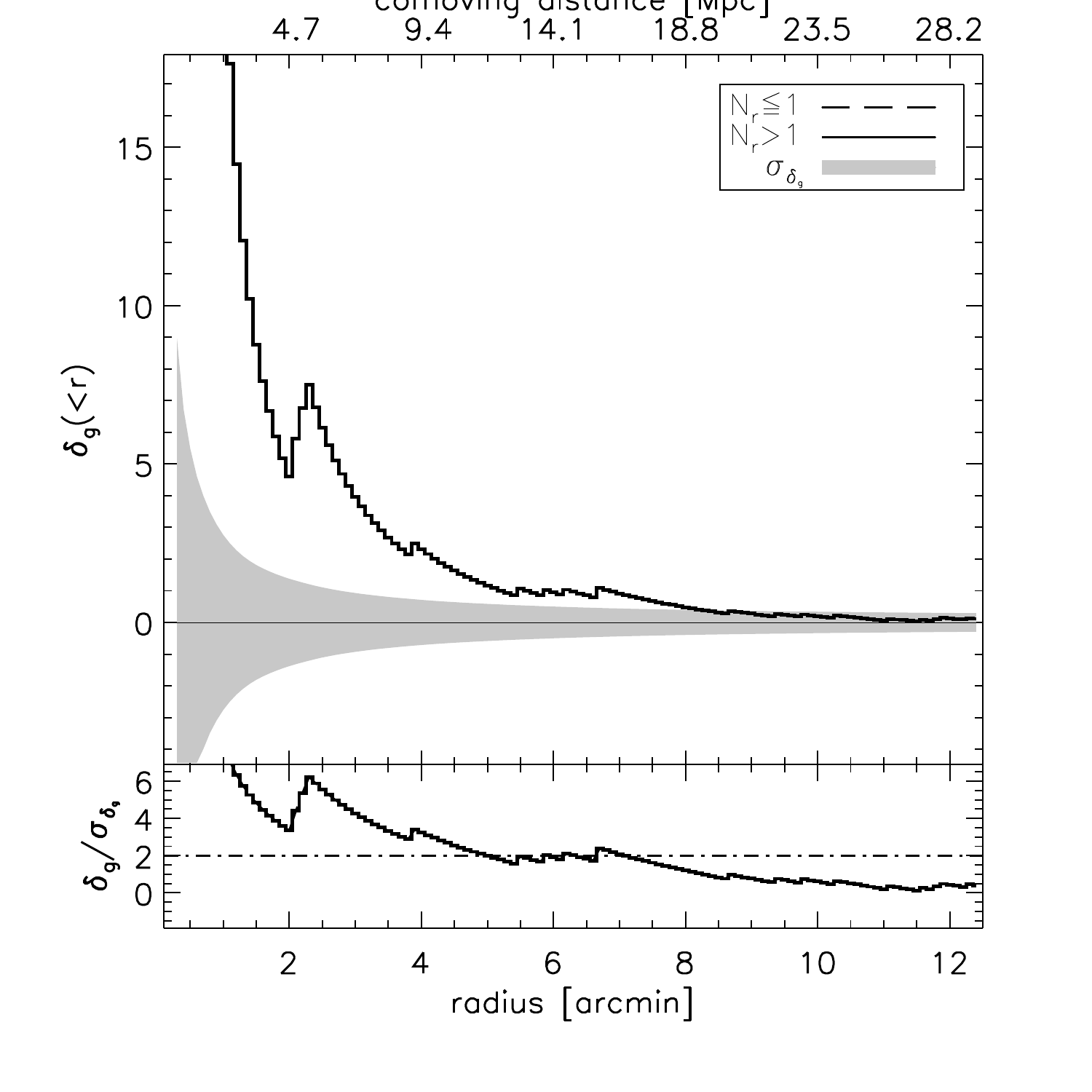}\\
\caption{
Each row corresponds to one SMG.
\textit{Left panel}: $\delta_{\rm g}$ versus radius, centered at the target SMG (see \s{sec:sss} ). The errors shown are Poisson errors. The number of sources found in a circular area ($N_\mathrm{r}$), and the  $\delta_{\rm g}=0$ line are indicated. Also indicated are the points with Poisson probability $p\leq0.05$, as listed in the legend (see \s{sec:sss} \ for details).
\textit{Middle panel:} Voronoi tessellation analysis (VTA; blue lines) shown around each target SMG indicated by the thick red circle. Overdense VTA cells are colored yellow. The area and density of the cells depends on the number of galaxies in the given redshift range ($\Delta z_\mathrm{phot}$), indicated above the panel. The overdensity center is shown by the red cross, and the radius within which it was computed is indicated by the black circle (see \s{sec:vta} \ for details).
Regions outlined by grey lines and curves indicate the masked areas in the COSMOS photometric catalogue. 
\textit{Right panel}: The top panel shows $\delta_{\rm g}$ versus radius, centered at the overdensity (marked by the red cross in the middle panel). The SMG's projected distance from the center is indicated by the red downward pointing arrow. The horizontal line marks the value $\delta_{\rm g}=0$. The meaning of various symbols/linestyles is listed in the legend. The bottom panel shows the significance of the overdensity ($\delta_\mathrm{g}/\sigma_{\delta_\mathrm{g}=0}$) as a function of radius. The dashed horizontal line designates the threshold beyond which the false detection probability is $\leq20\%$ (see \s{sec:lss0} \ for details).
}
\label{figure:voronoi}
\end{center}
\end{figure*}

\addtocounter{figure}{-1}
\begin{figure*}
\begin{center}
\includegraphics[width=0.31\textwidth]{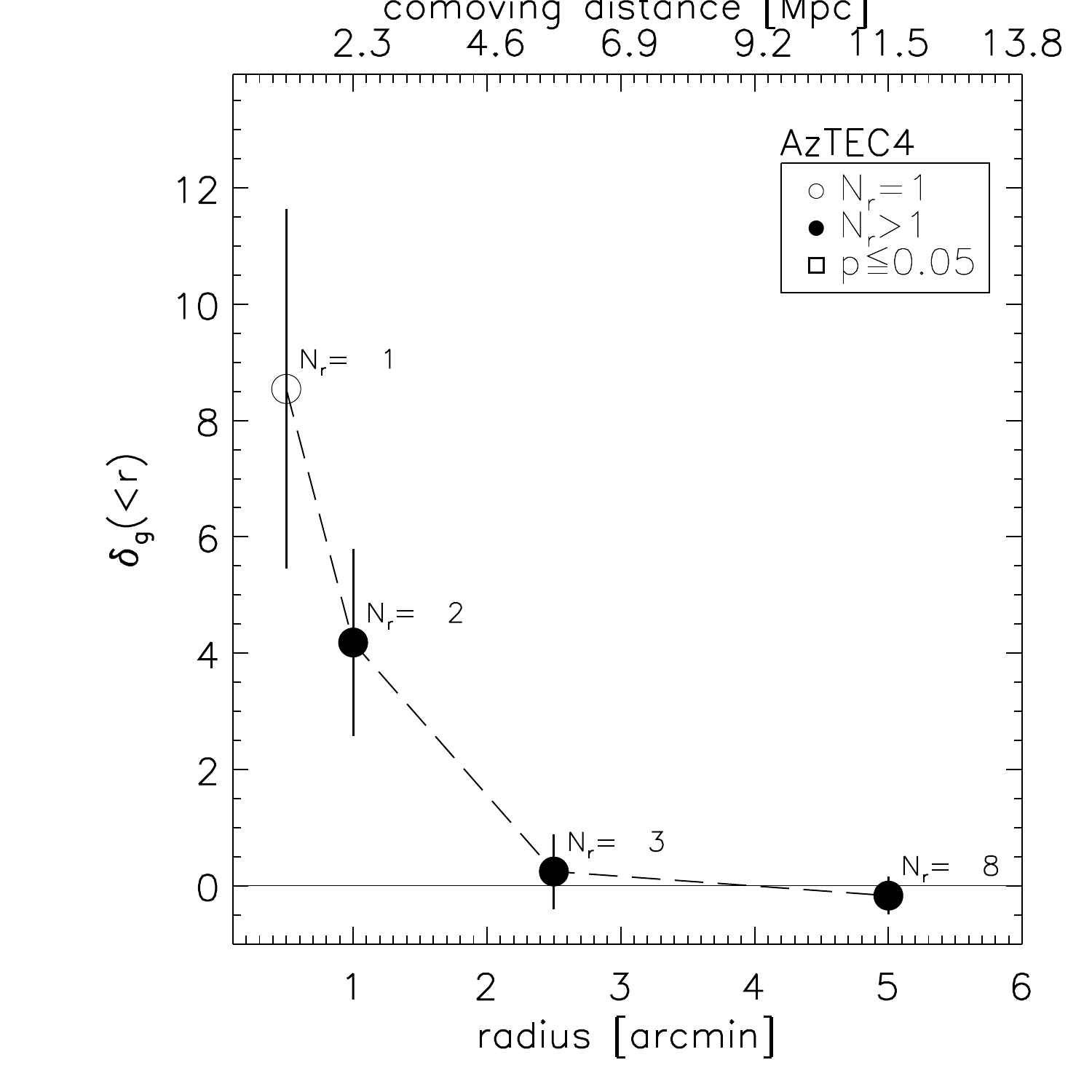}
\includegraphics[width=0.31\textwidth]{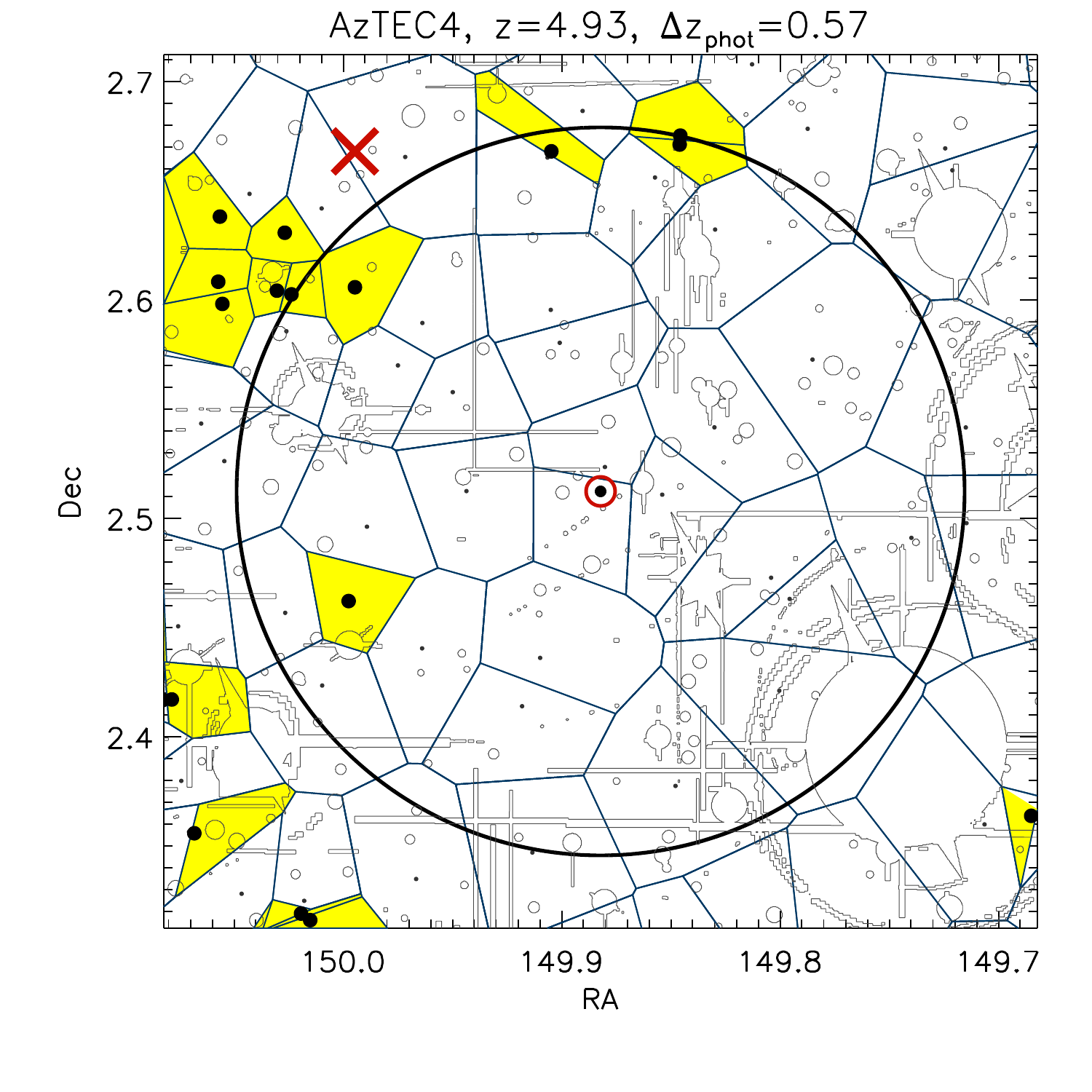}
\includegraphics[width=0.31\textwidth]{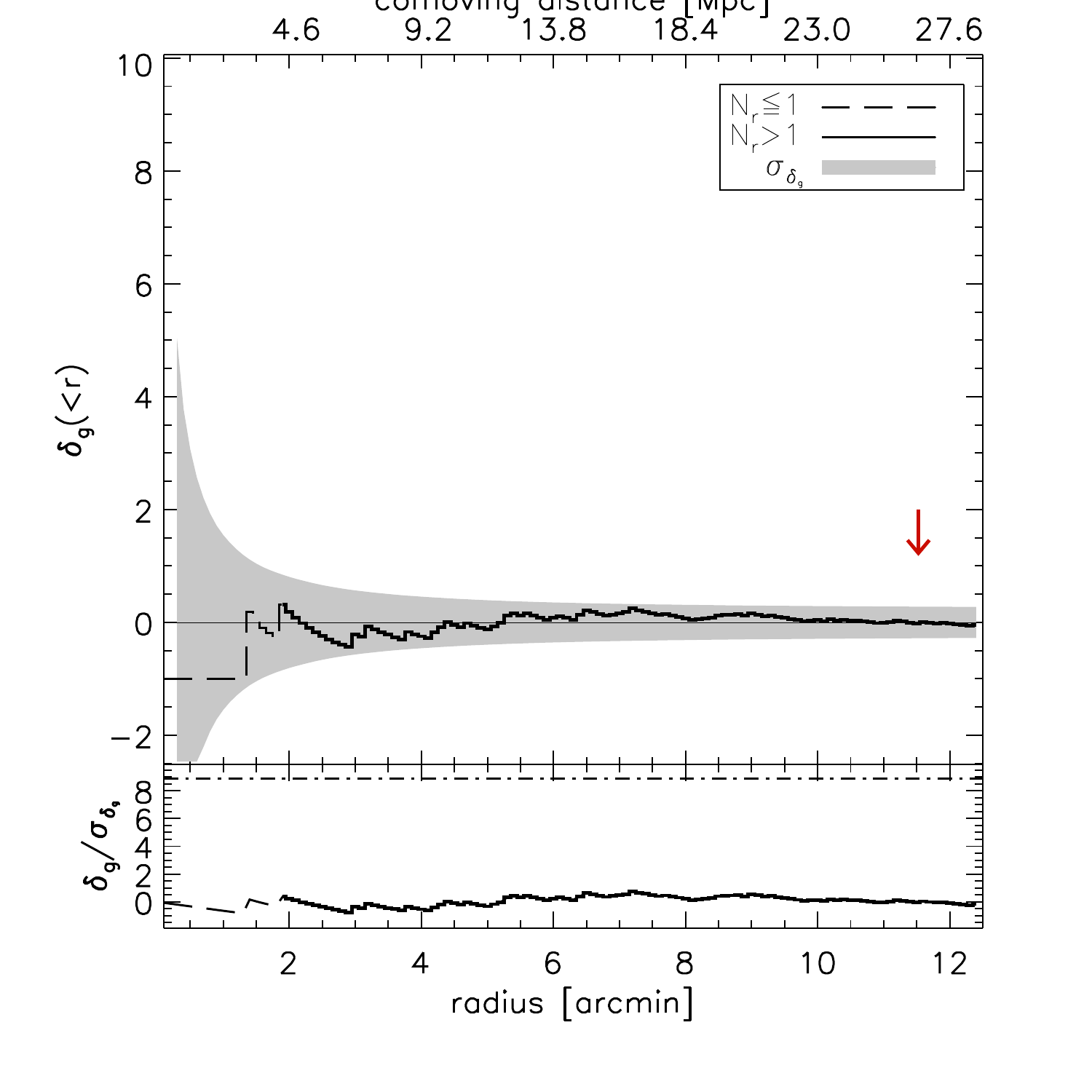}\\
\includegraphics[width=0.31\textwidth]{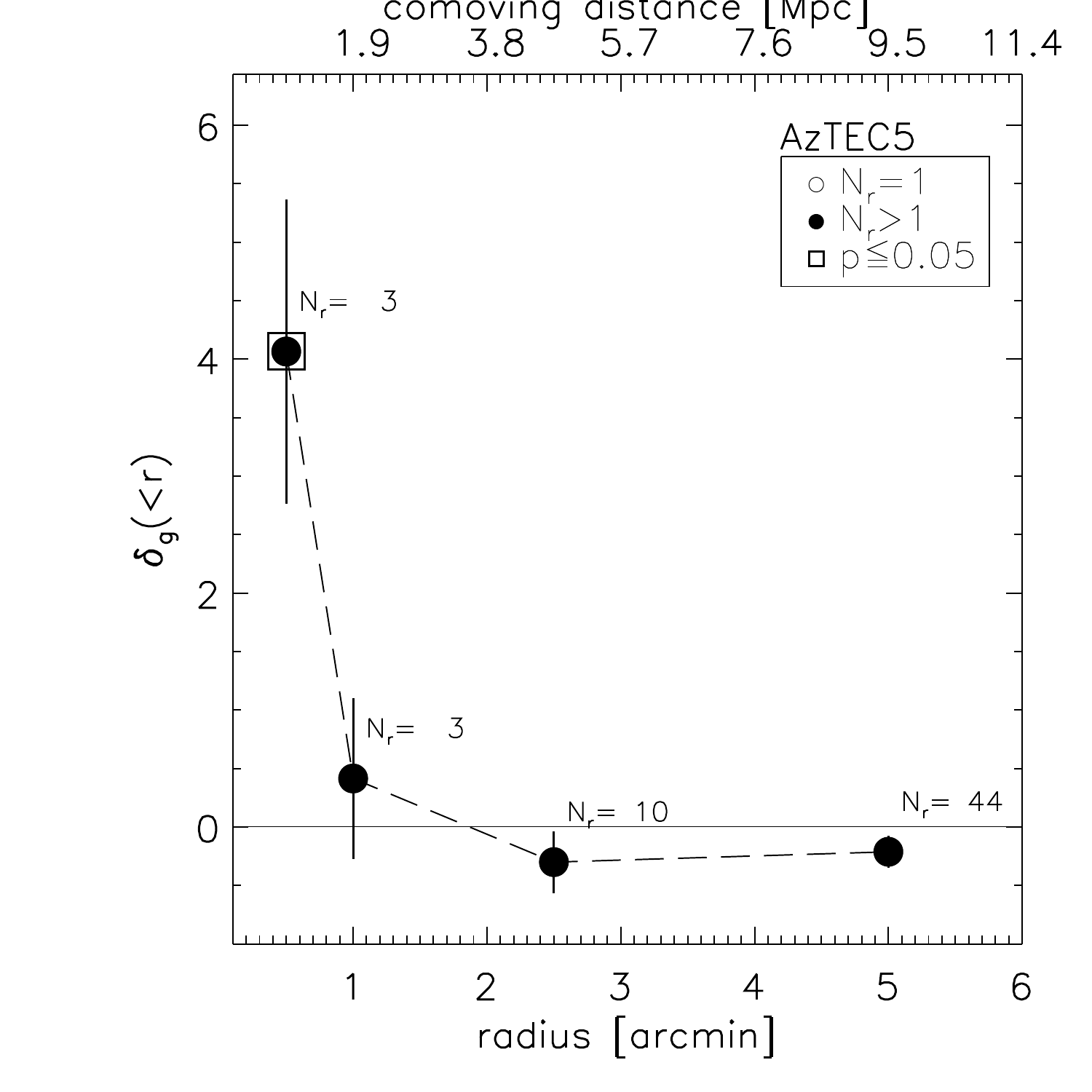}
\includegraphics[width=0.31\textwidth]{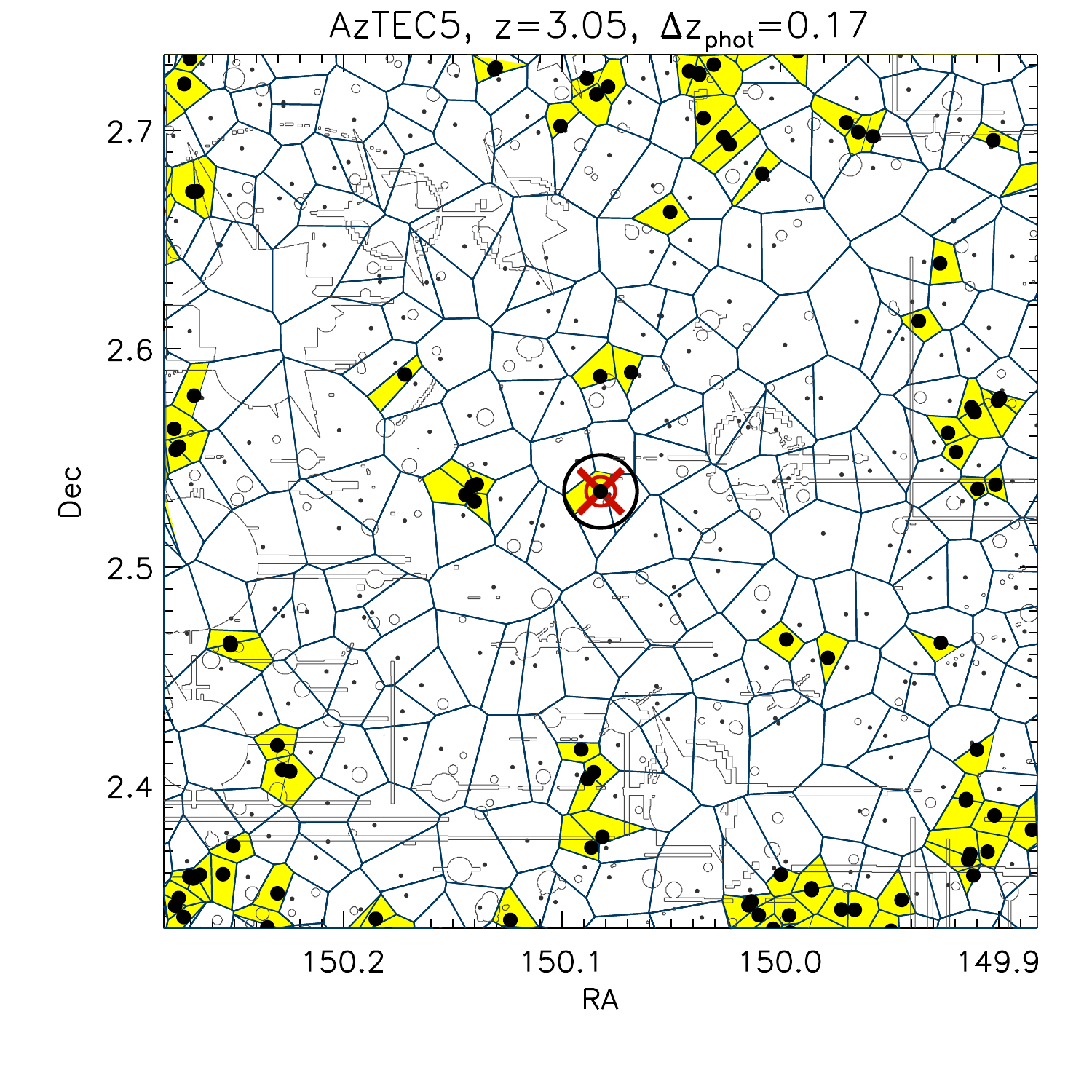}
\includegraphics[width=0.31\textwidth]{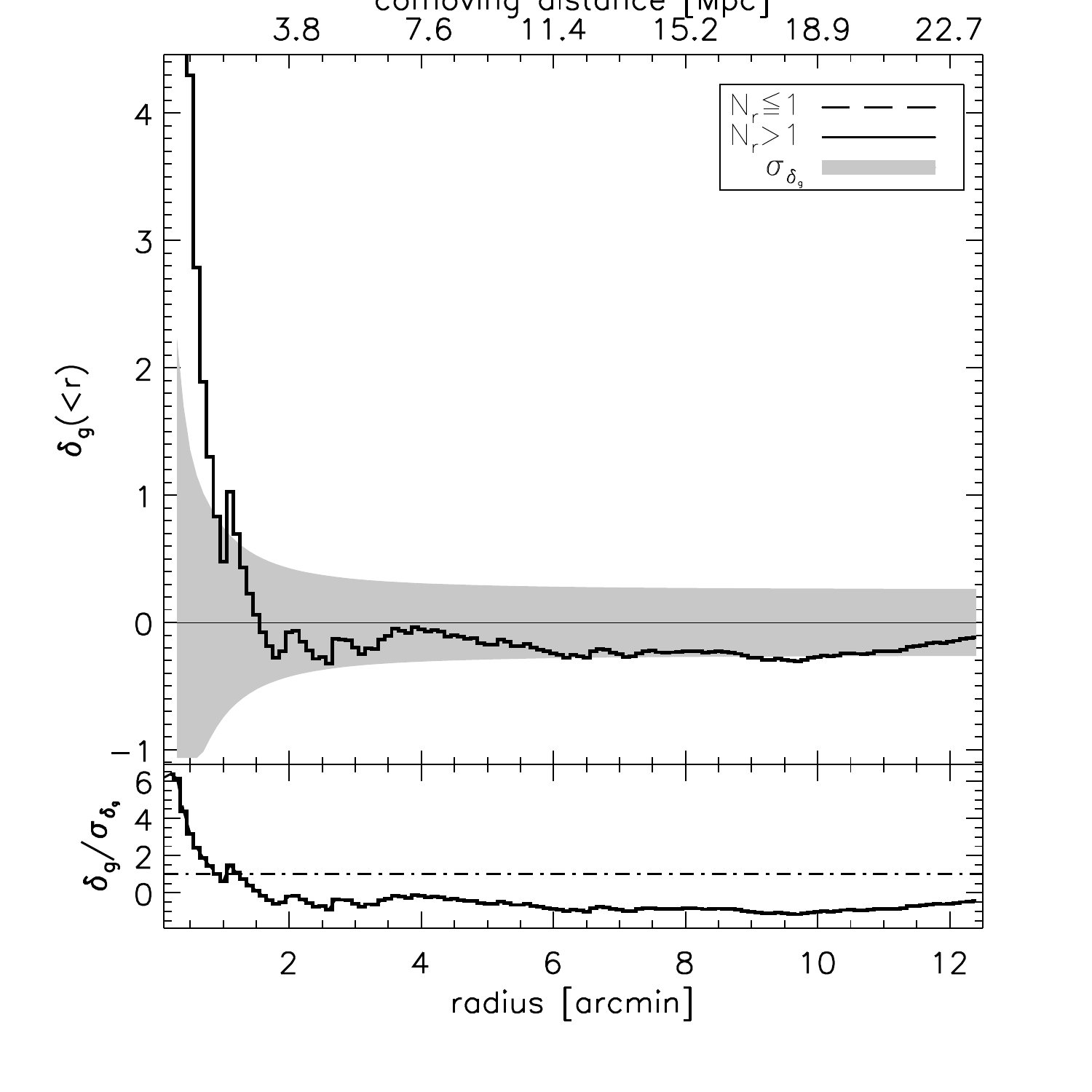}\\
\includegraphics[width=0.31\textwidth]{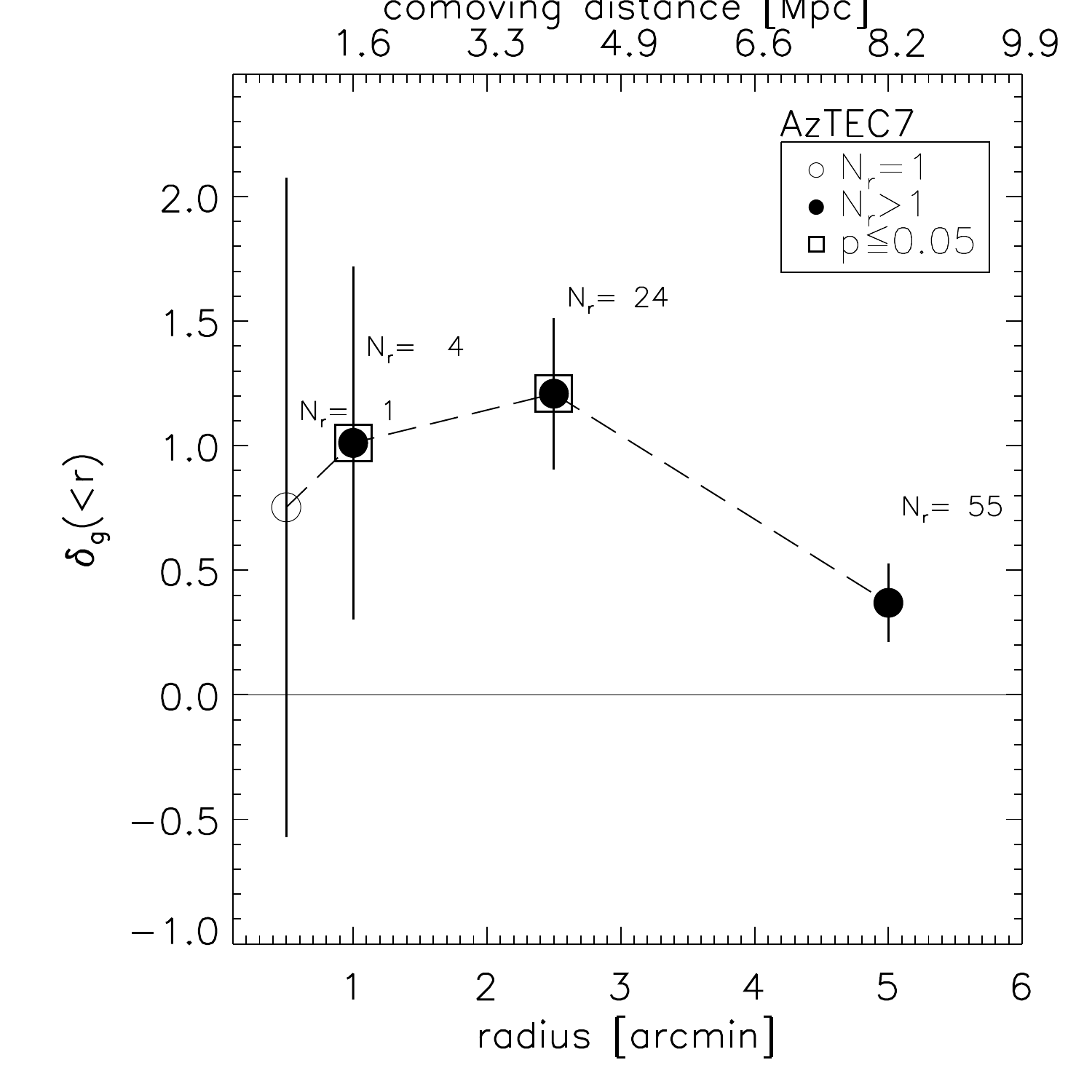}\
\includegraphics[width=0.31\textwidth]{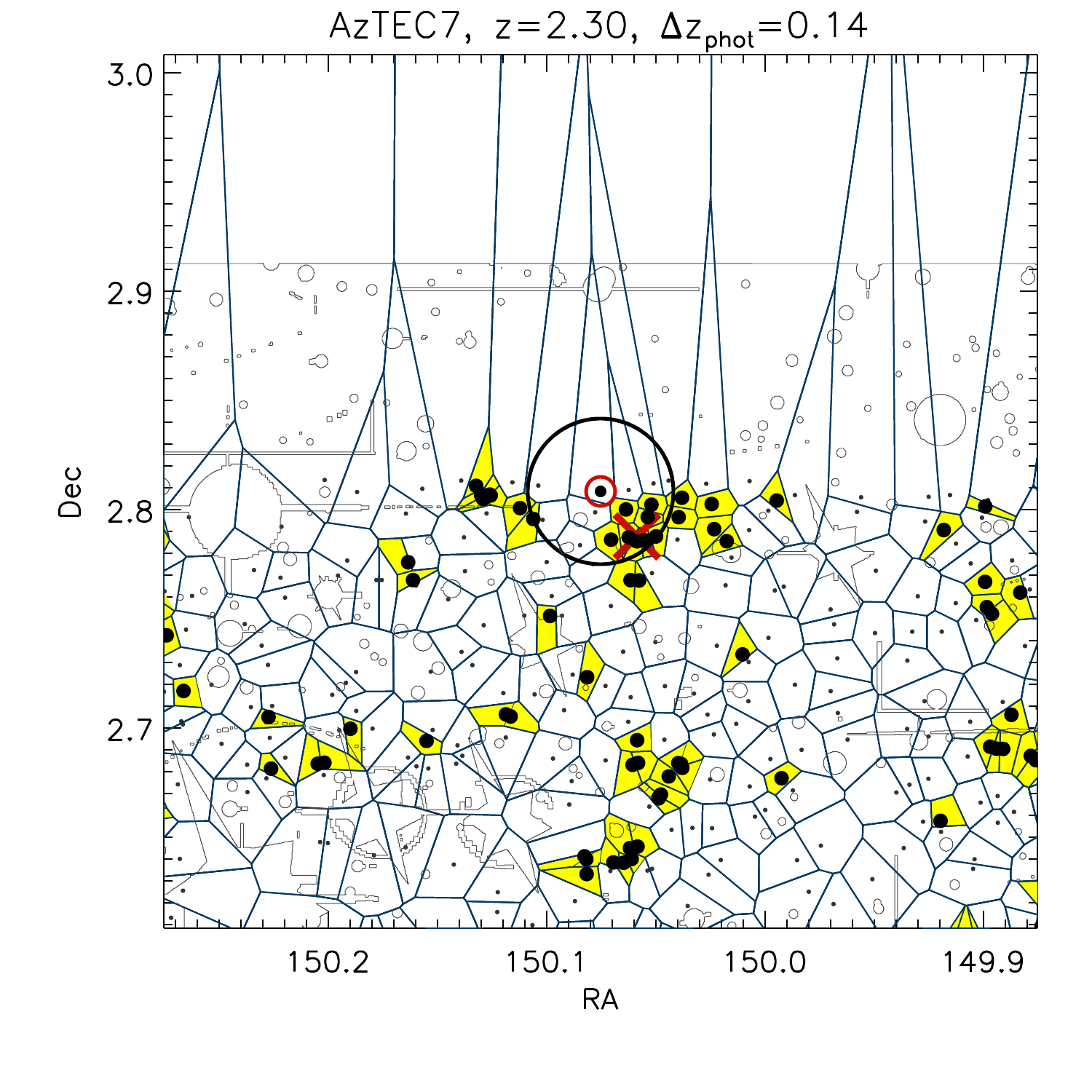}
\includegraphics[width=0.31\textwidth]{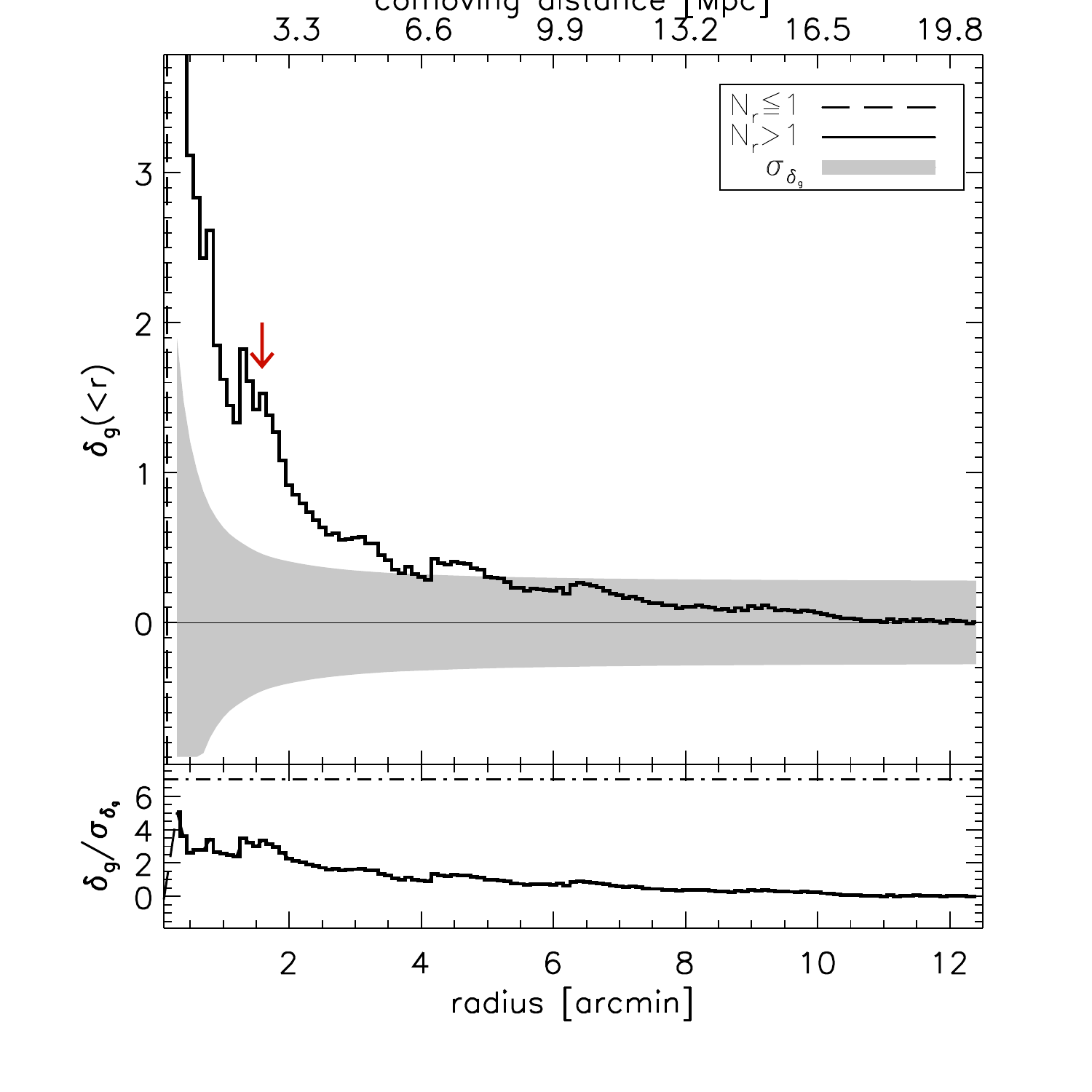}\\
\includegraphics[width=0.31\textwidth]{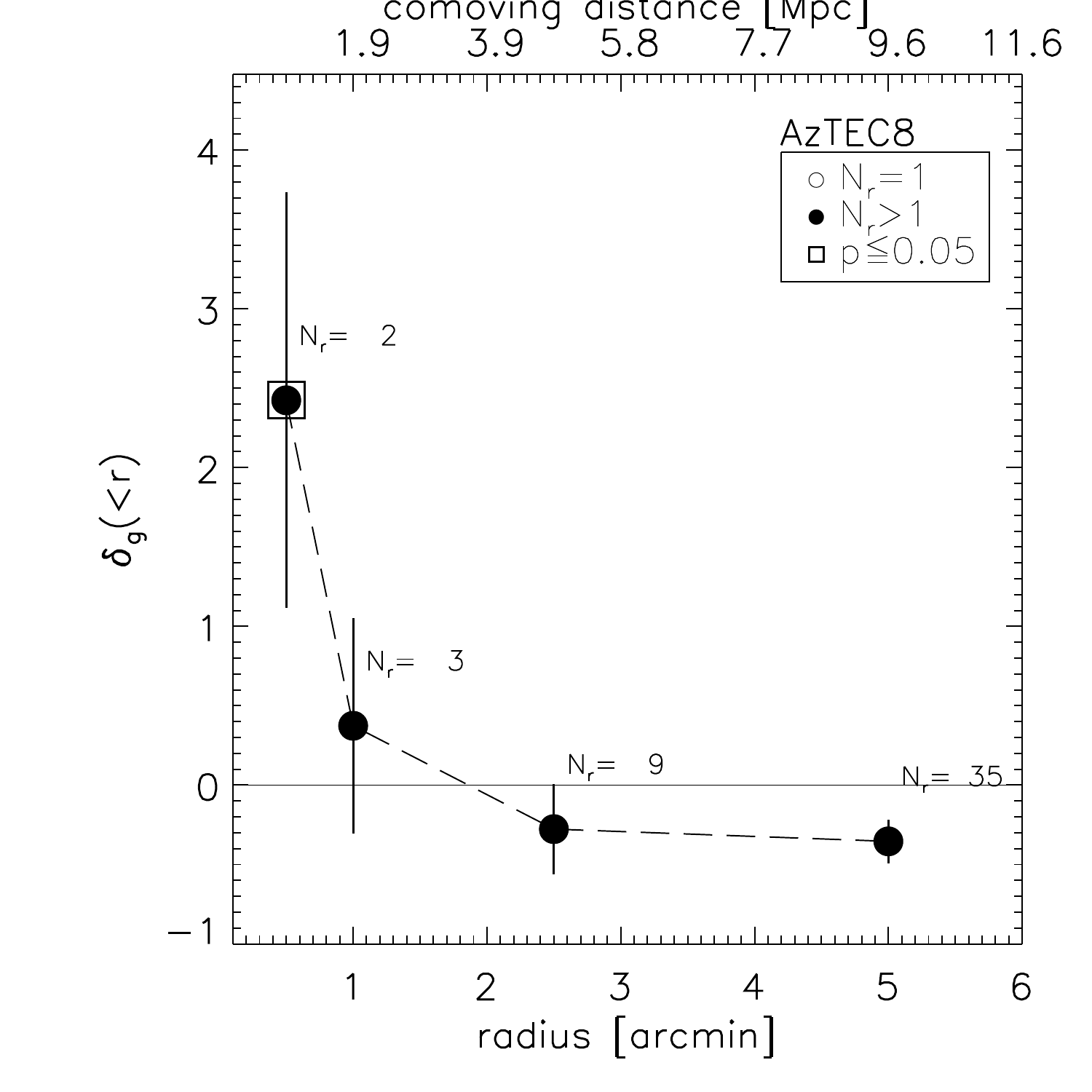}
\includegraphics[width=0.31\textwidth]{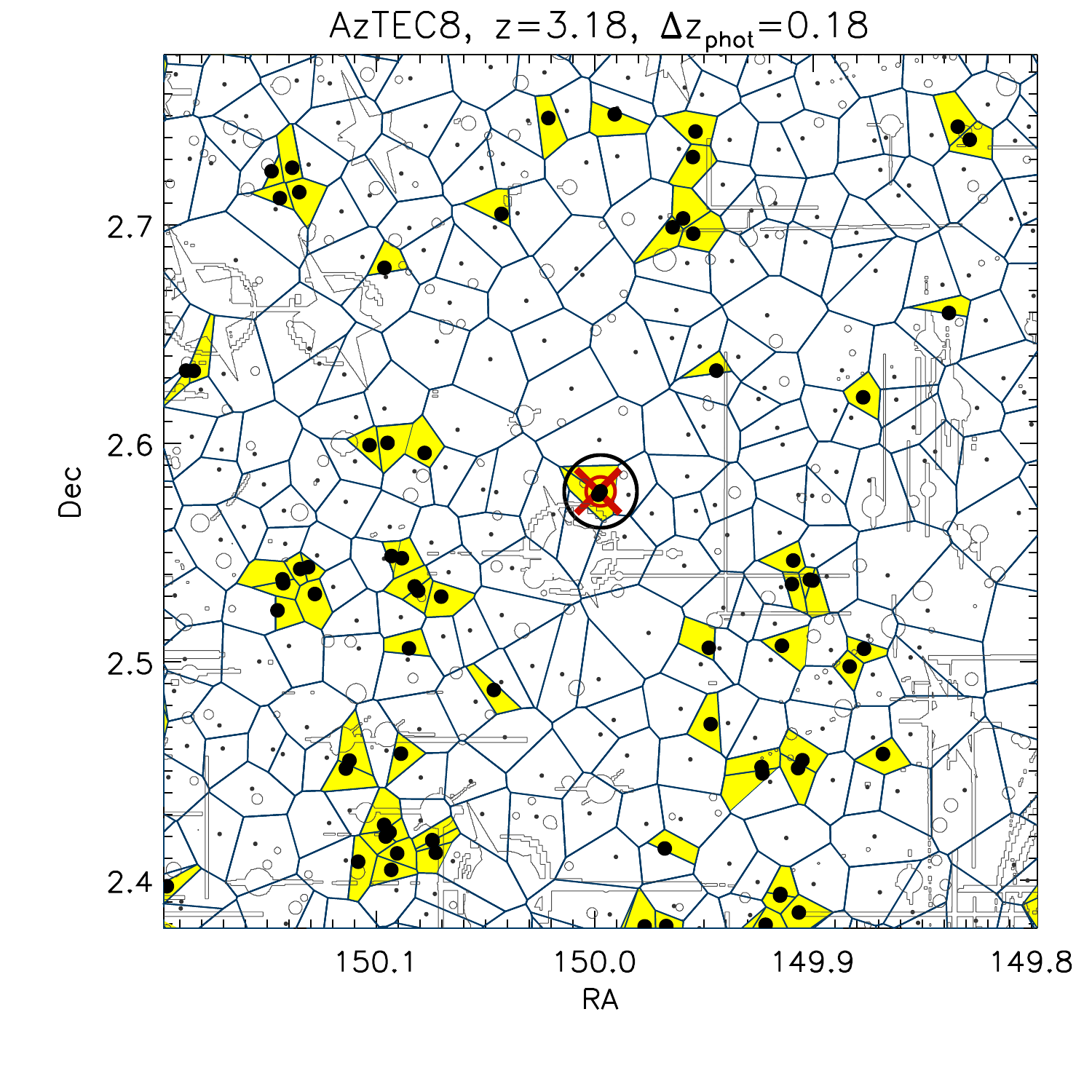}
\includegraphics[width=0.31\textwidth]{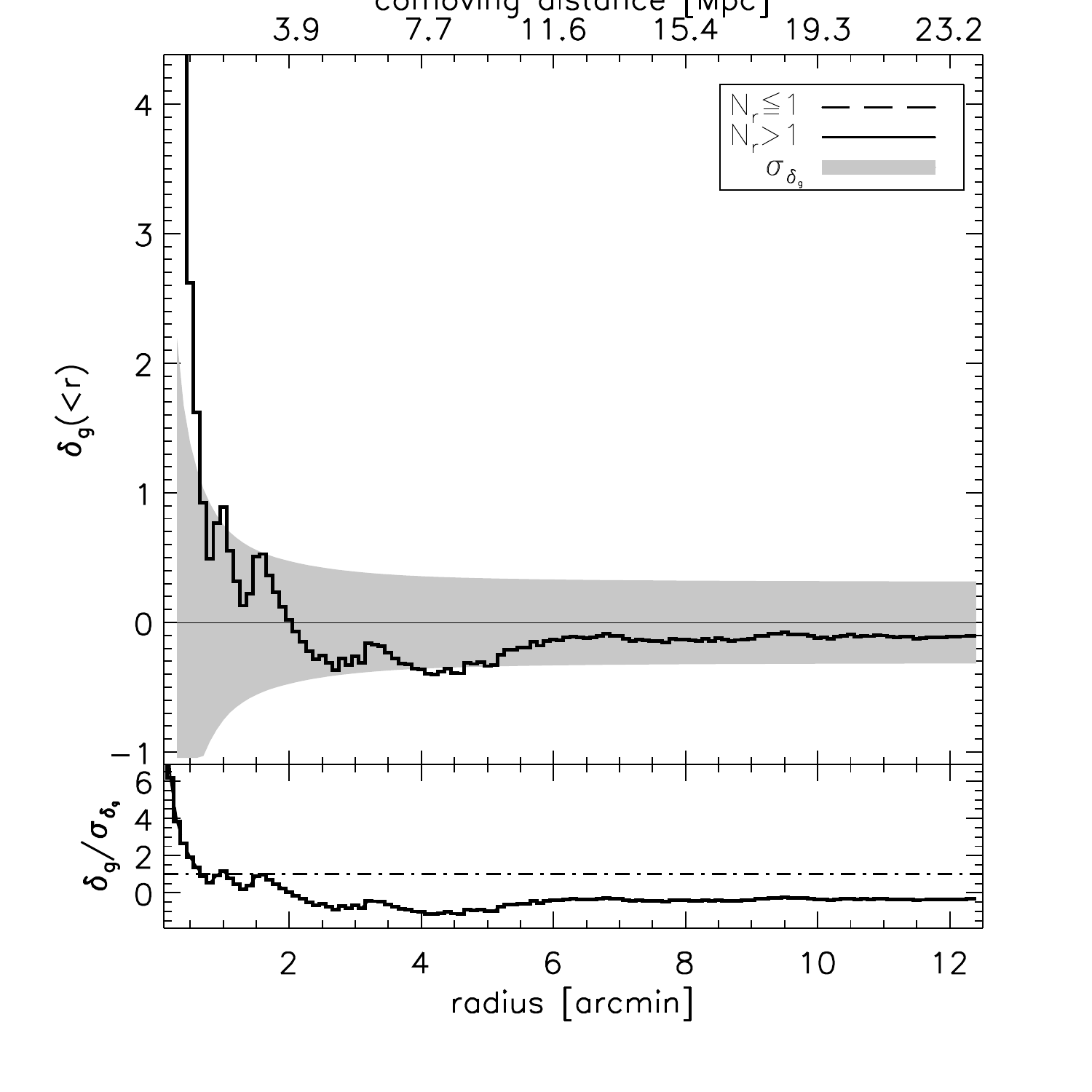}\\
\caption{continued.}
\end{center}
\end{figure*}

\addtocounter{figure}{-1}
\begin{figure*}
\begin{center}
\includegraphics[width=0.31\textwidth]{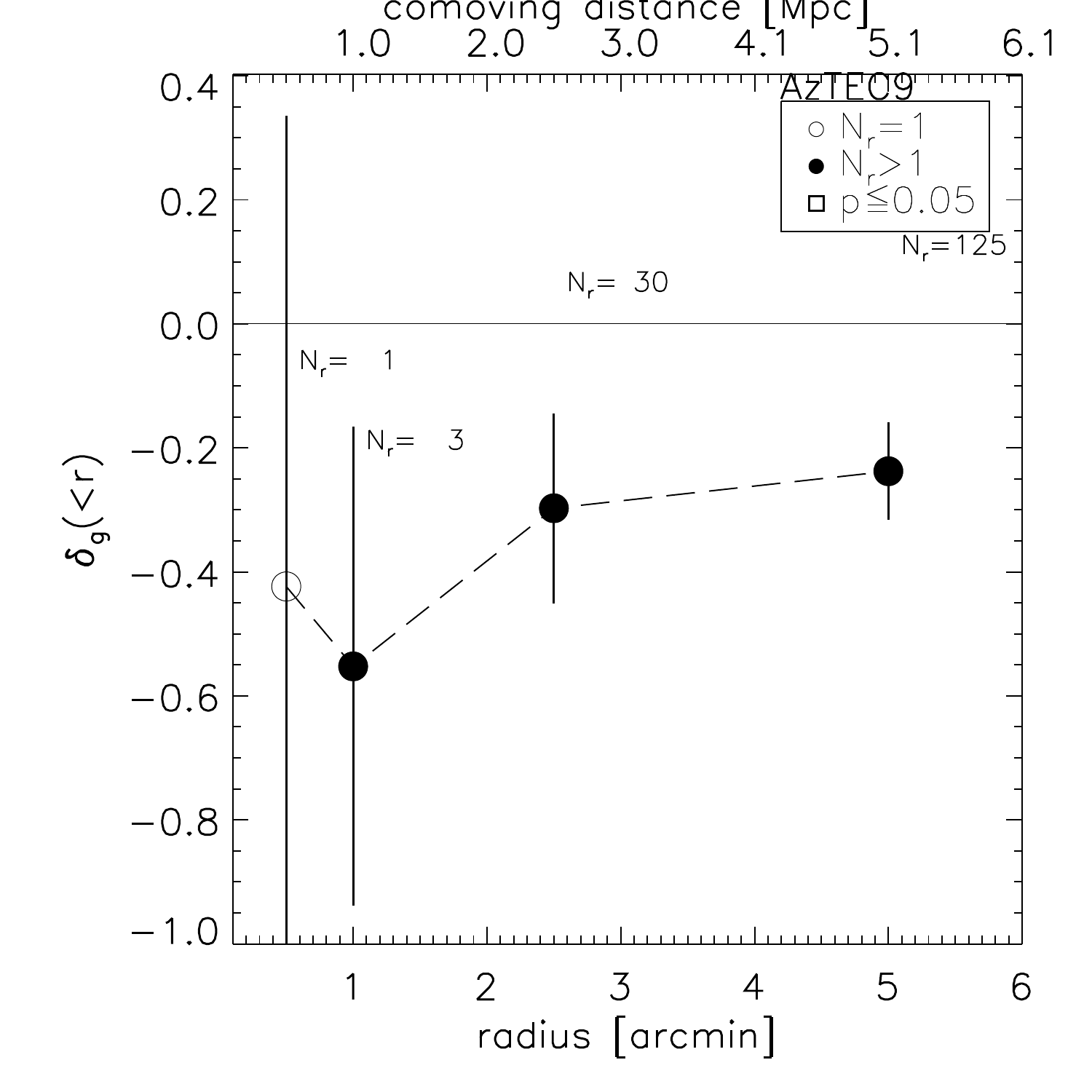}
\includegraphics[width=0.31\textwidth]{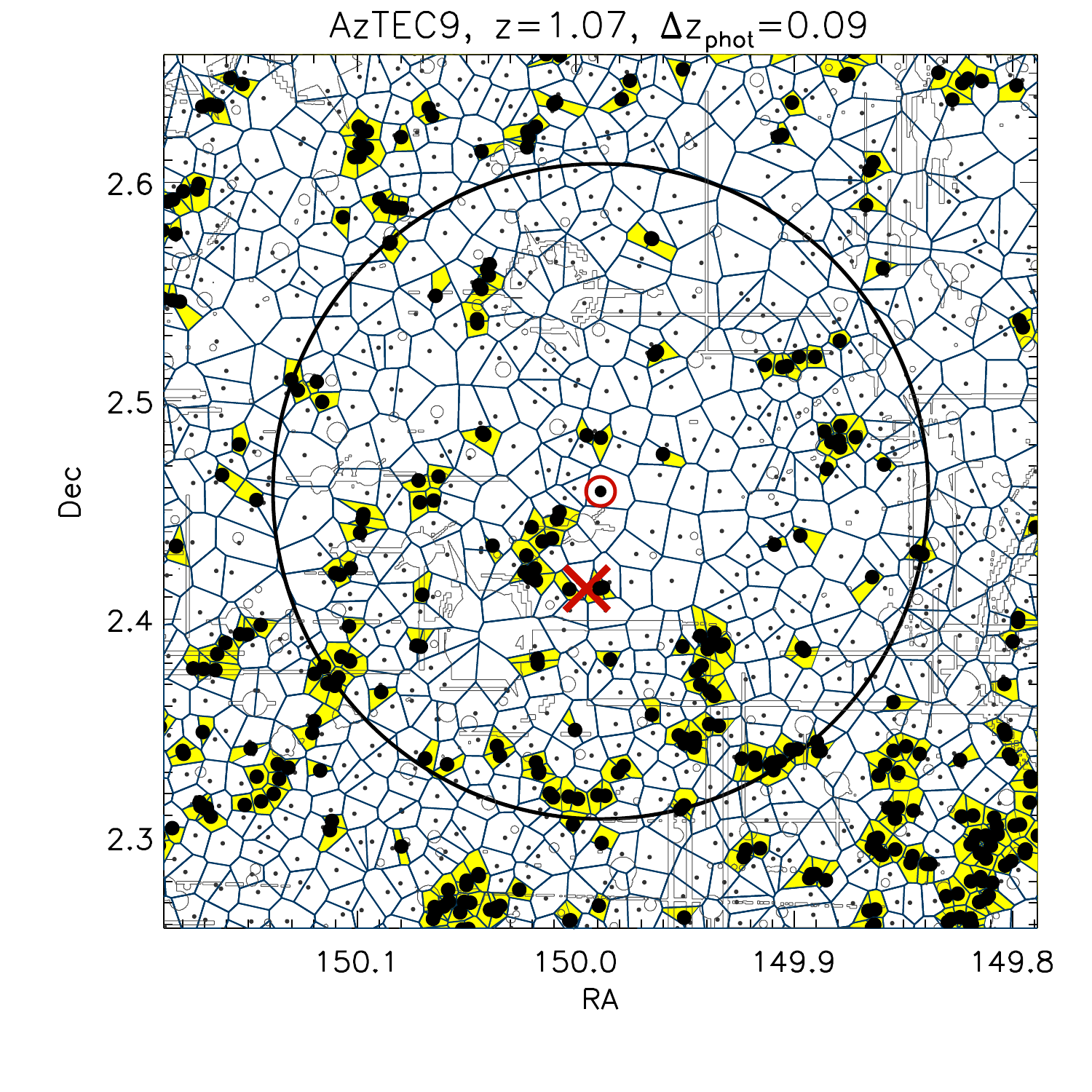}
\includegraphics[width=0.31\textwidth]{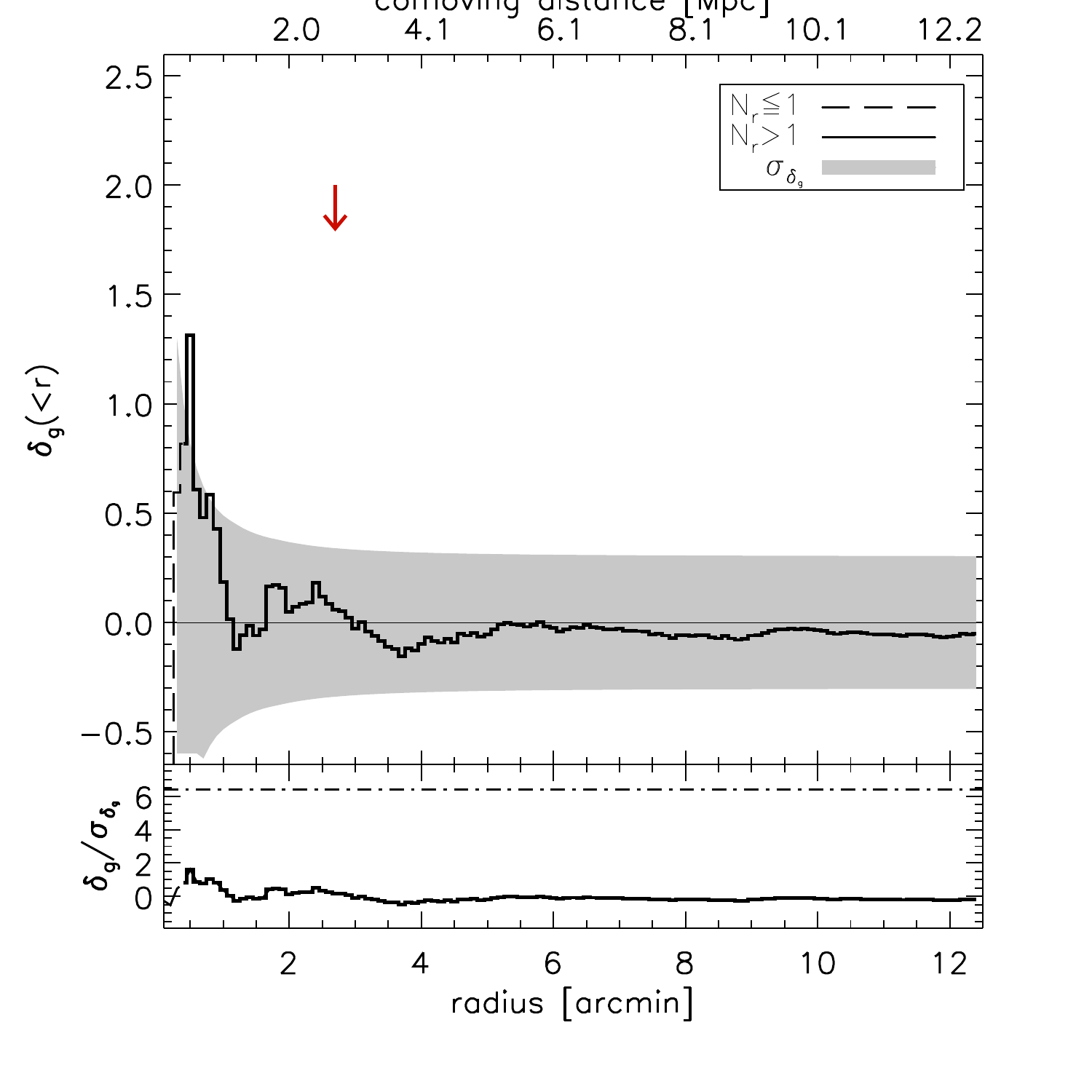}\\
\includegraphics[width=0.31\textwidth]{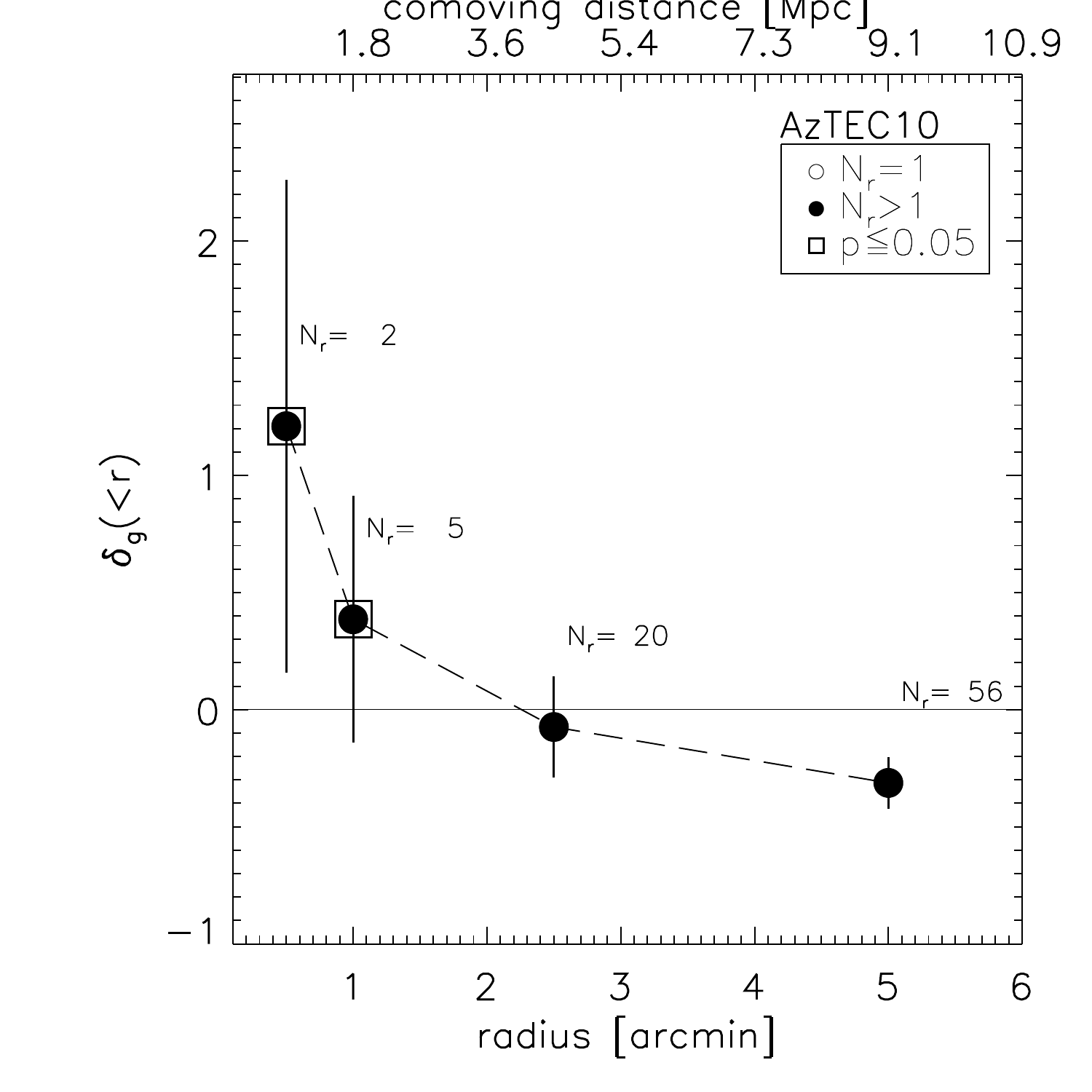}
\includegraphics[width=0.31\textwidth]{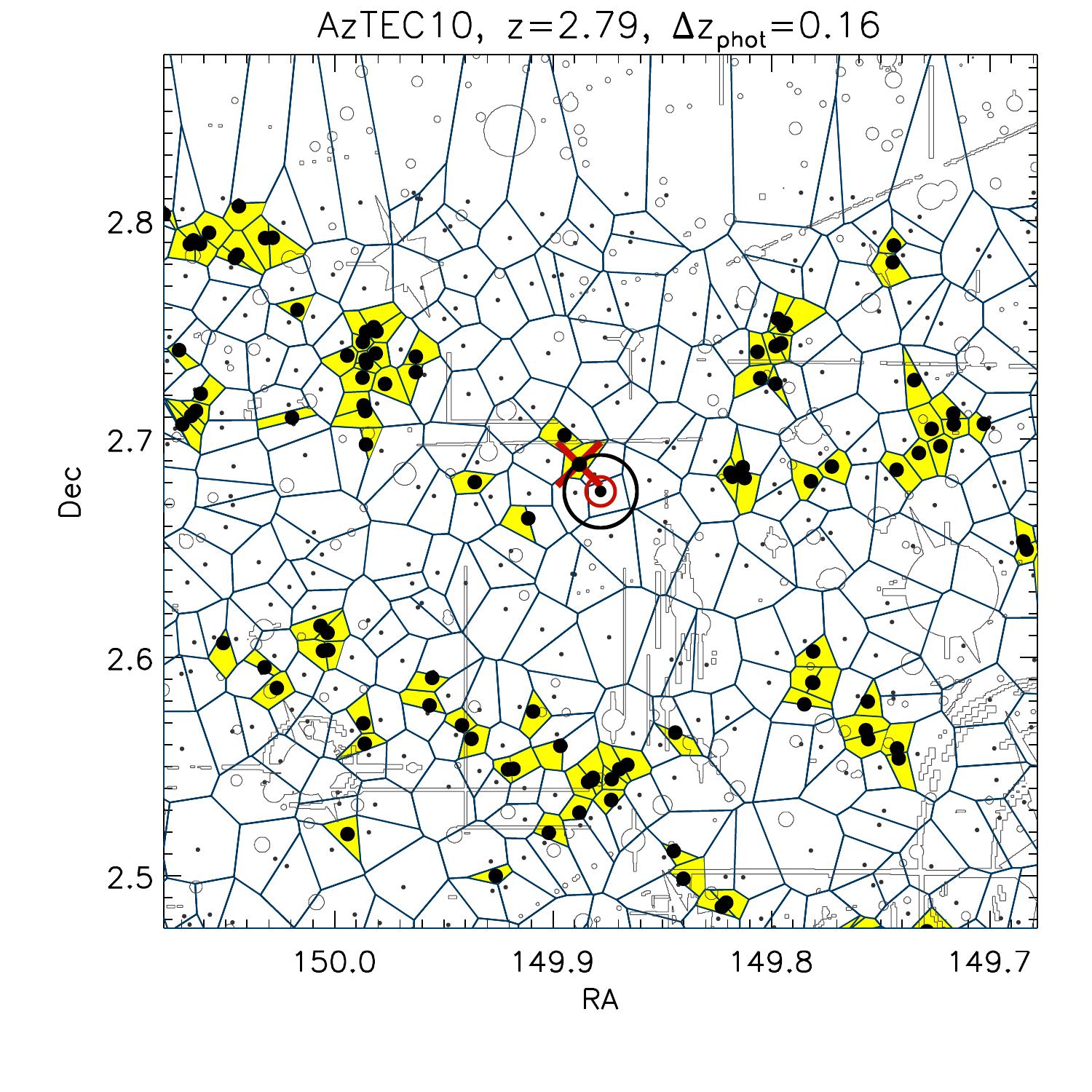}
\includegraphics[width=0.31\textwidth]{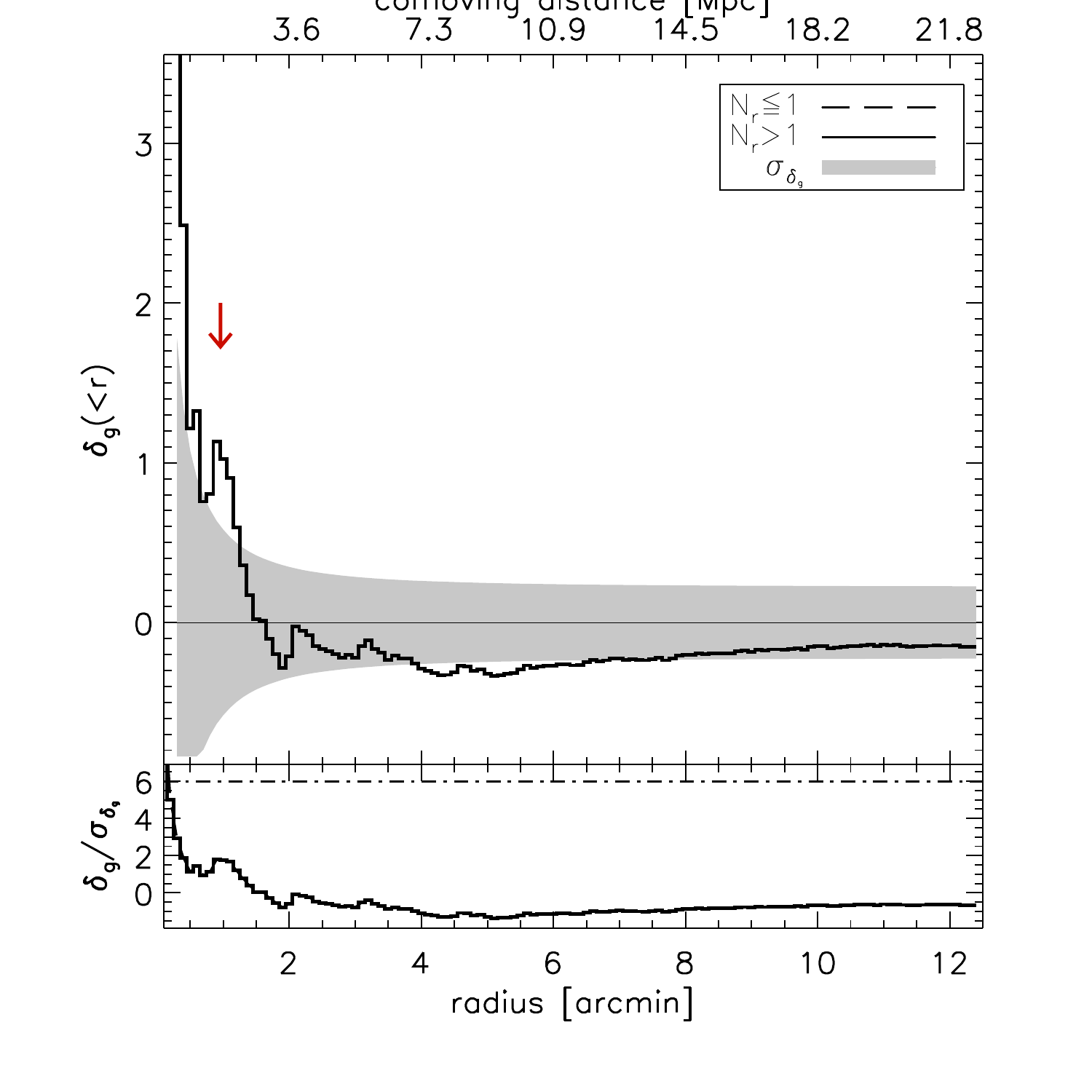}\\
\includegraphics[width=0.31\textwidth]{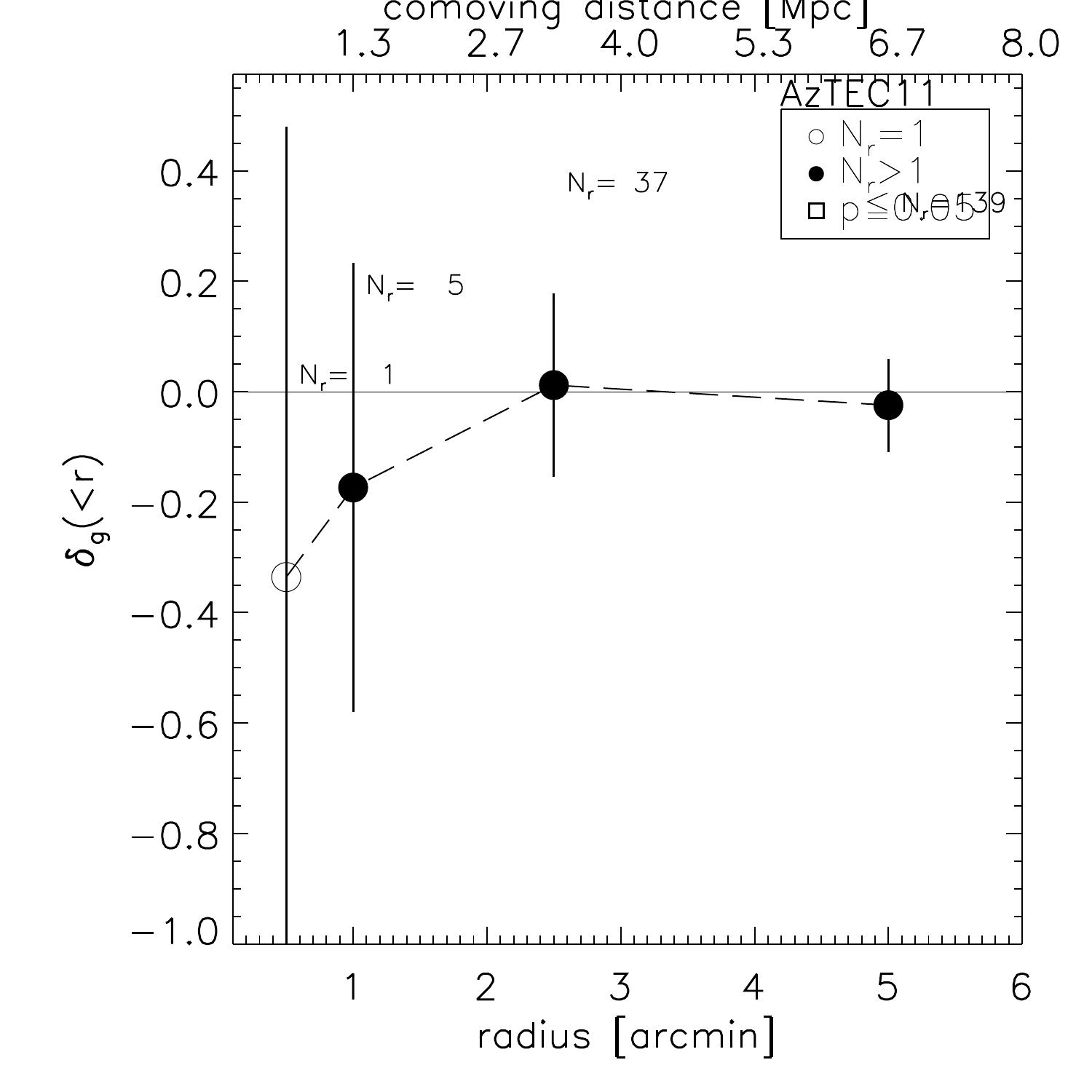}
\includegraphics[width=0.31\textwidth]{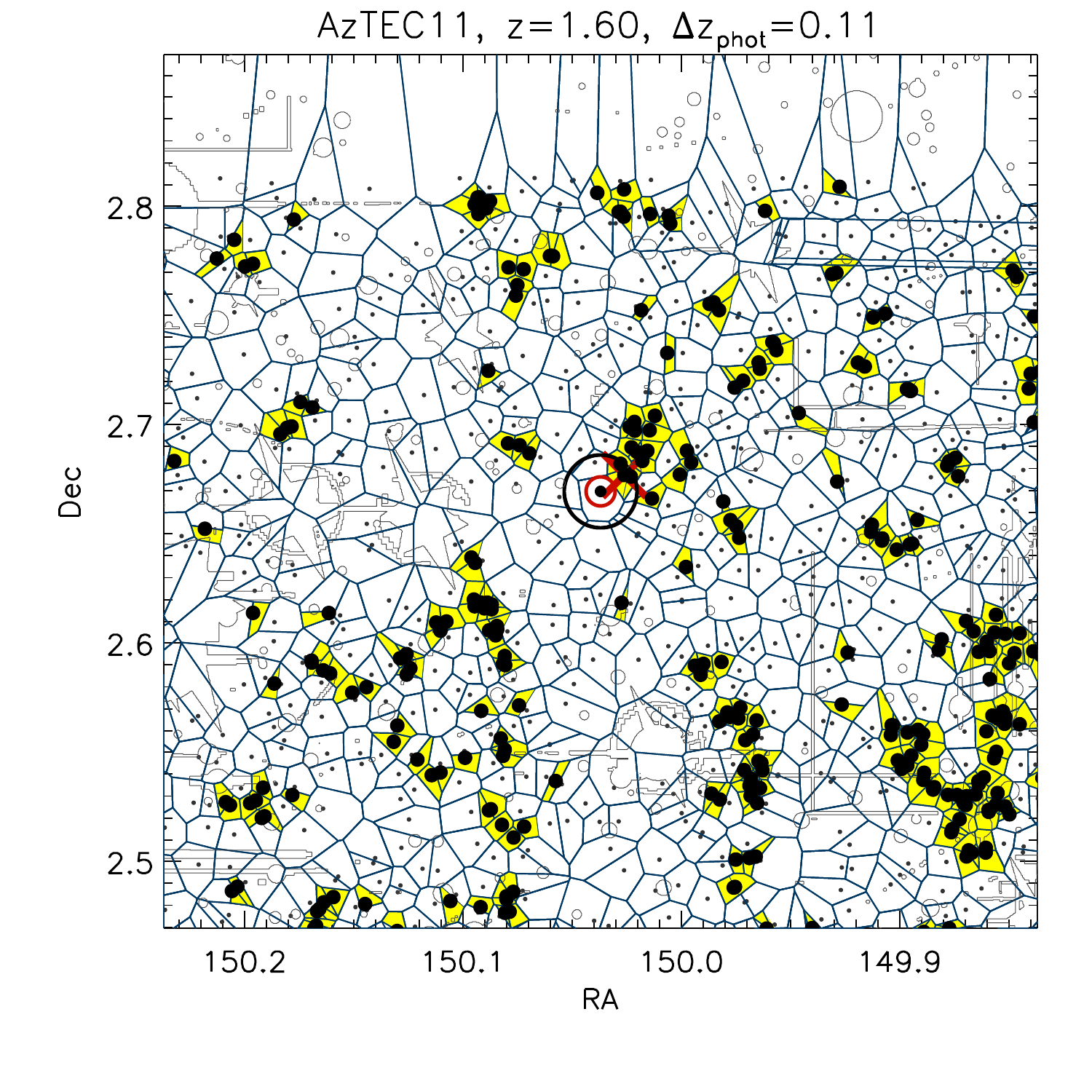}
\includegraphics[width=0.31\textwidth]{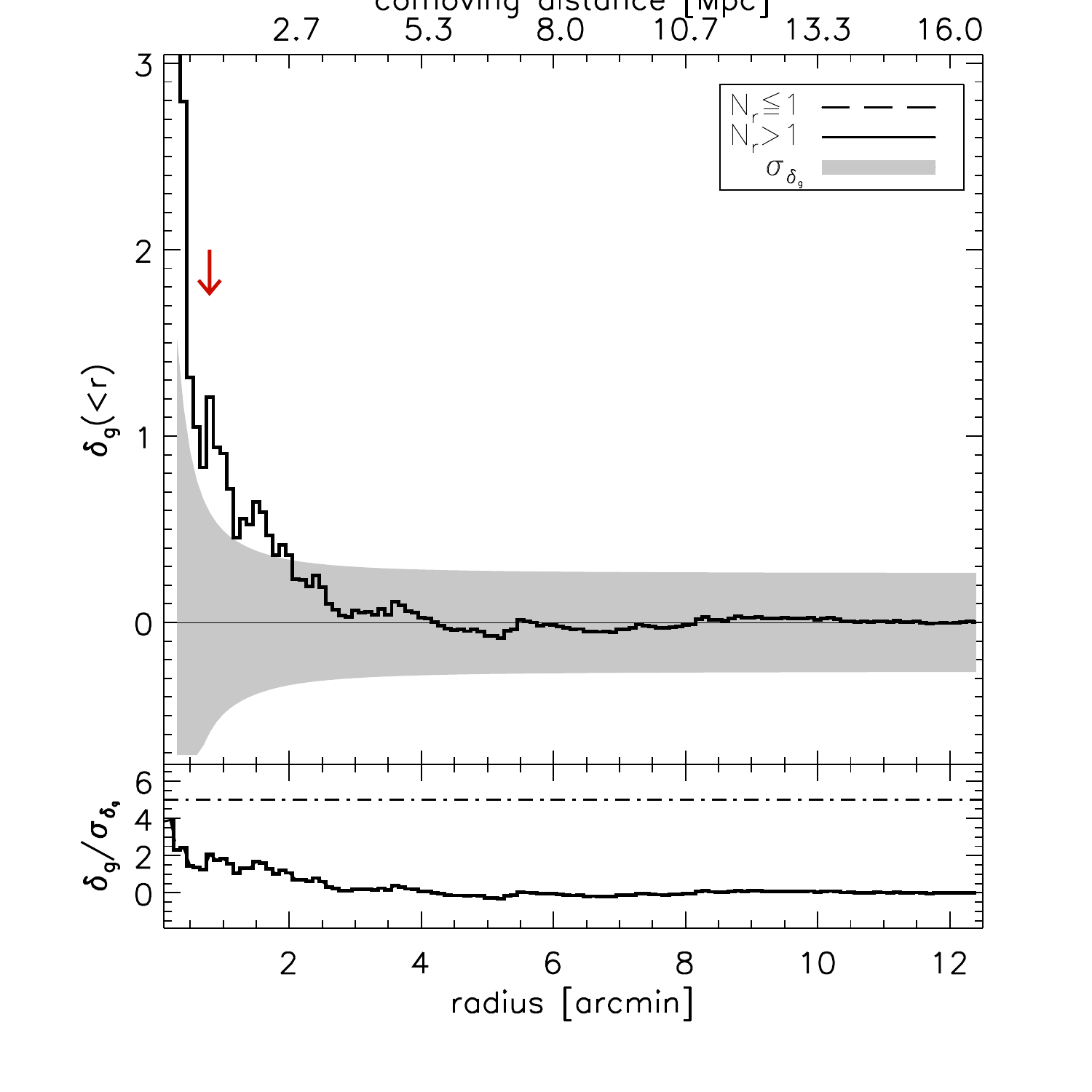}\\
\includegraphics[width=0.31\textwidth]{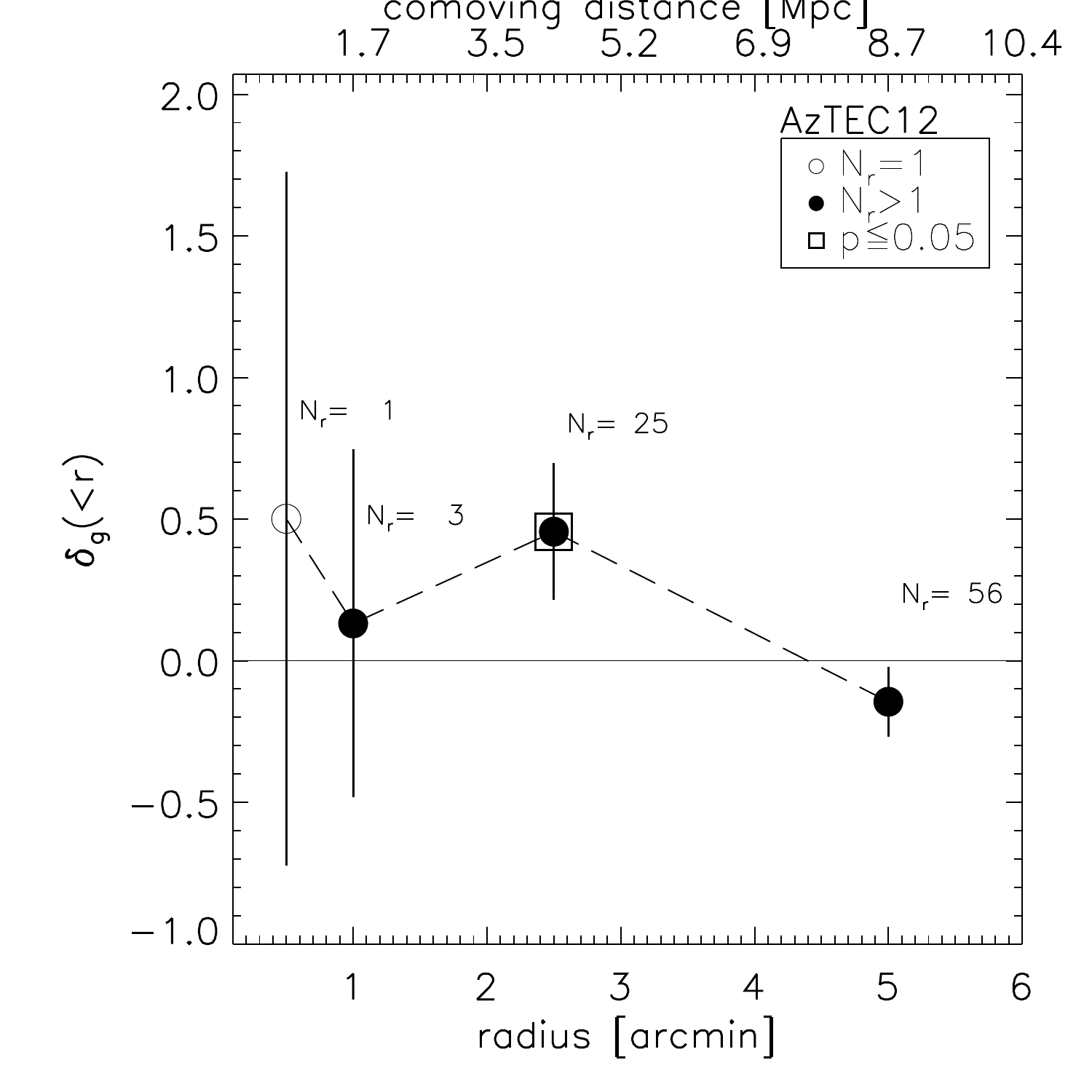}
\includegraphics[width=0.31\textwidth]{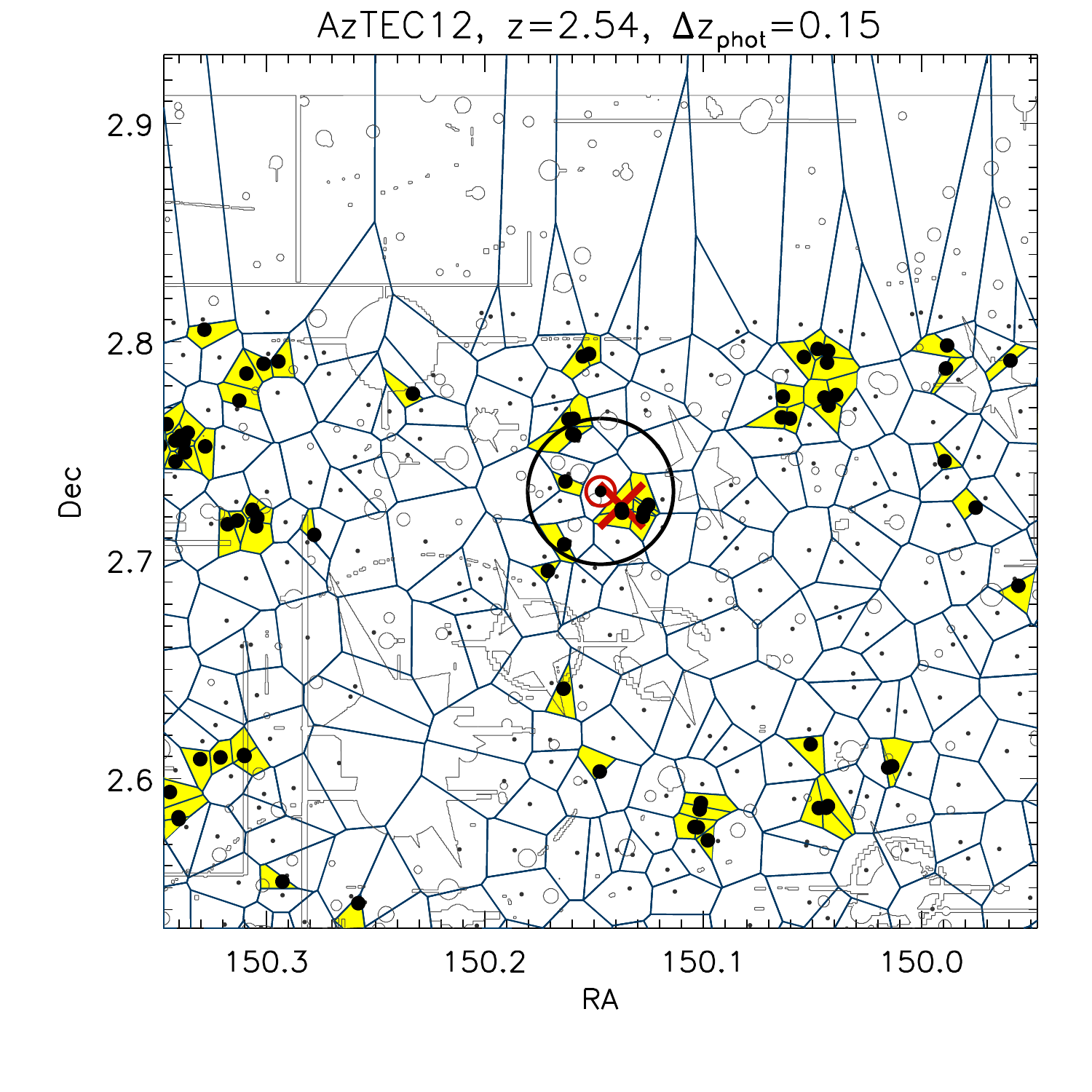}
\includegraphics[width=0.31\textwidth]{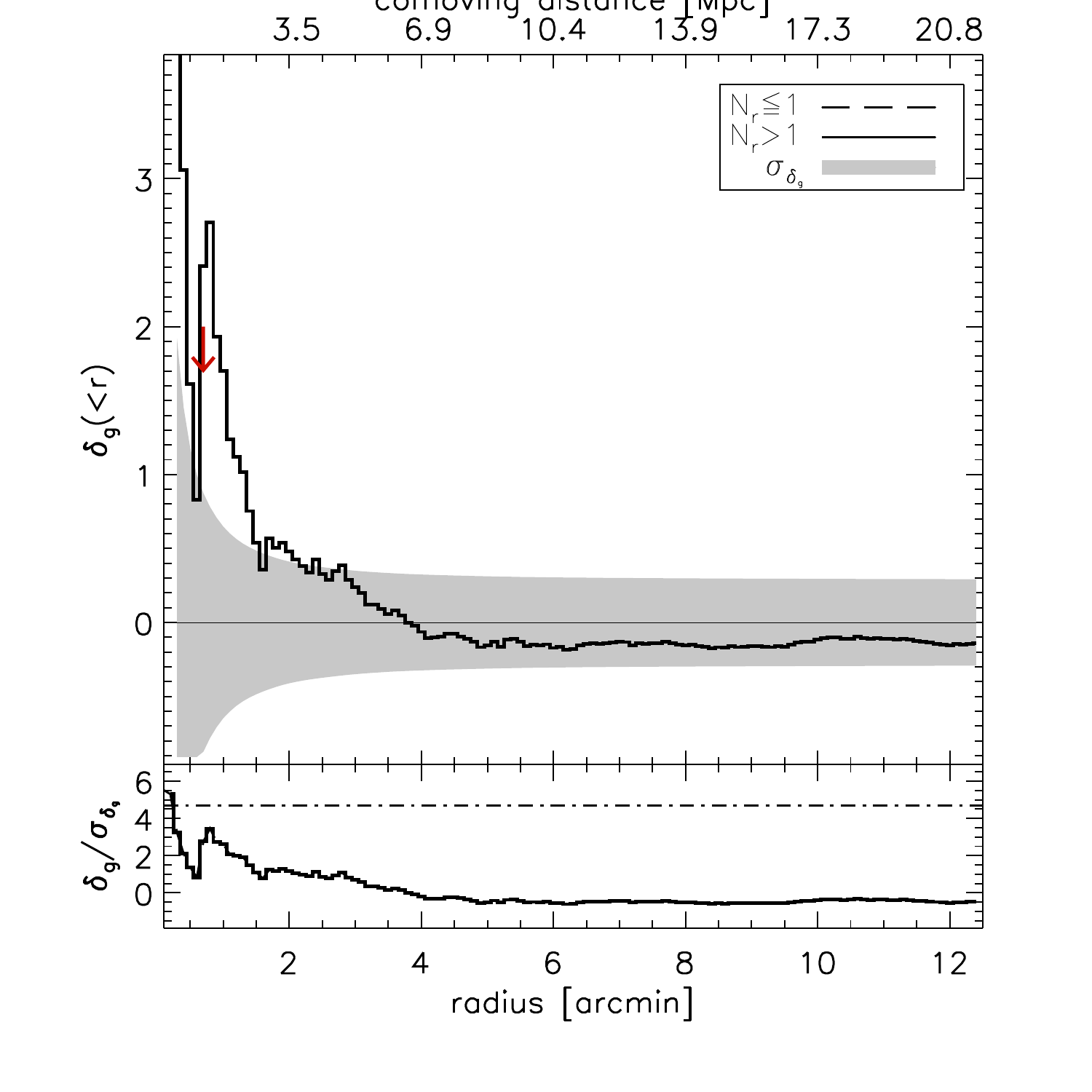}\\
\caption{continued.}
\end{center}
\end{figure*}

\addtocounter{figure}{-1}
\begin{figure*}
\begin{center}
\includegraphics[width=0.31\textwidth]{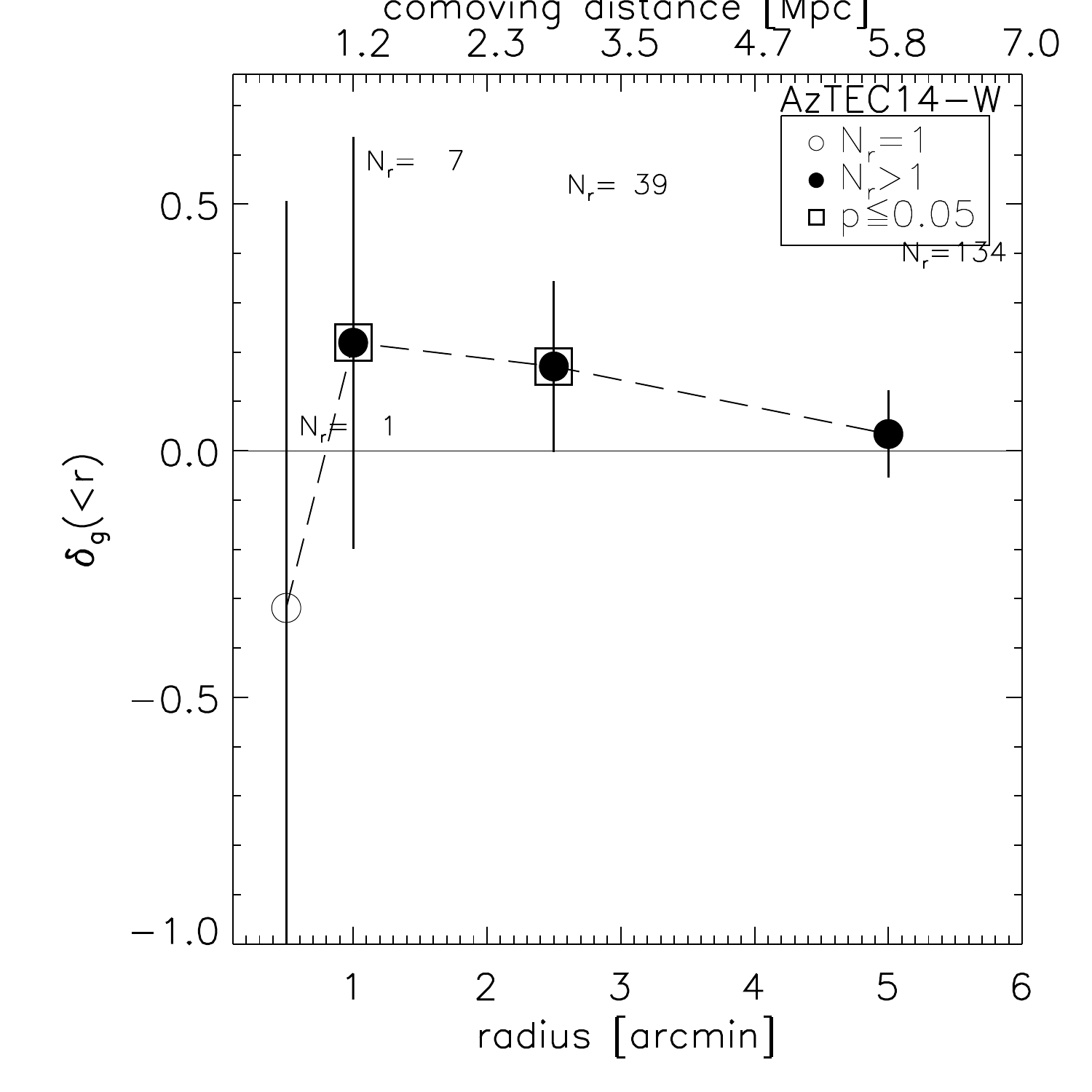}
\includegraphics[width=0.31\textwidth]{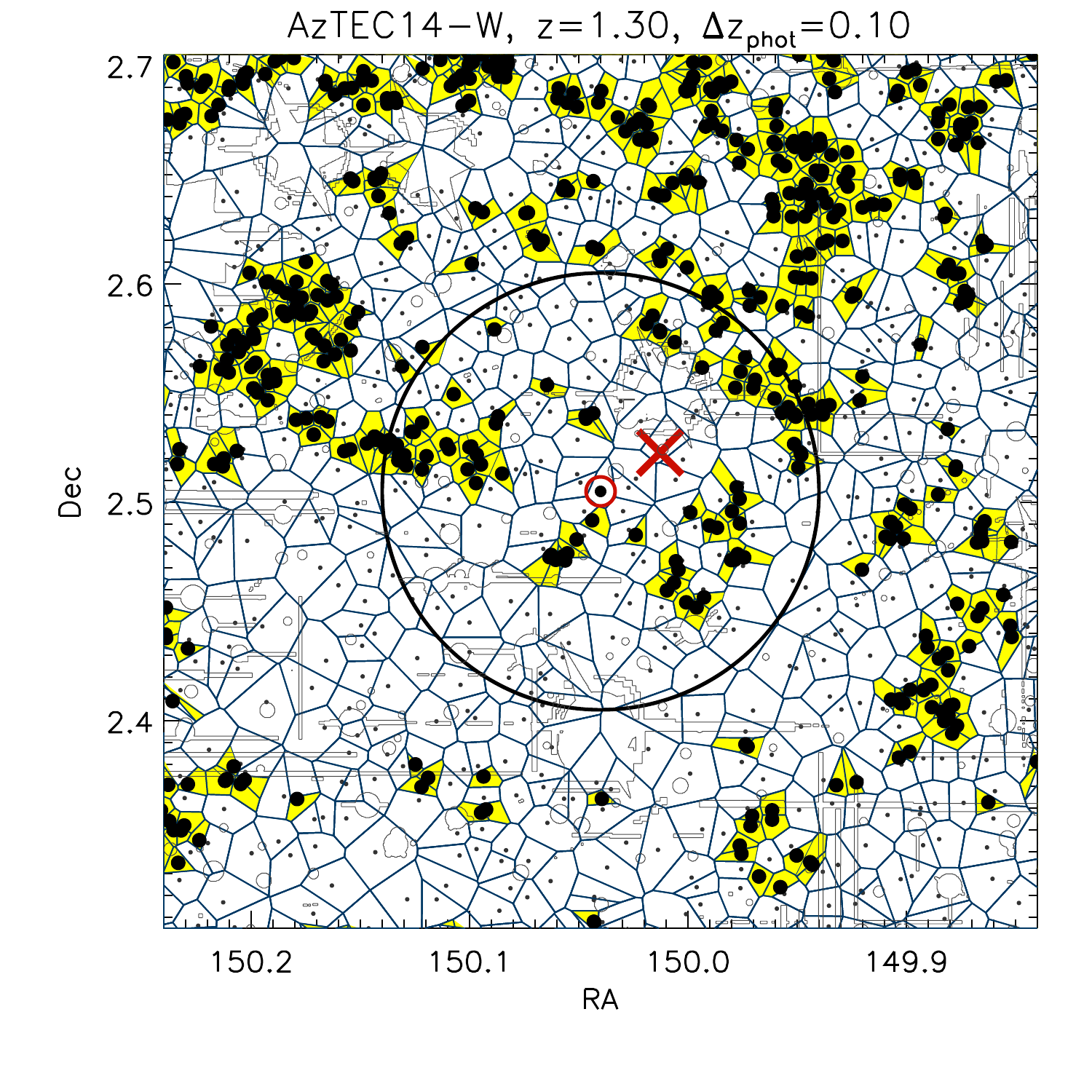}
\includegraphics[width=0.31\textwidth]{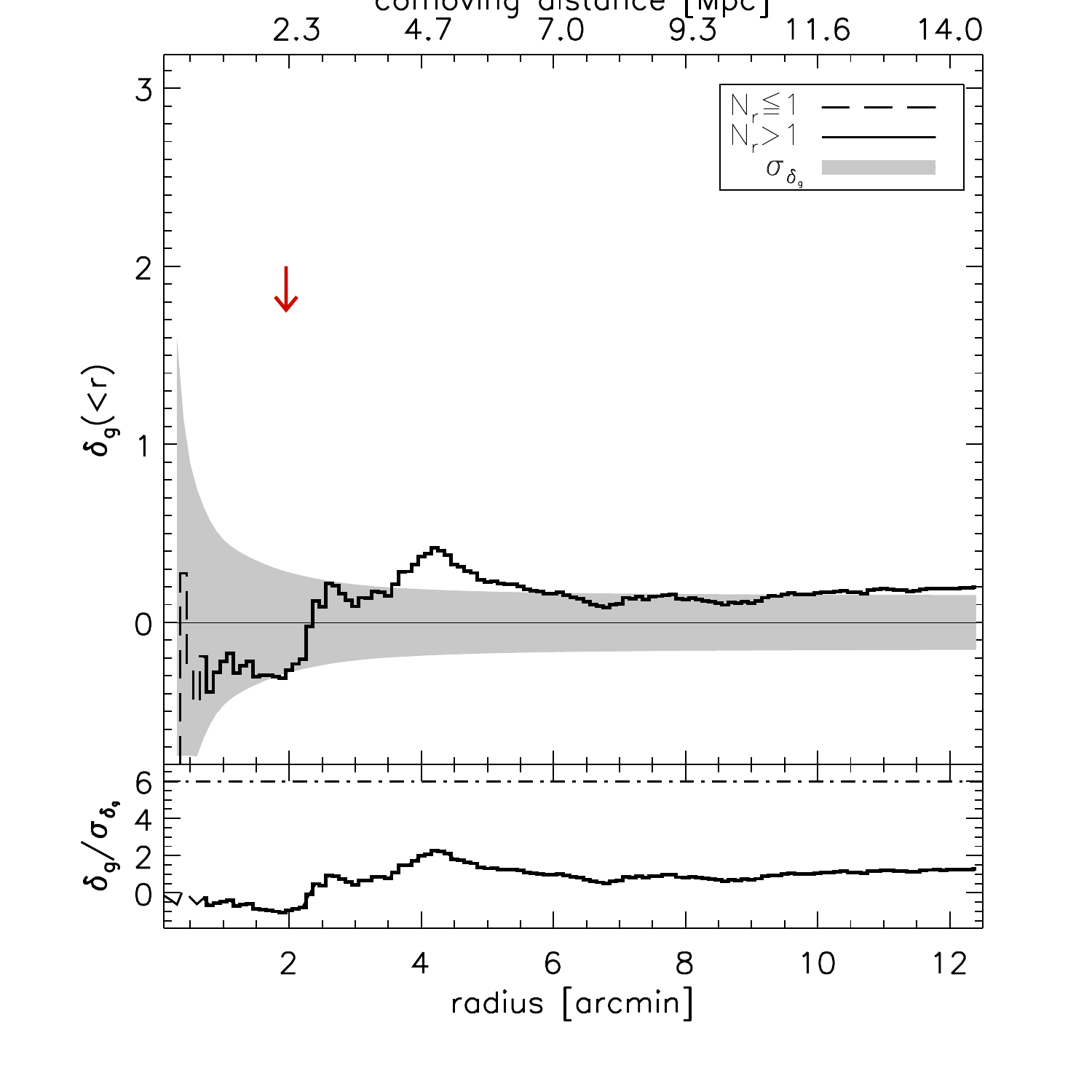}\\
\includegraphics[width=0.31\textwidth]{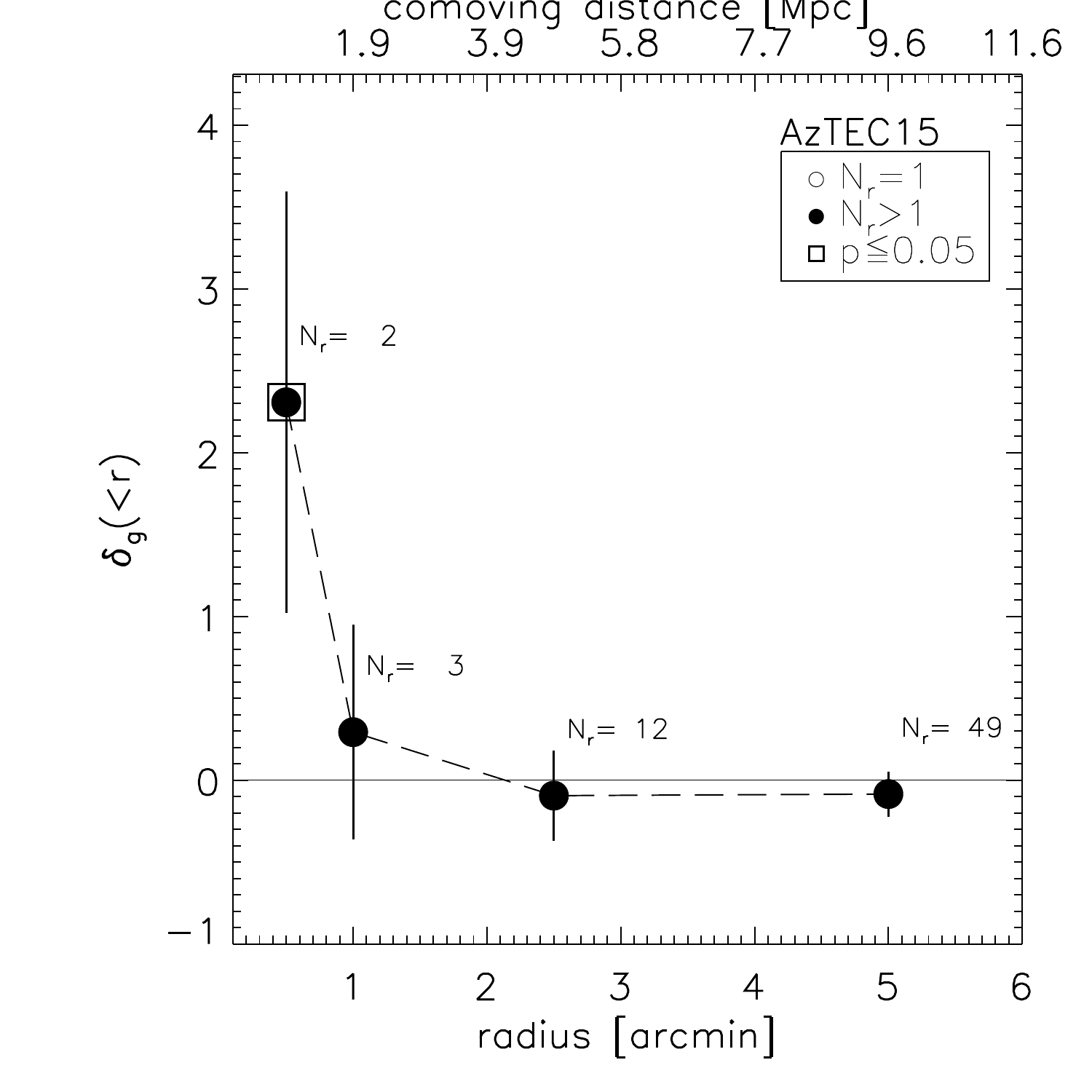}
\includegraphics[width=0.31\textwidth]{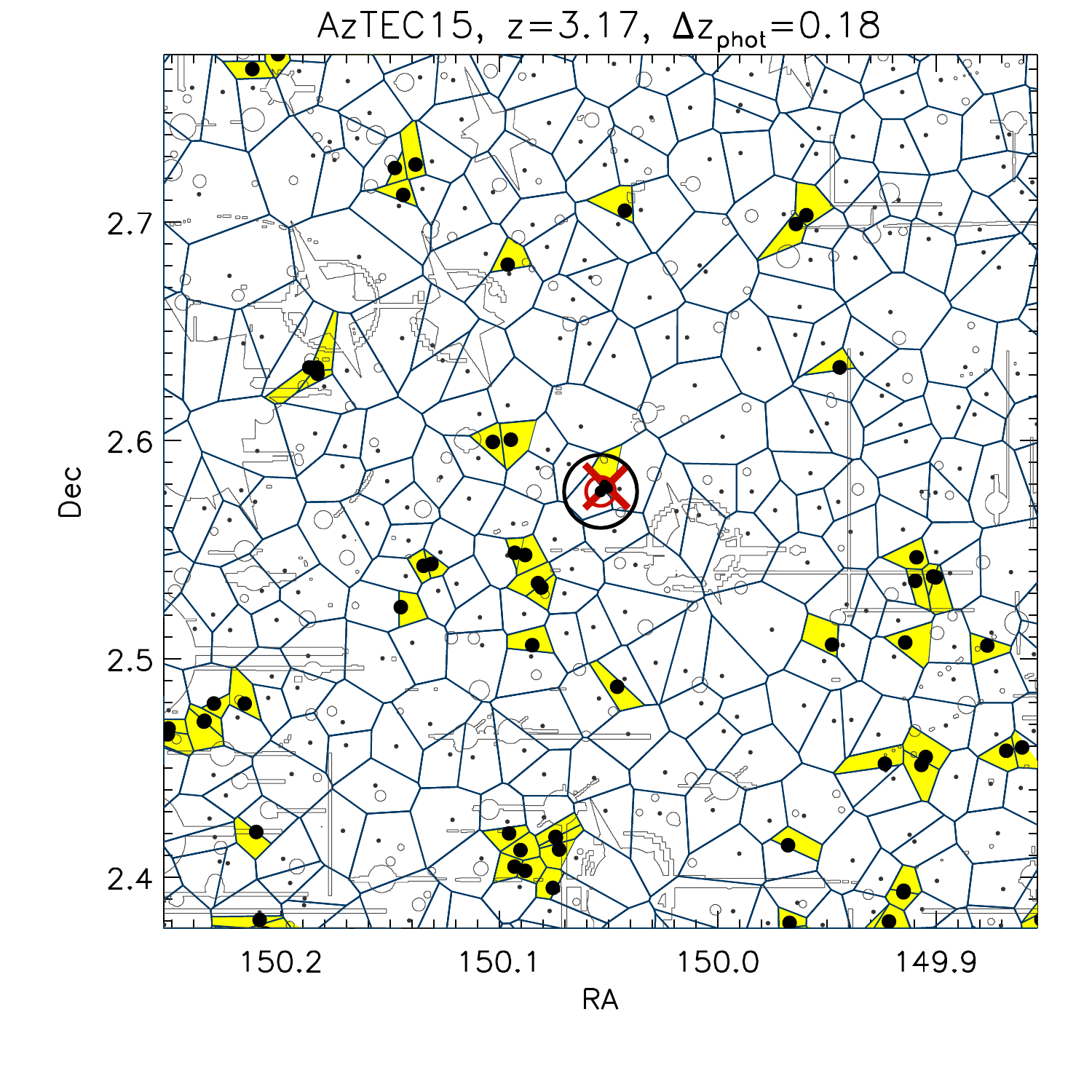}
\includegraphics[width=0.31\textwidth]{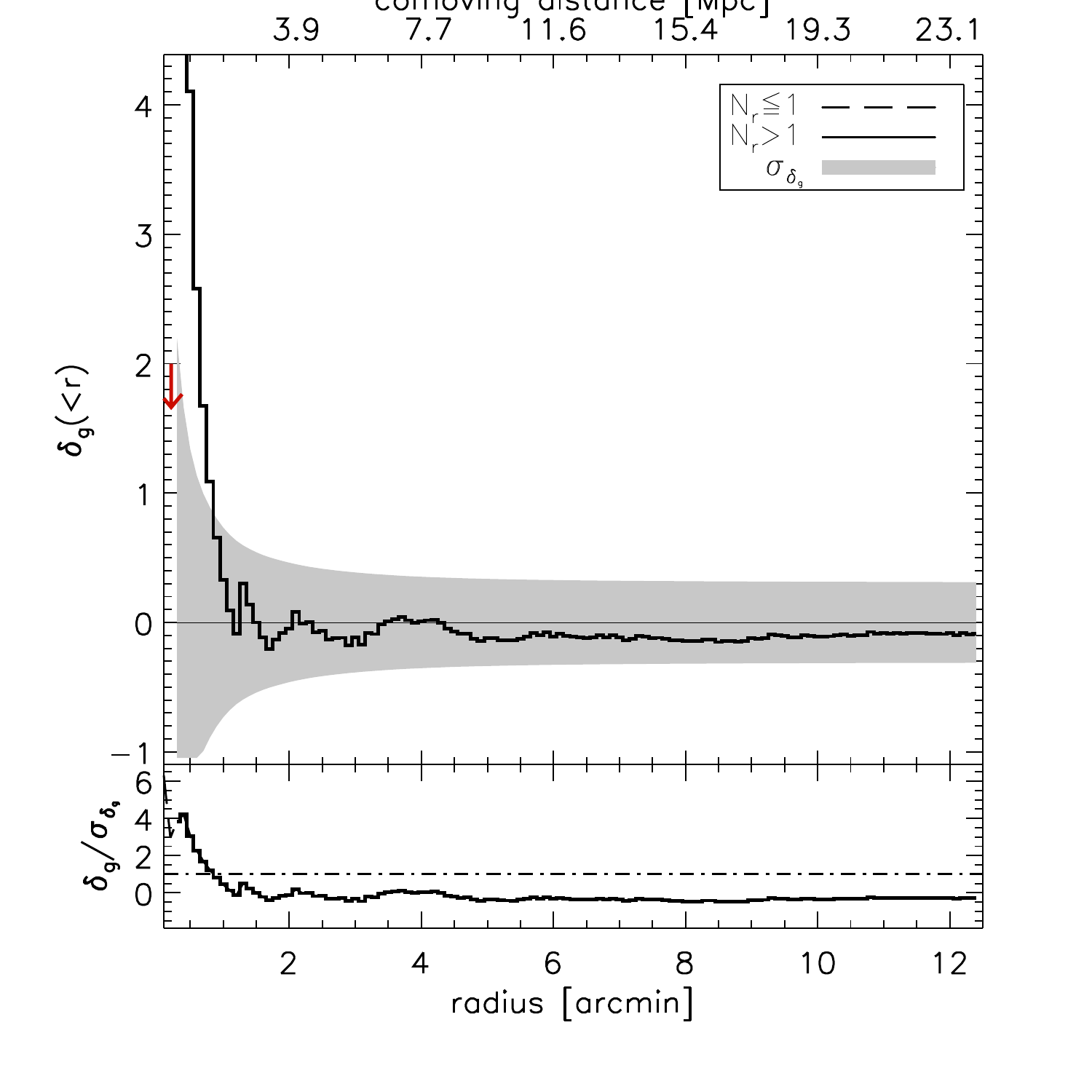}\\
\includegraphics[width=0.31\textwidth]{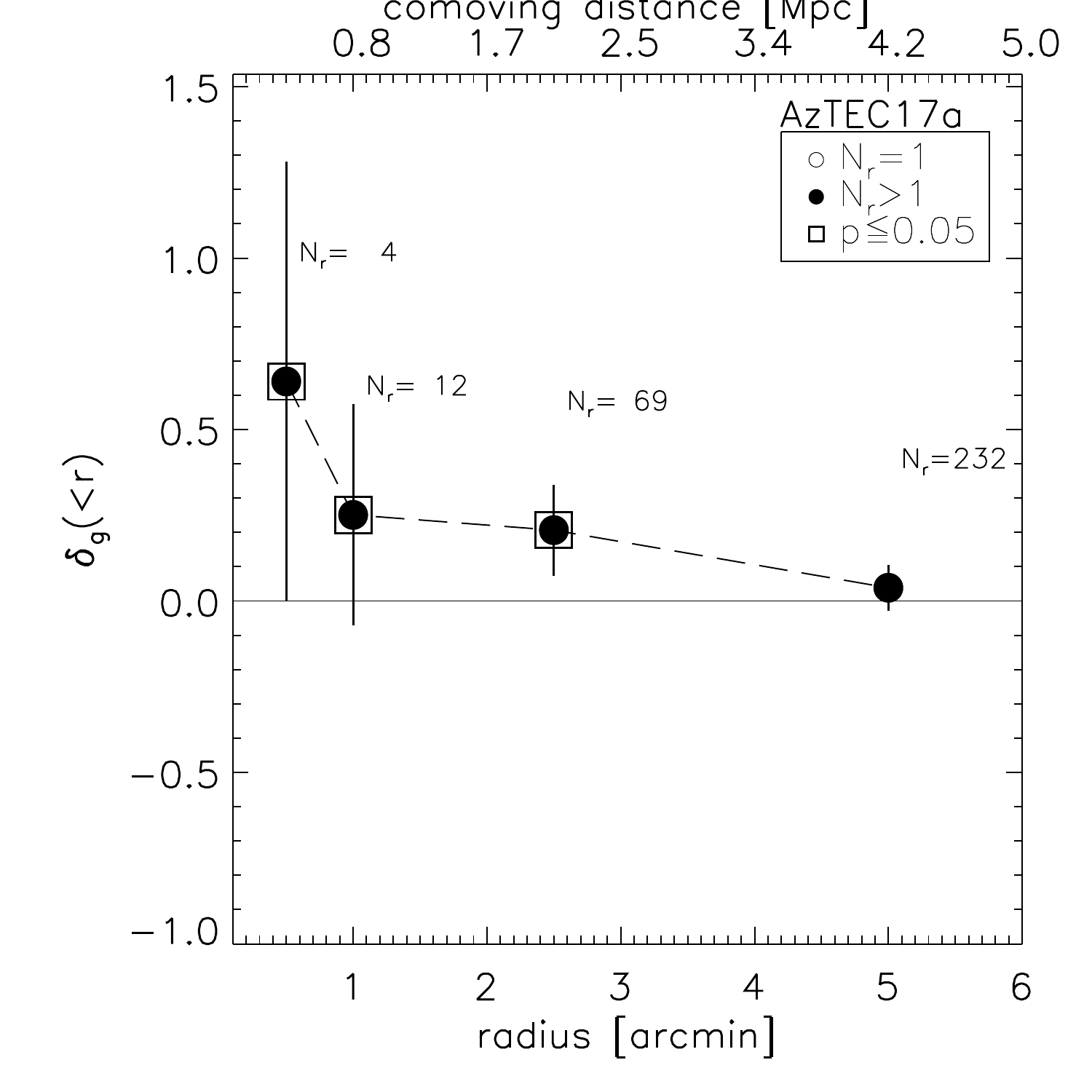}
\includegraphics[width=0.31\textwidth]{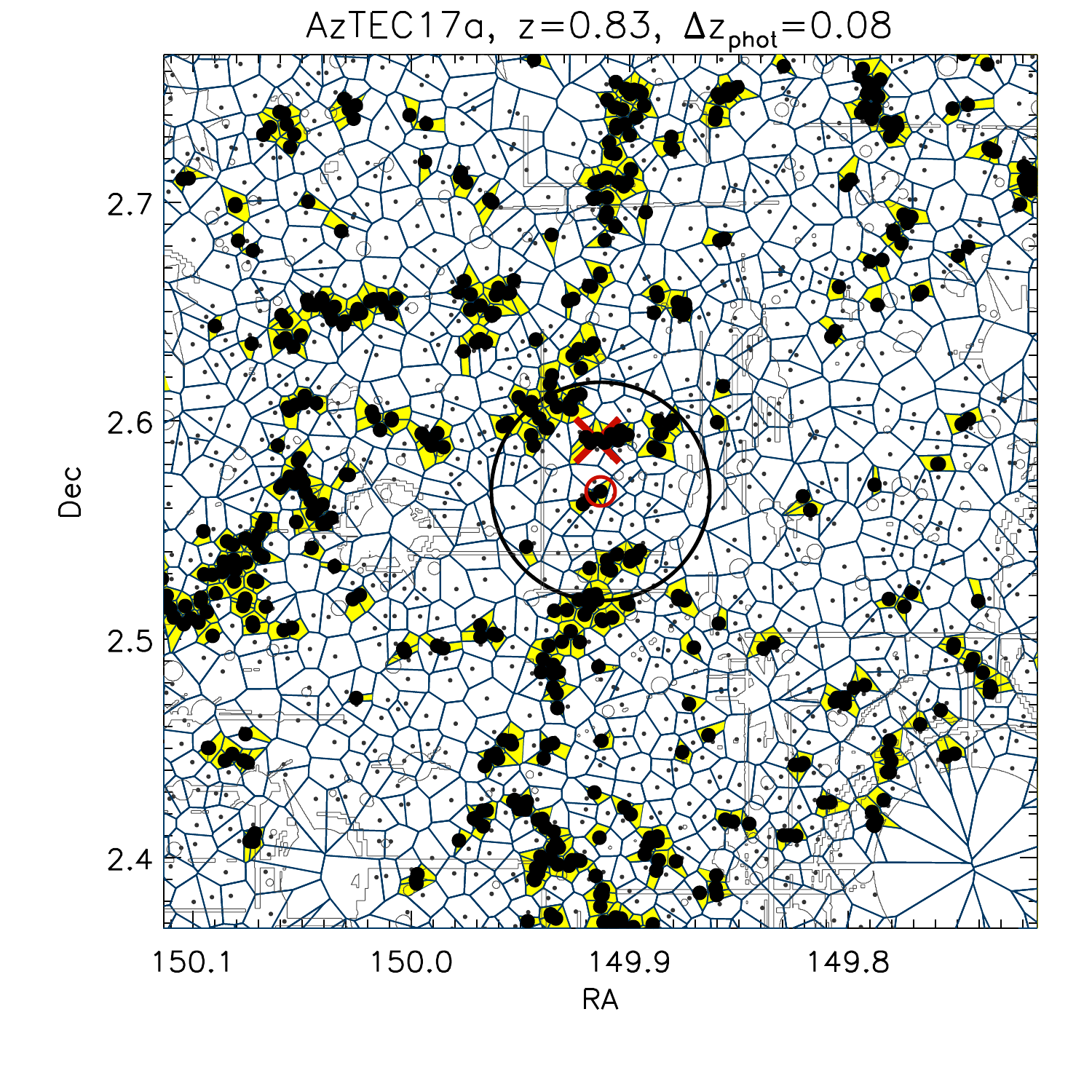}
\includegraphics[width=0.31\textwidth]{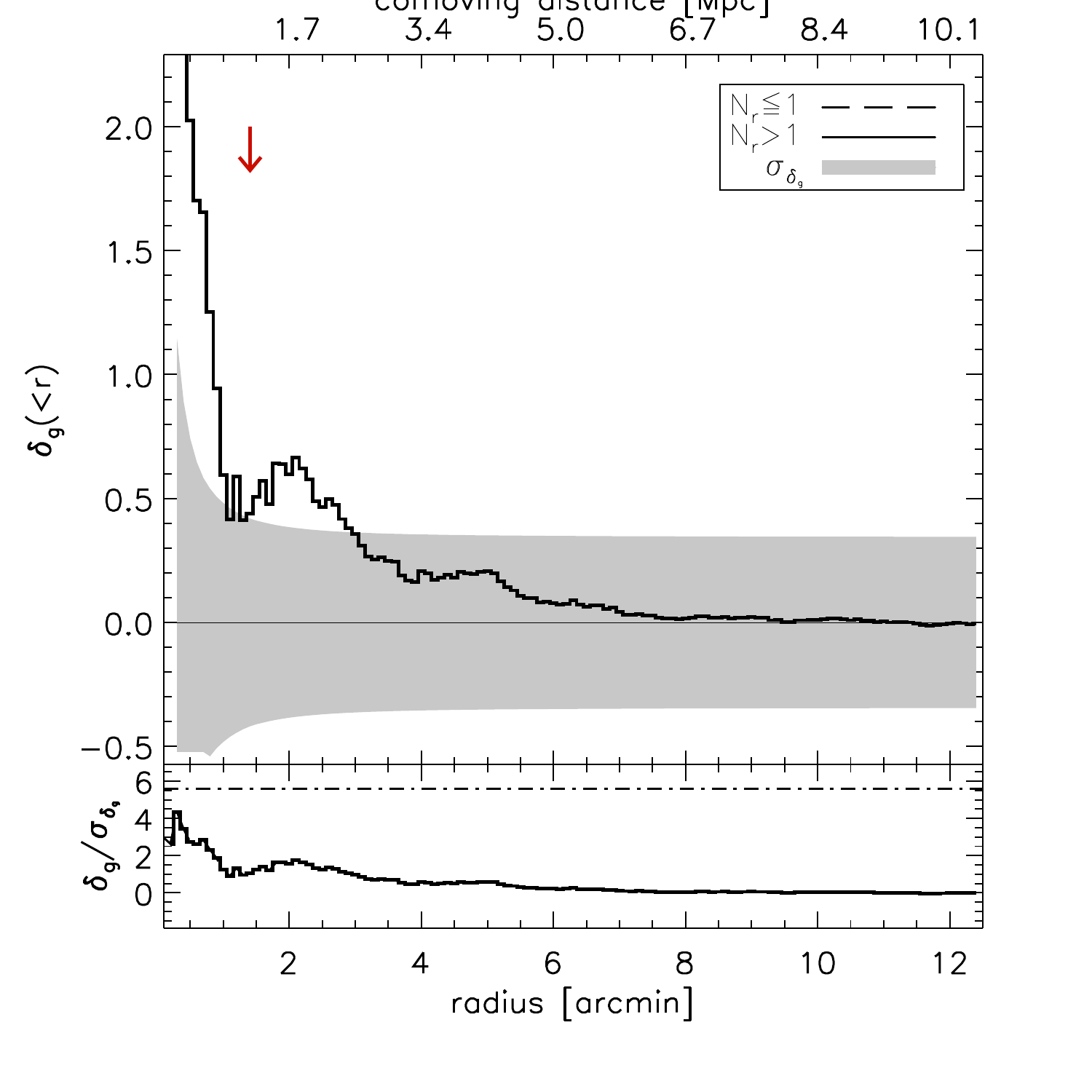}\\
\includegraphics[width=0.31\textwidth]{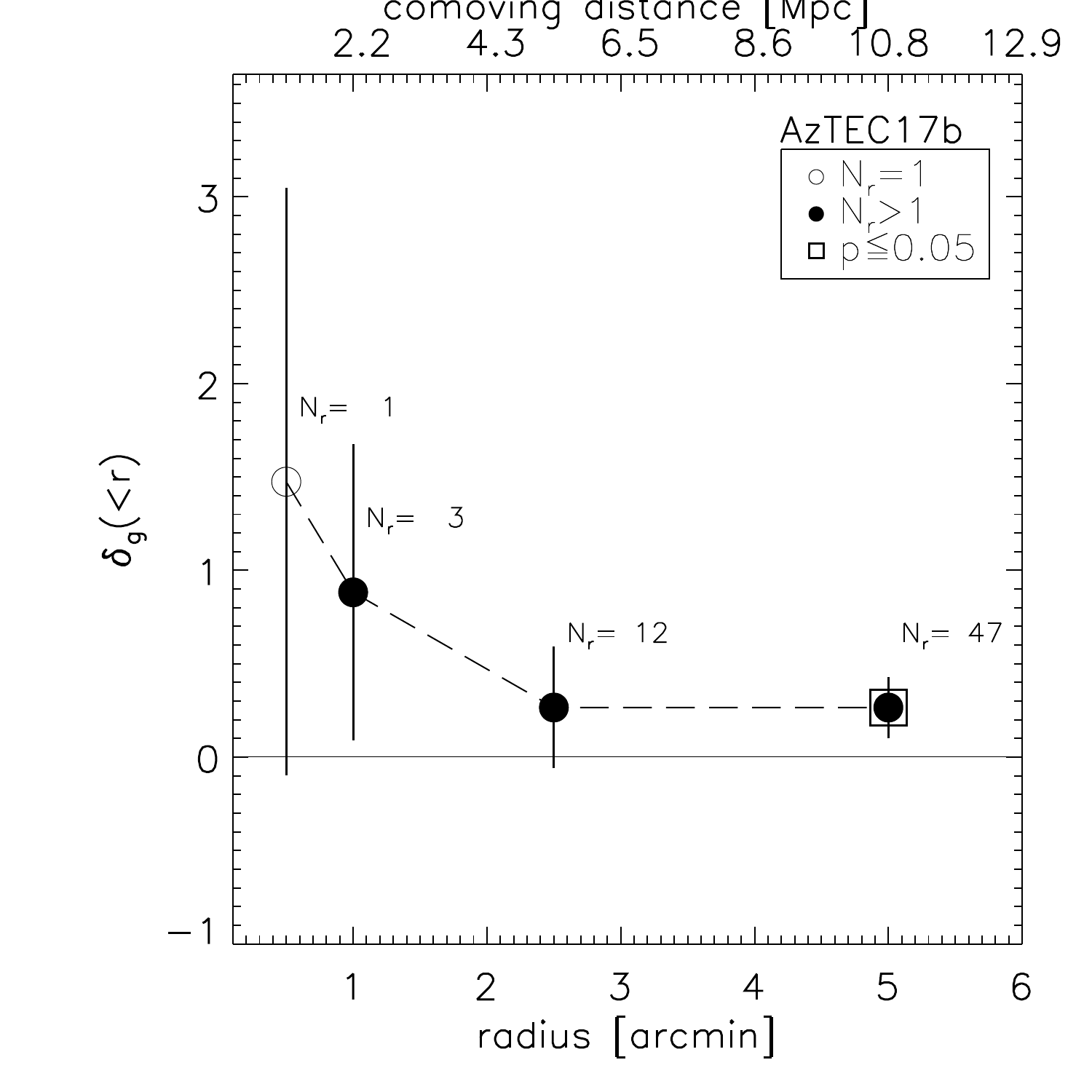}
\includegraphics[width=0.31\textwidth]{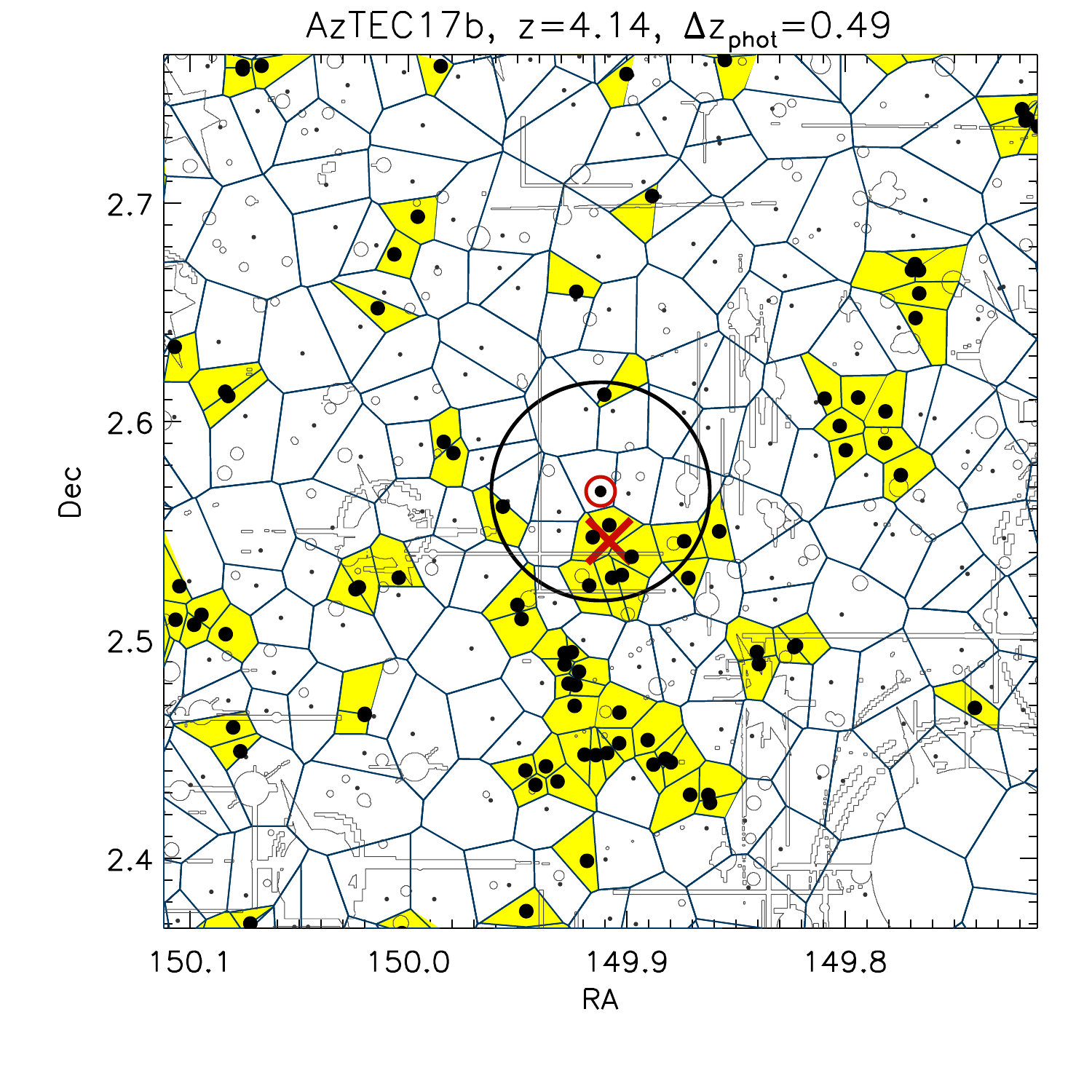}
\includegraphics[width=0.31\textwidth]{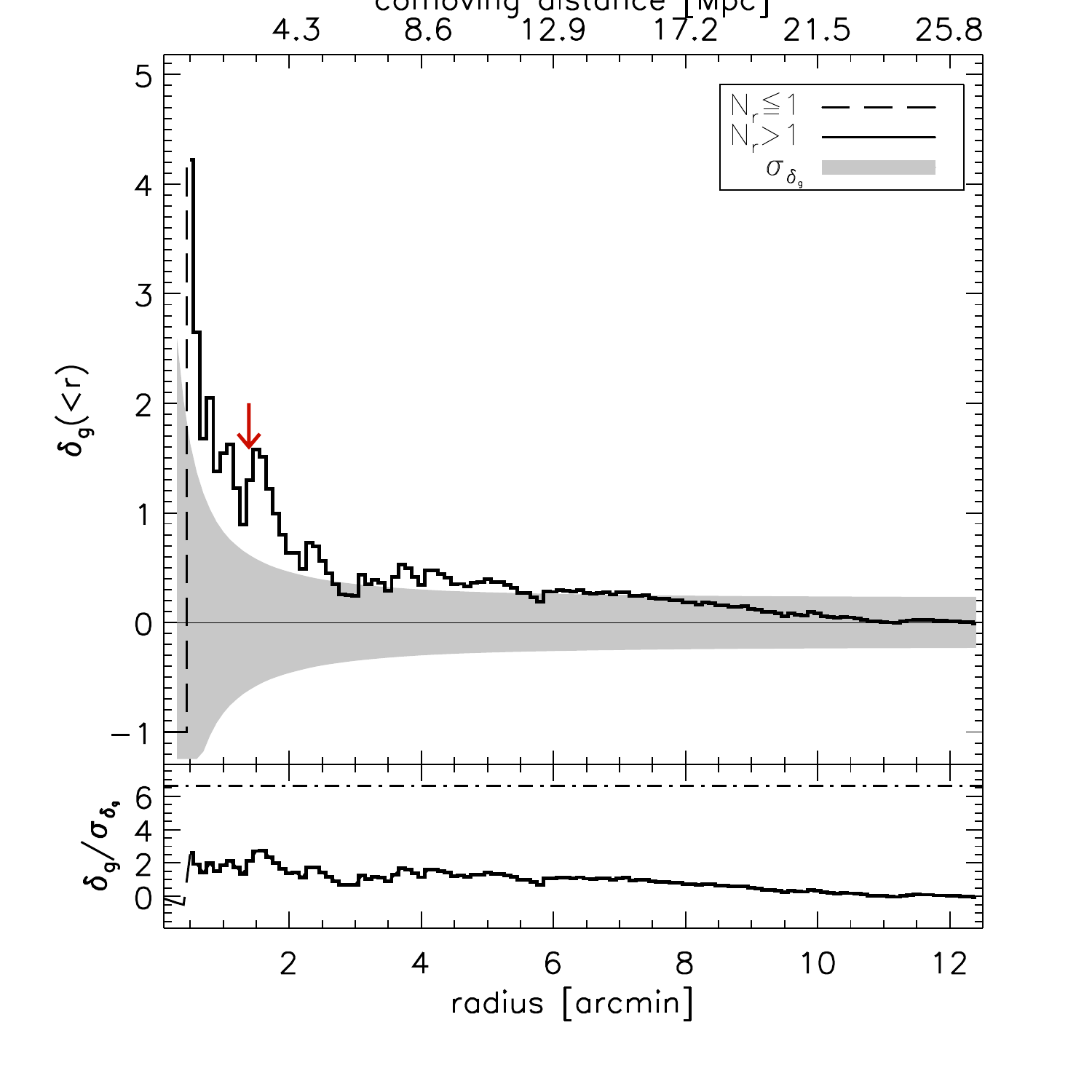}\\
\caption{continued.}
\end{center}
\end{figure*}

\addtocounter{figure}{-1}
\begin{figure*}
\begin{center}
\includegraphics[width=0.31\textwidth]{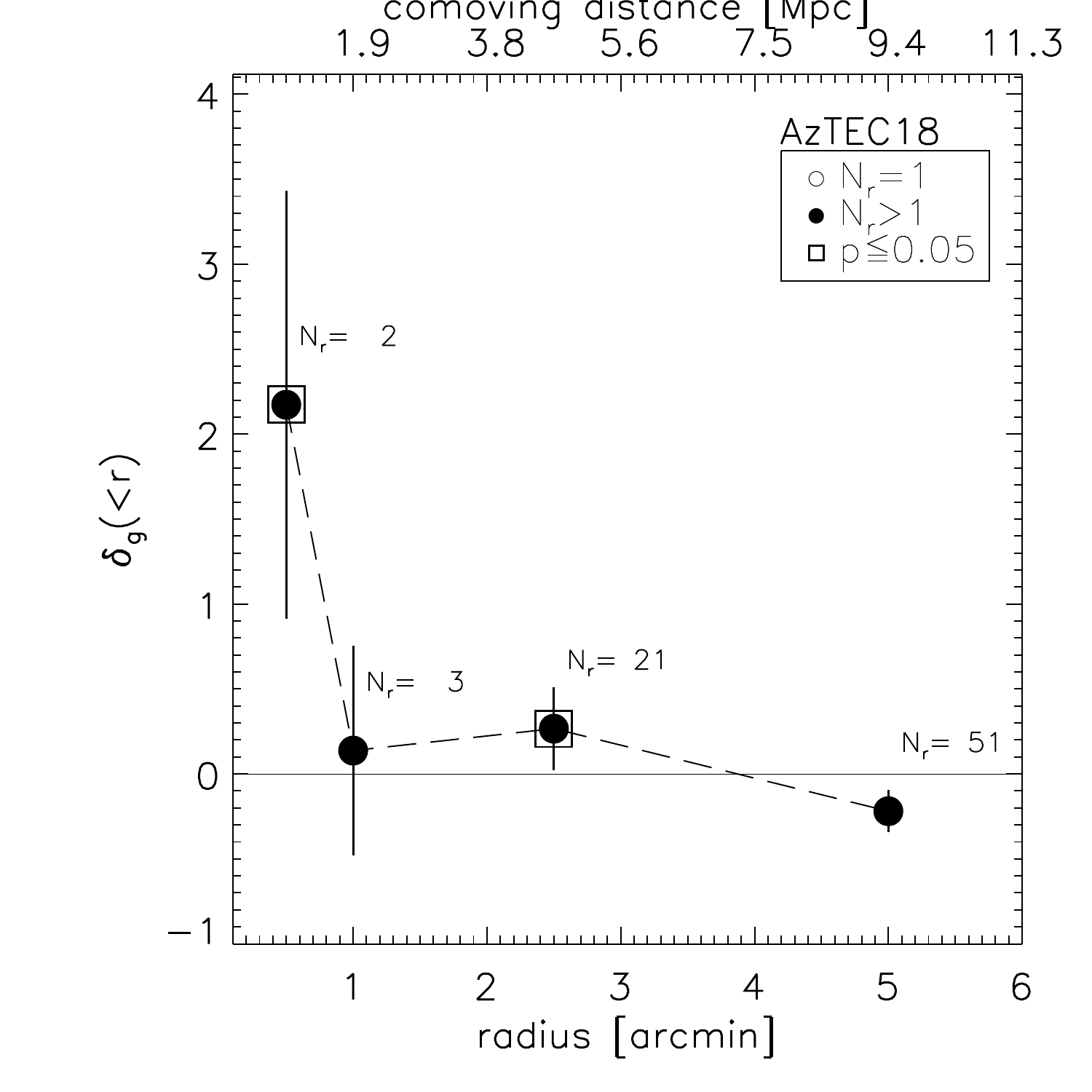}
\includegraphics[width=0.31\textwidth]{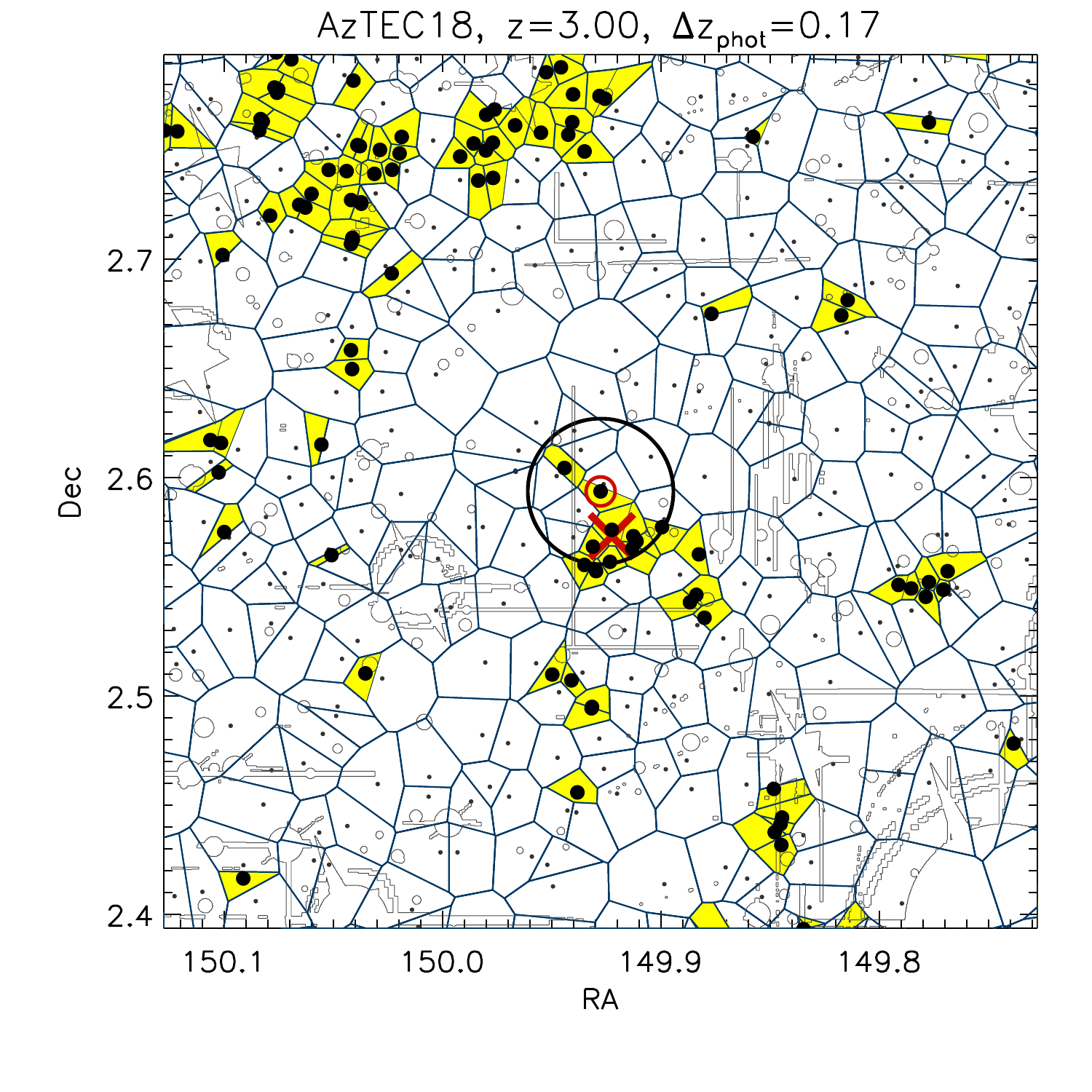}
\includegraphics[width=0.31\textwidth]{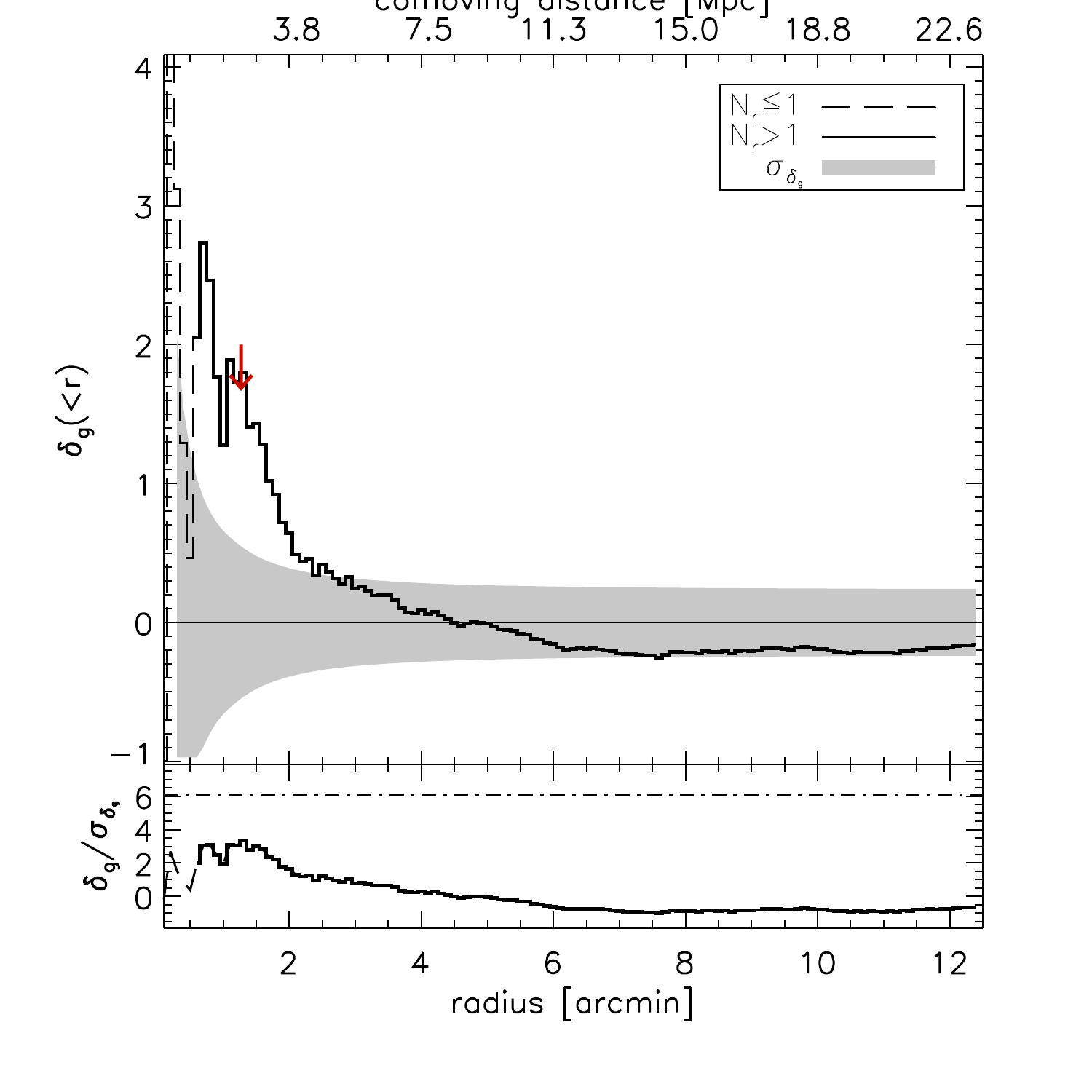}\\
\includegraphics[width=0.31\textwidth]{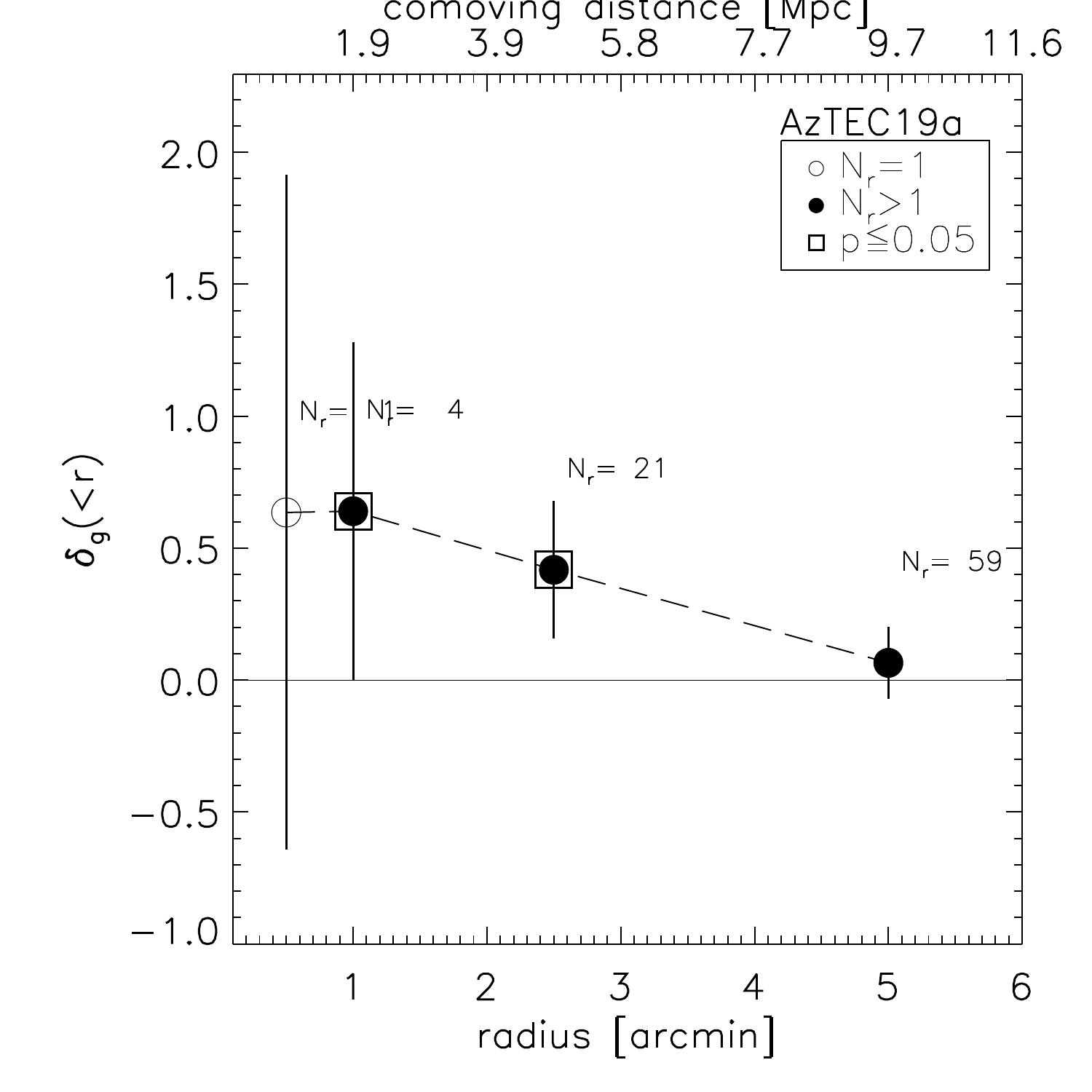}
\includegraphics[width=0.31\textwidth]{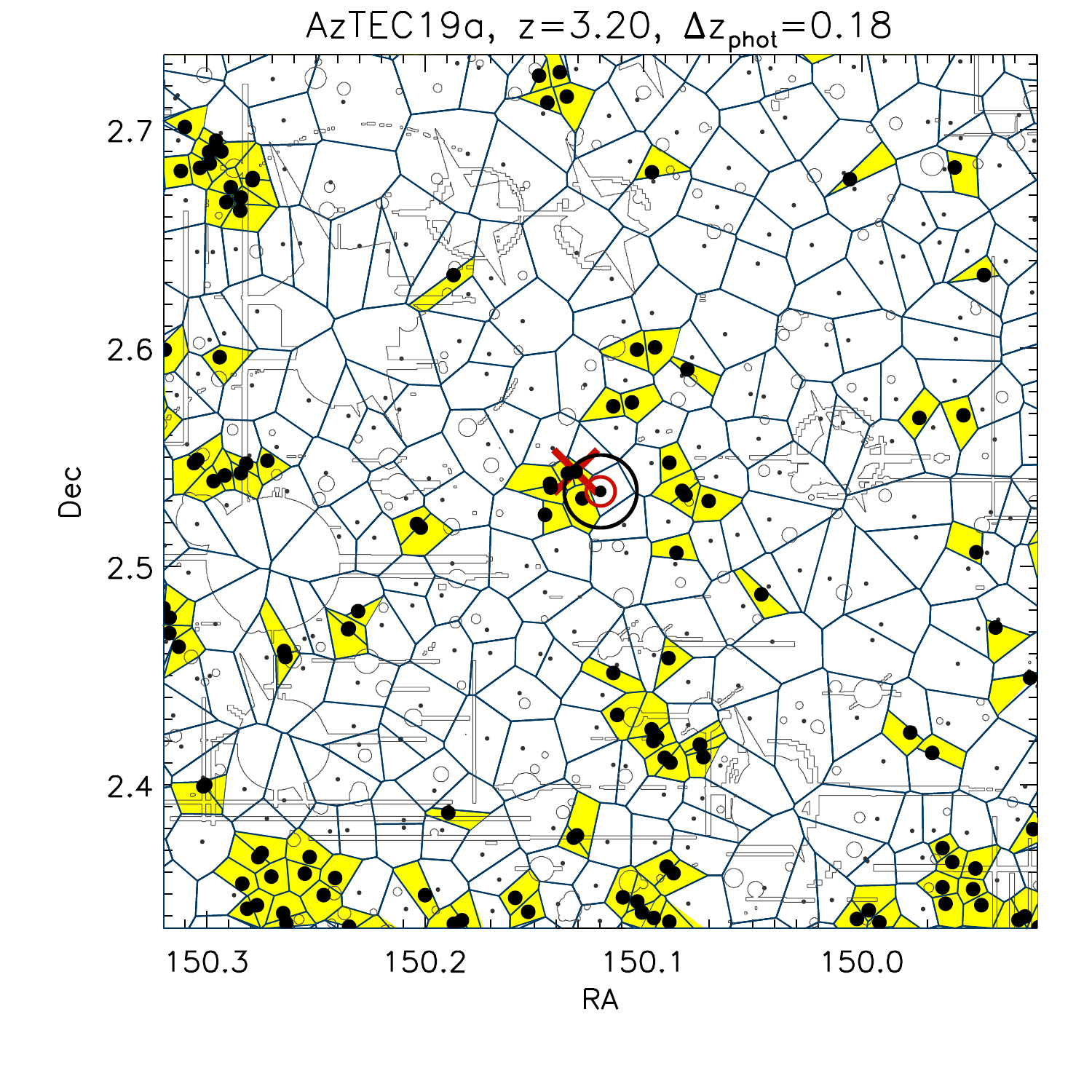}
\includegraphics[width=0.31\textwidth]{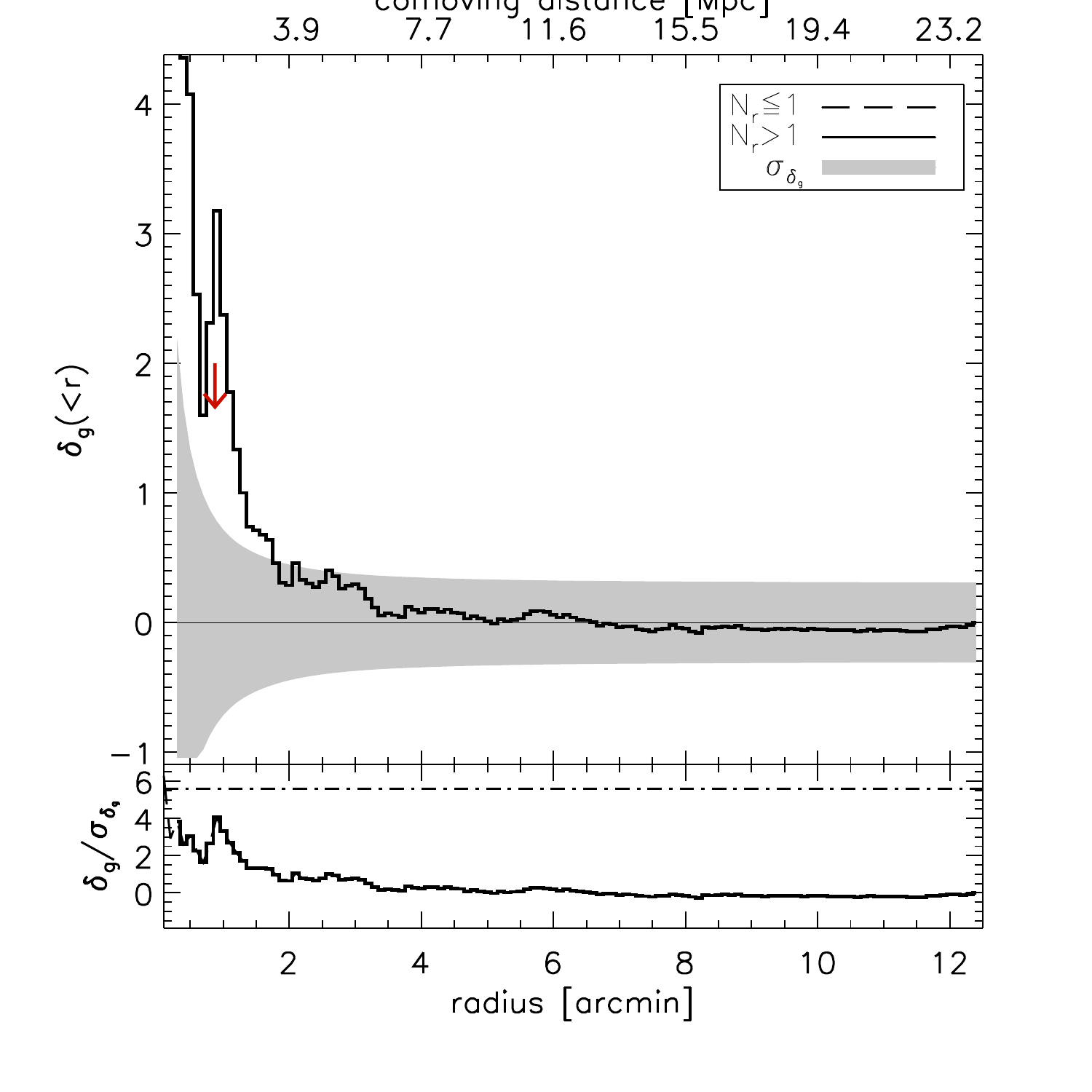}\\
\includegraphics[width=0.31\textwidth]{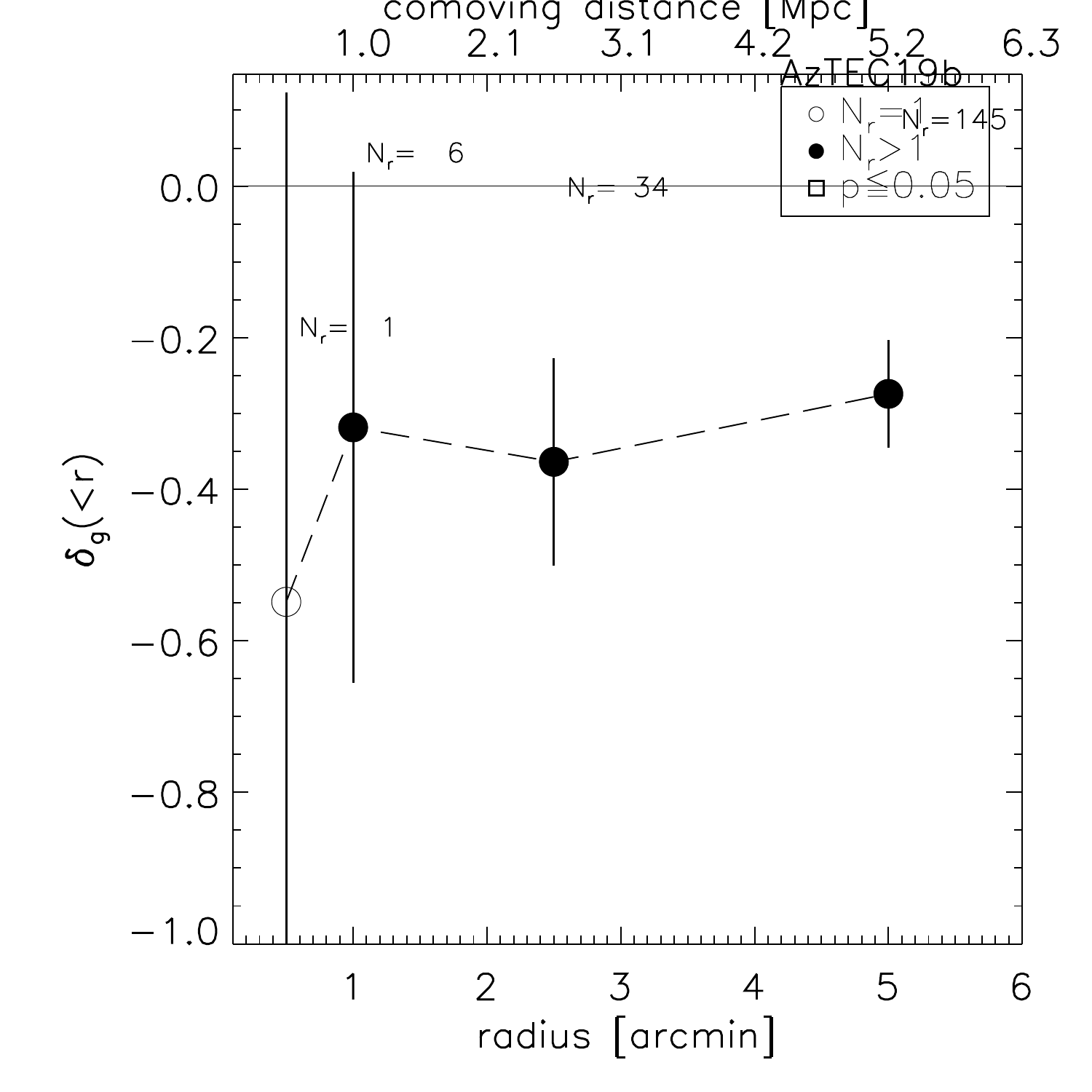}
\includegraphics[width=0.31\textwidth]{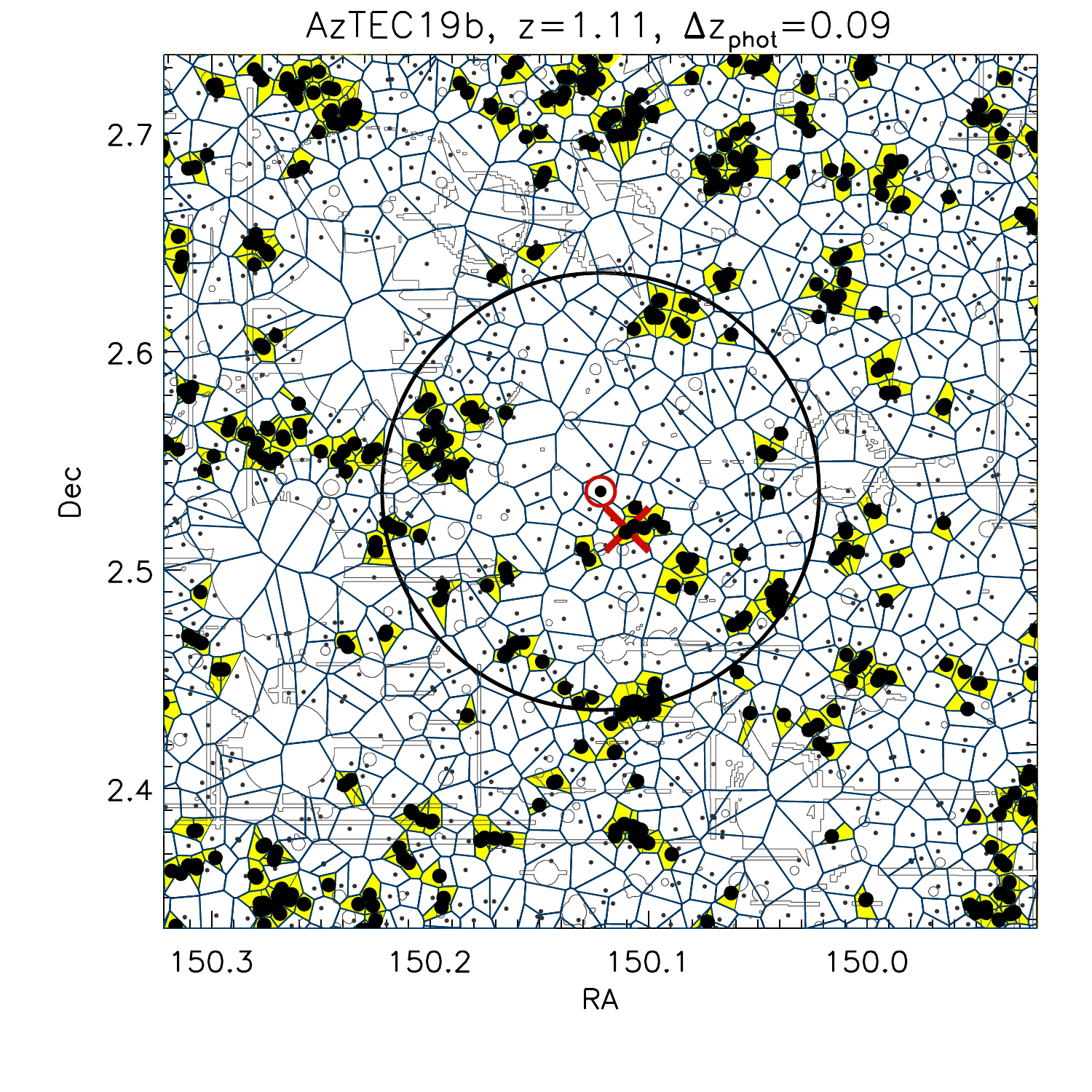}
\includegraphics[width=0.31\textwidth]{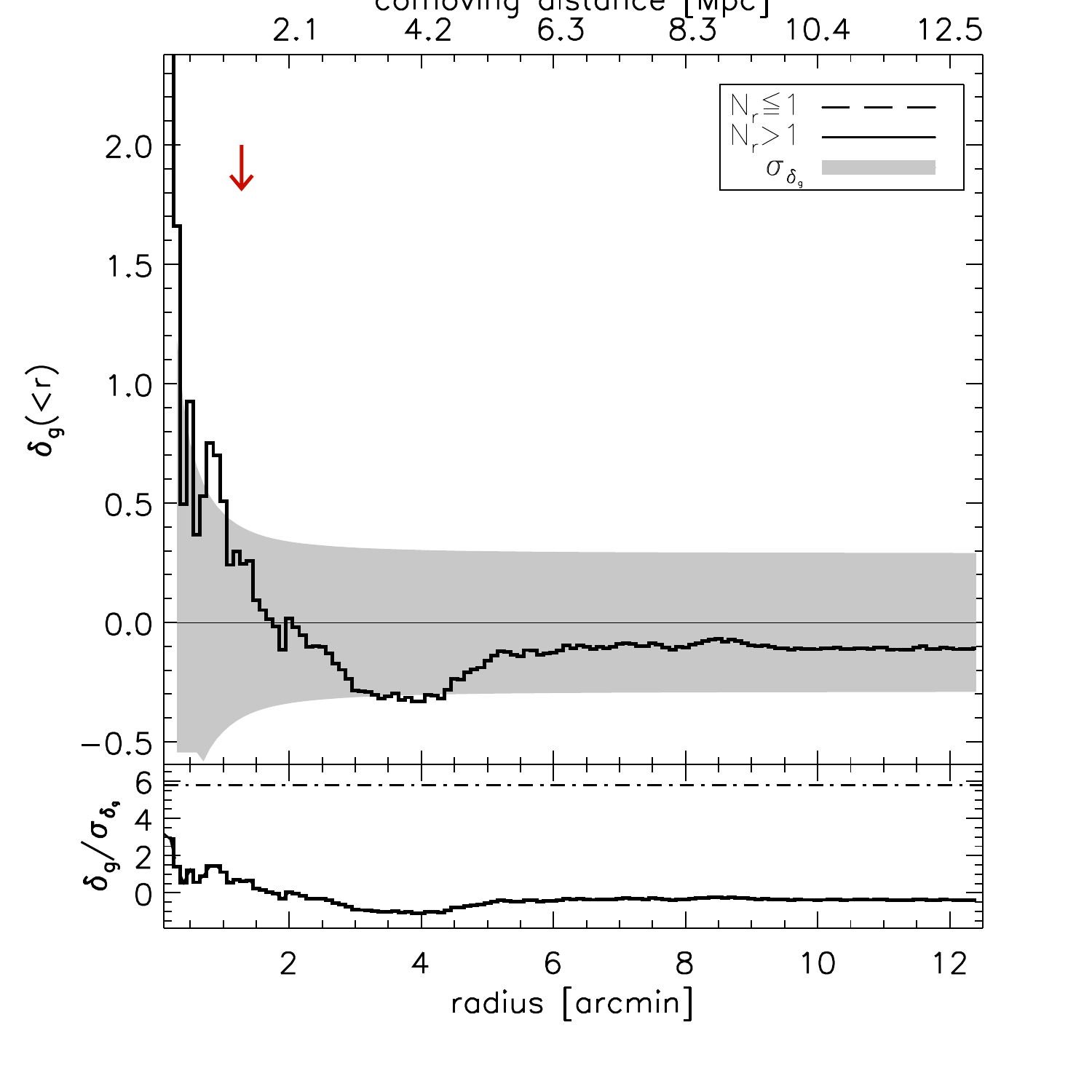}\\
\includegraphics[width=0.31\textwidth]{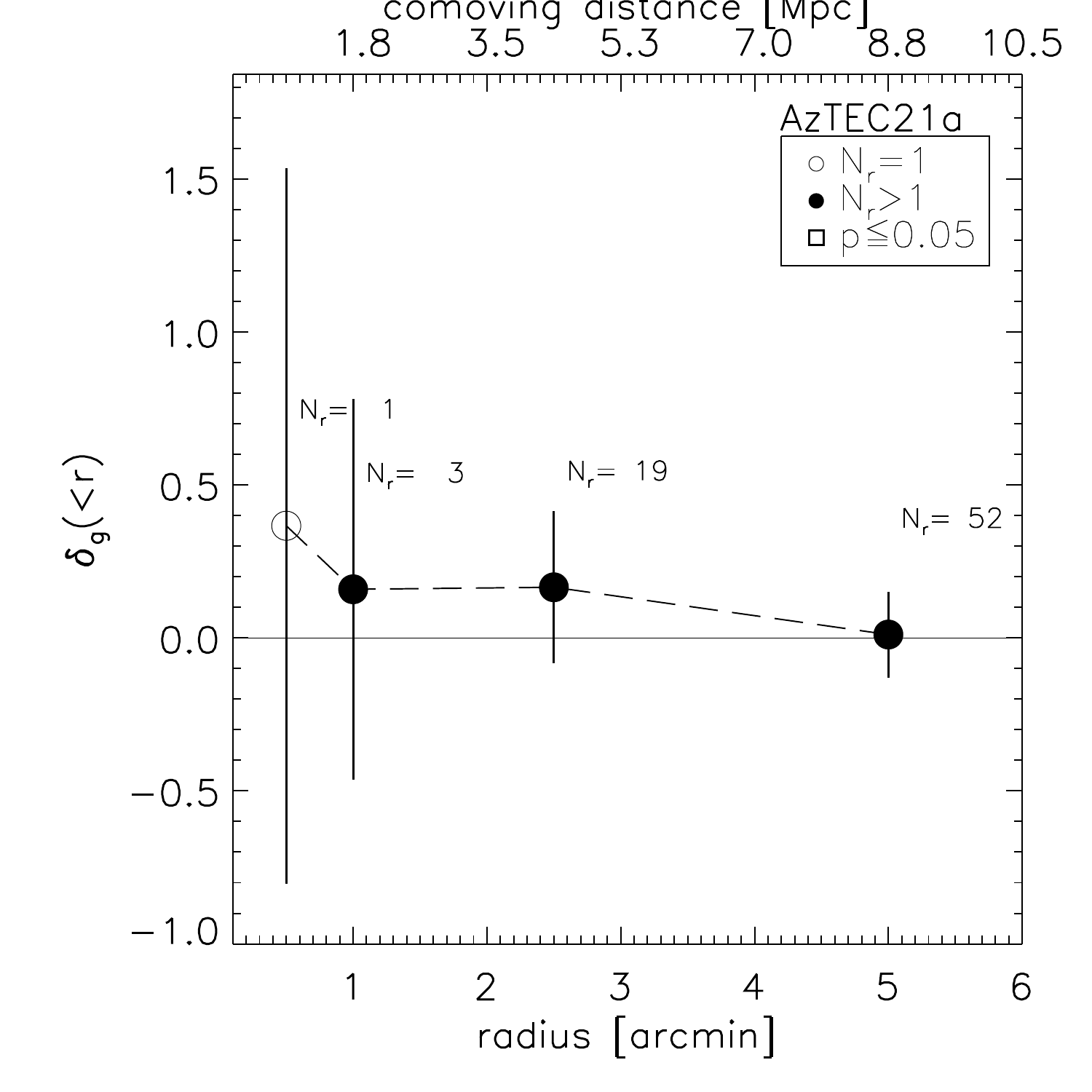}
\includegraphics[width=0.31\textwidth]{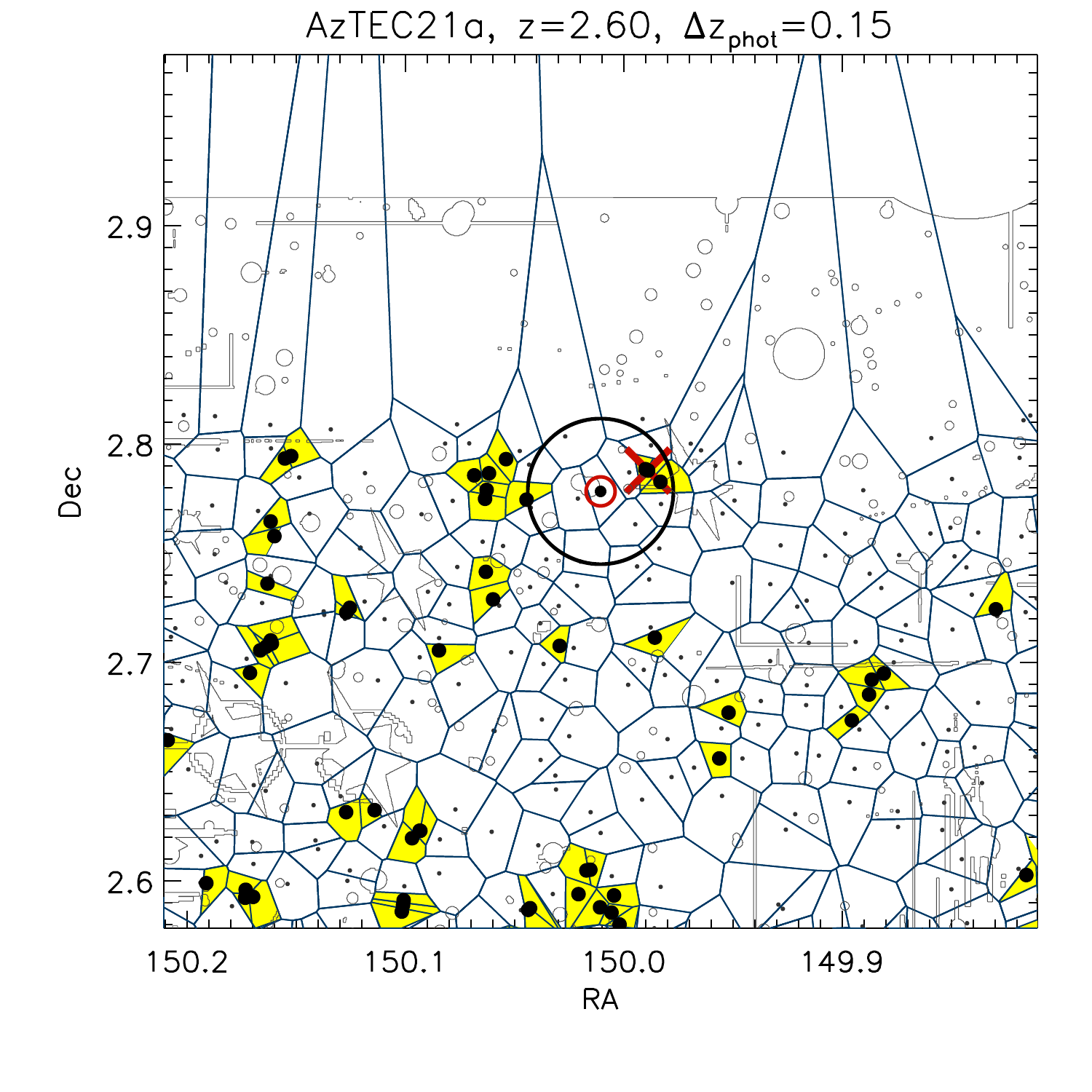}
\includegraphics[width=0.31\textwidth]{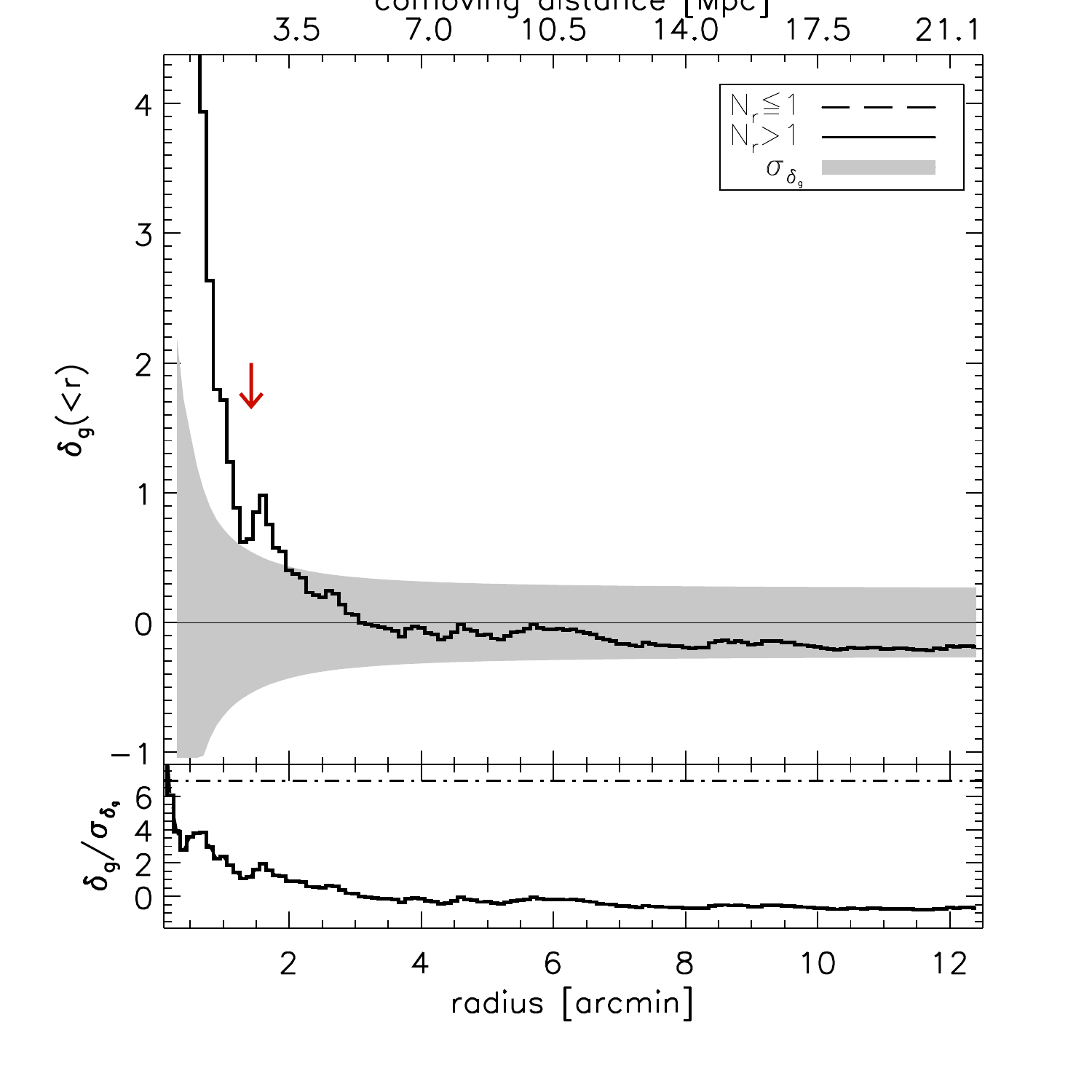}\\
\caption{continued.}
\end{center}
\end{figure*}

\addtocounter{figure}{-1}
\begin{figure*}
\begin{center}
\includegraphics[width=0.31\textwidth]{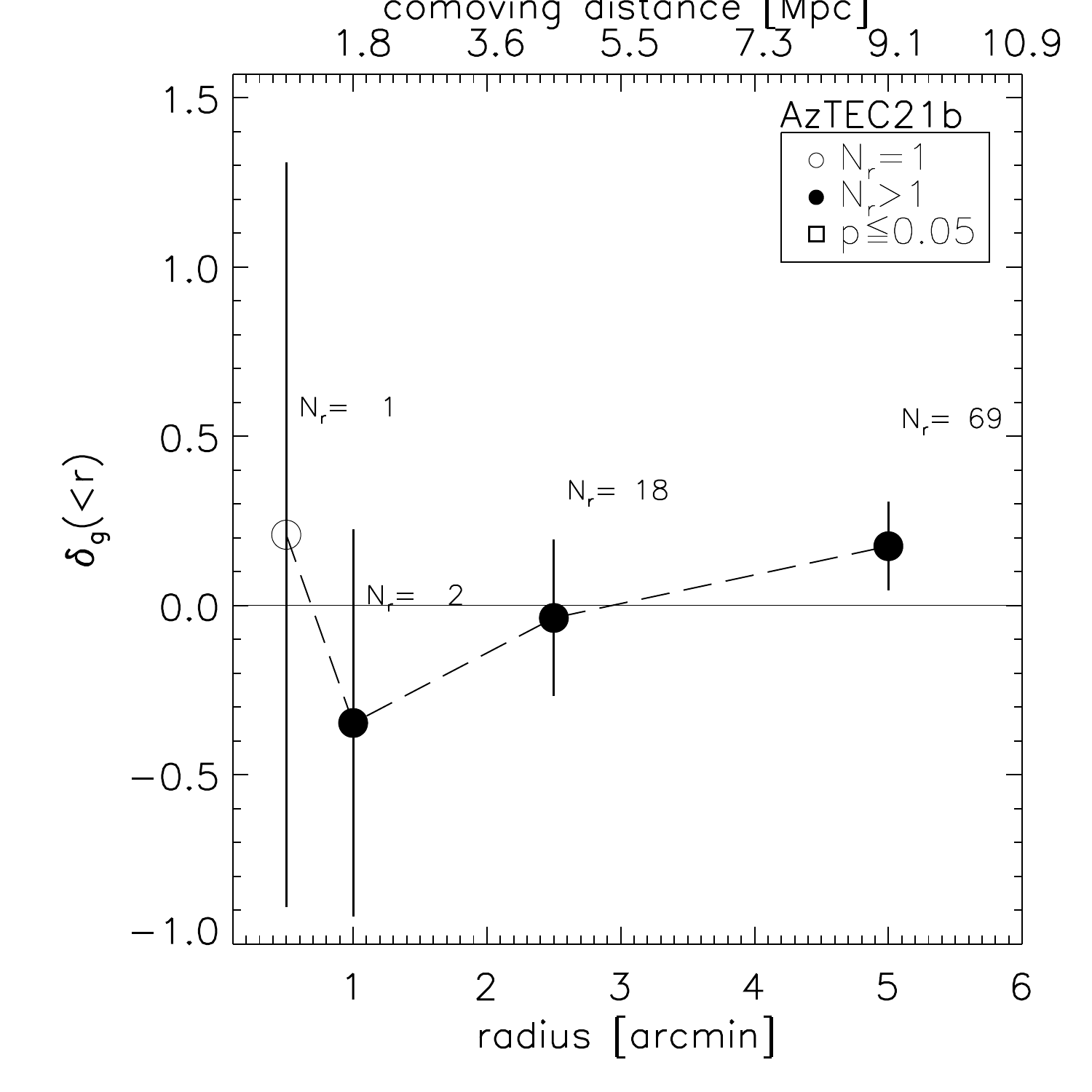}
\includegraphics[width=0.31\textwidth]{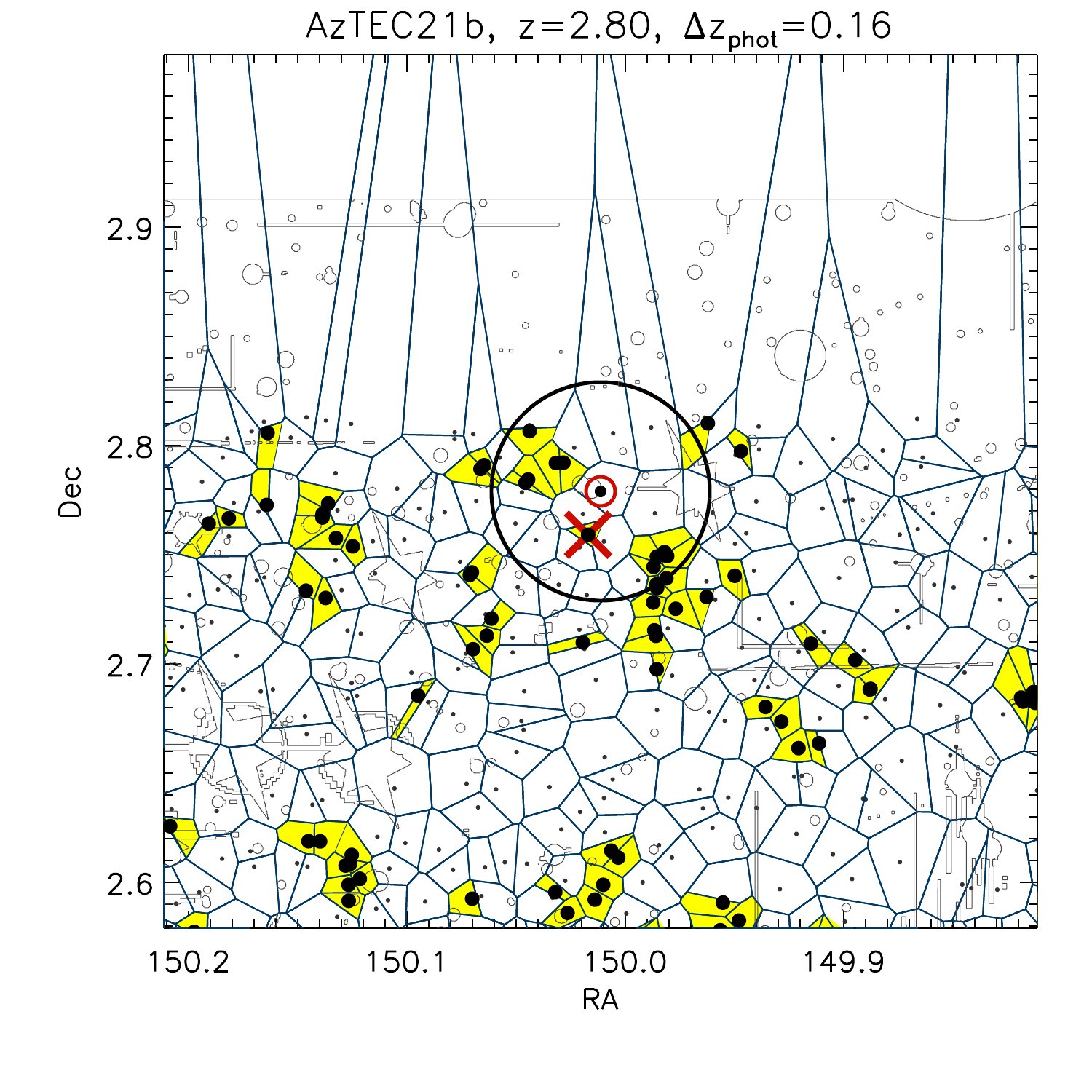}
\includegraphics[width=0.31\textwidth]{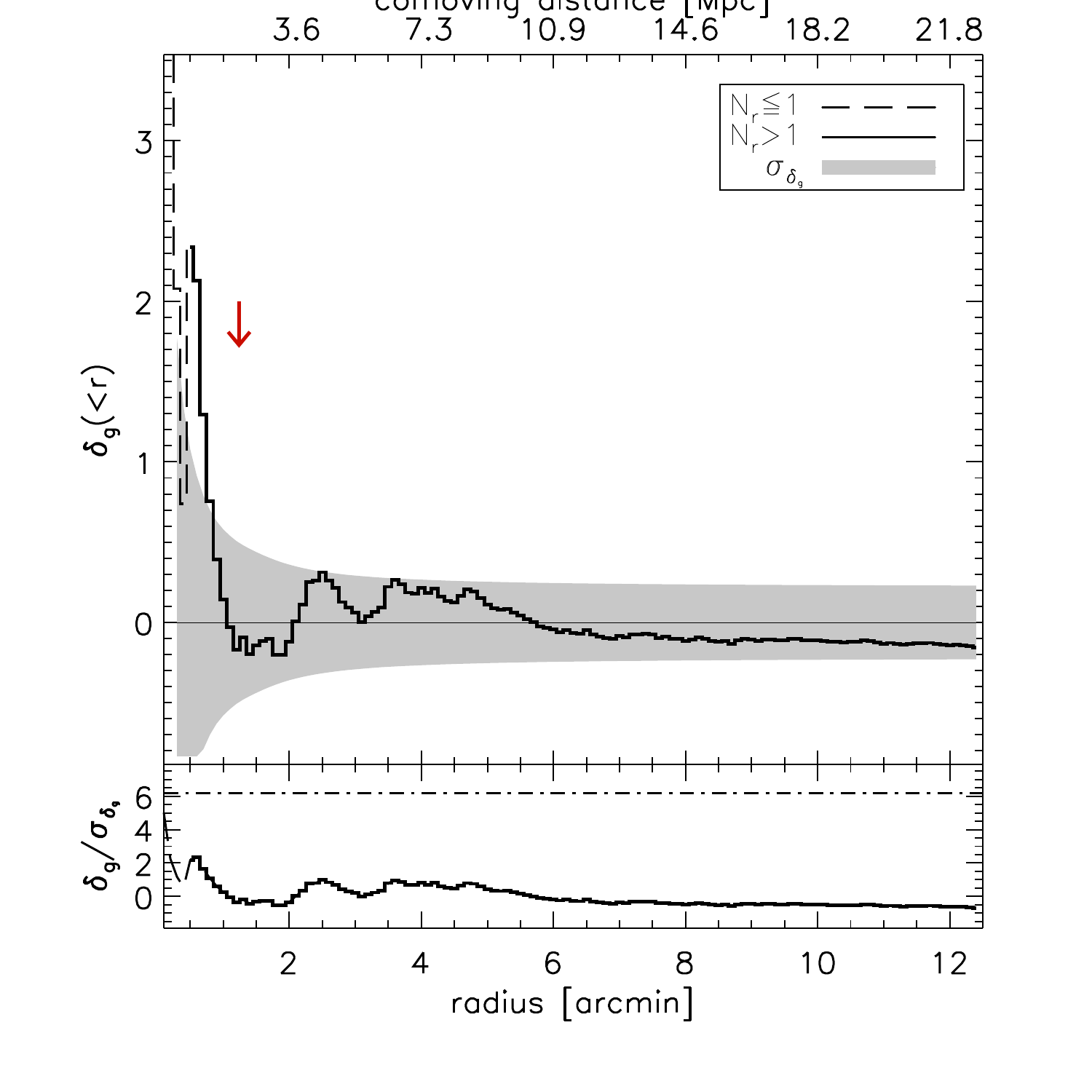}\\
\includegraphics[width=0.31\textwidth]{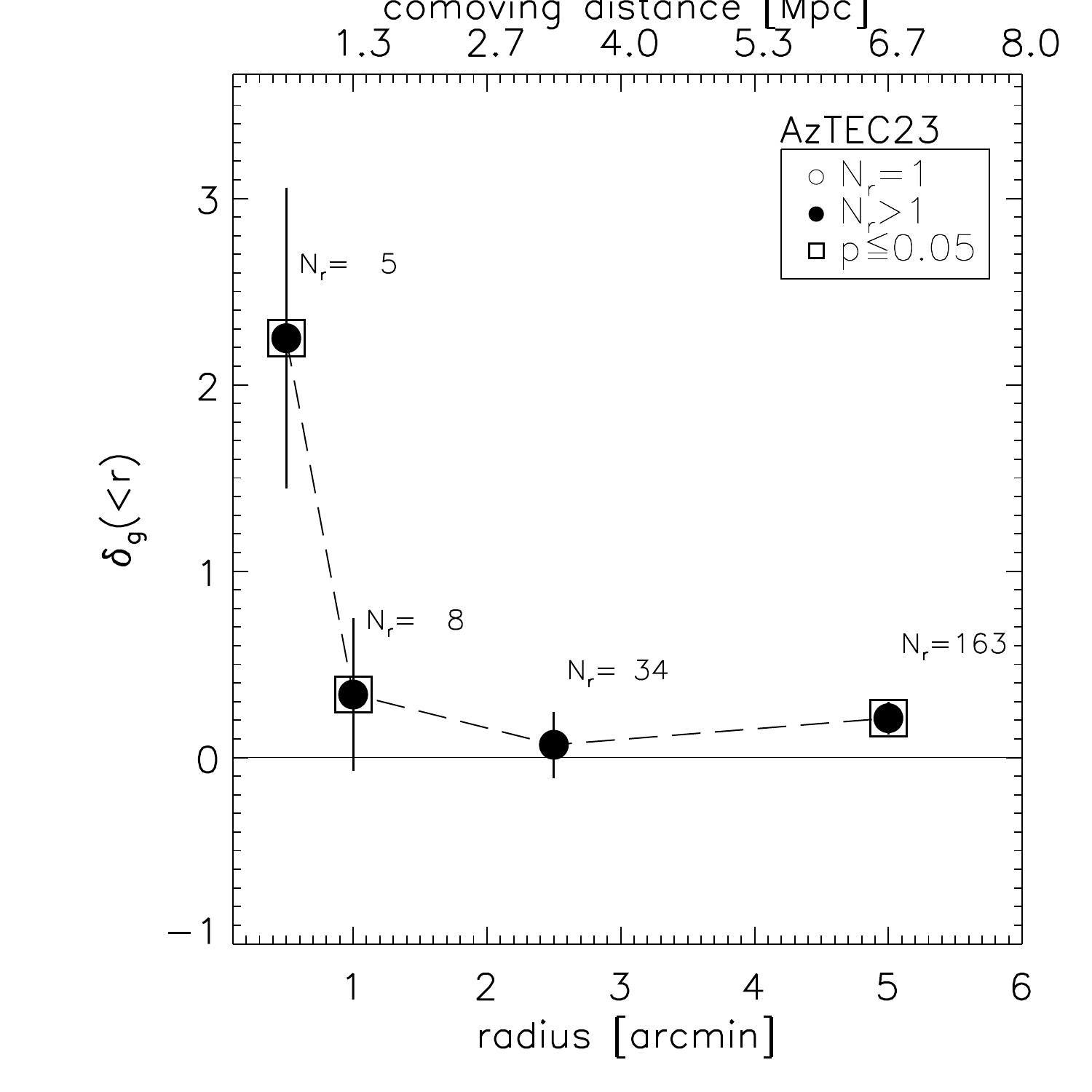}
\includegraphics[width=0.31\textwidth]{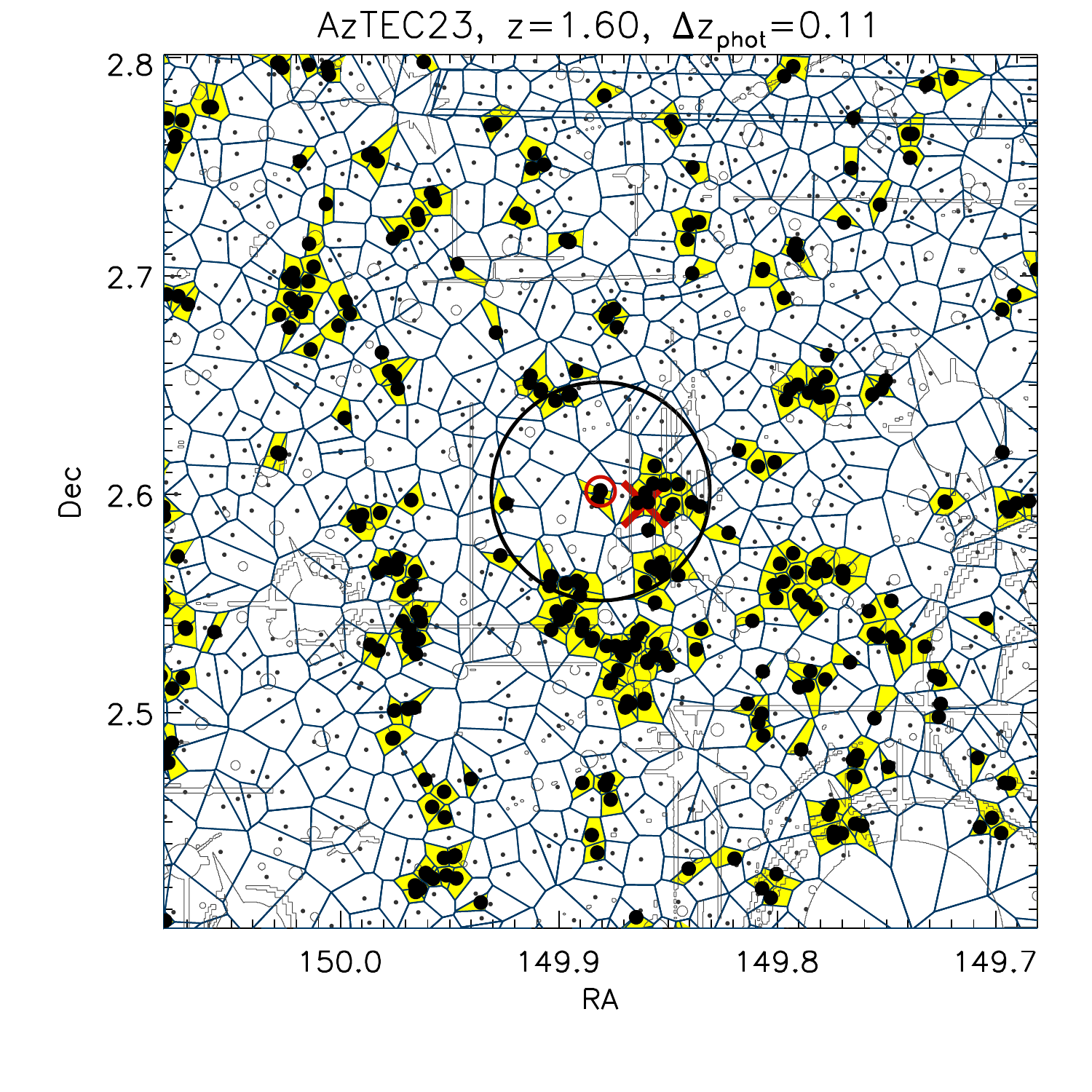}
\includegraphics[width=0.31\textwidth]{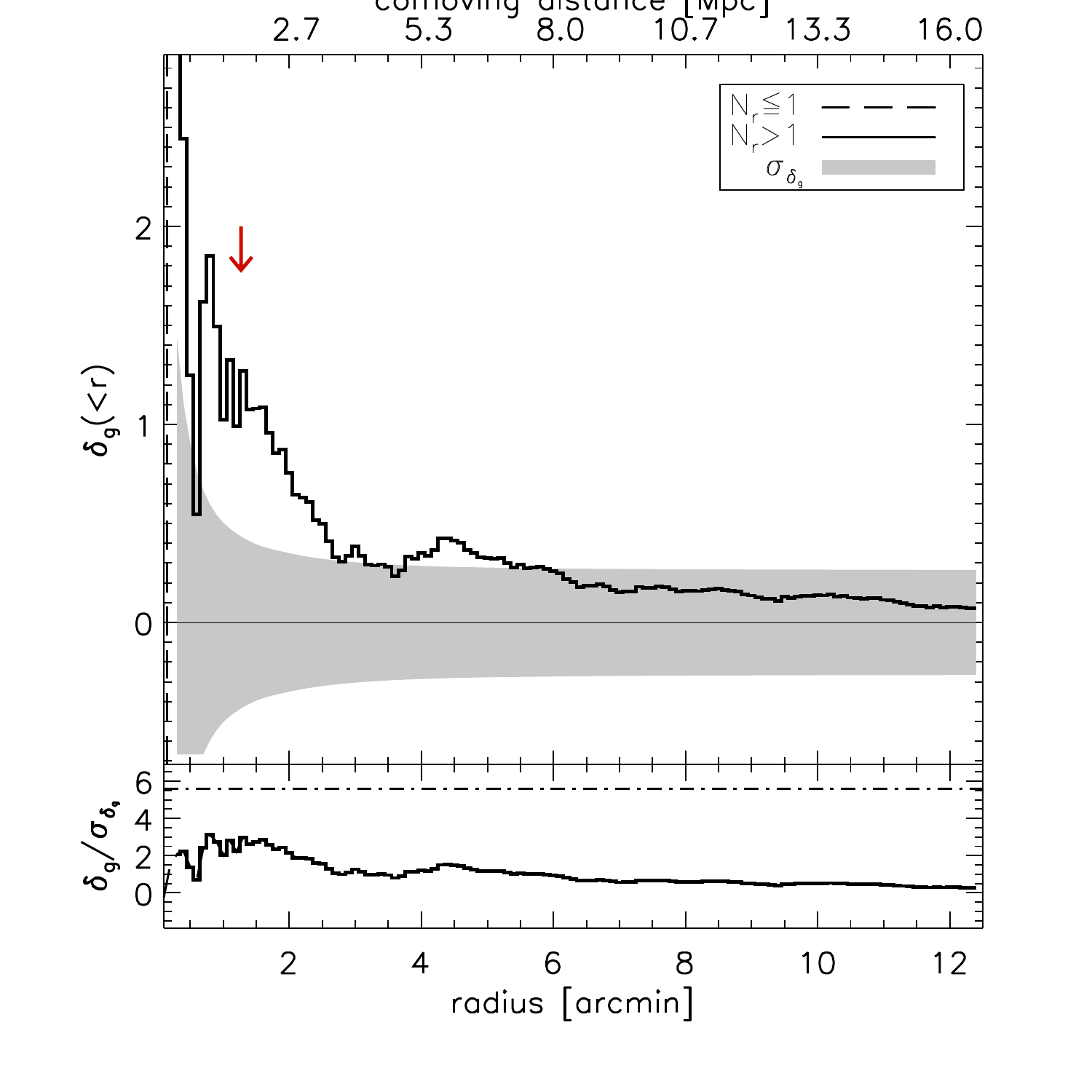}\\
\includegraphics[width=0.31\textwidth]{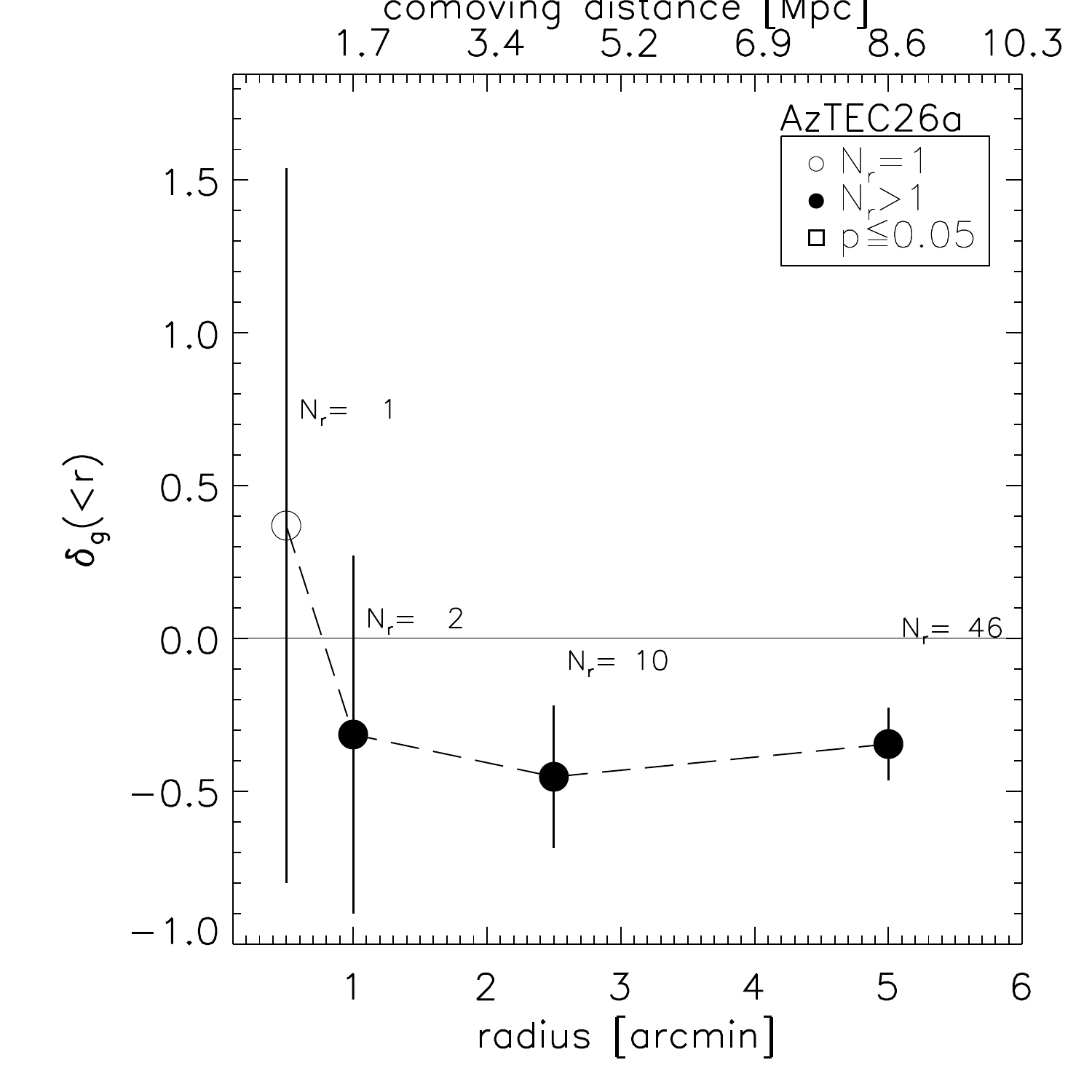}
\includegraphics[width=0.31\textwidth]{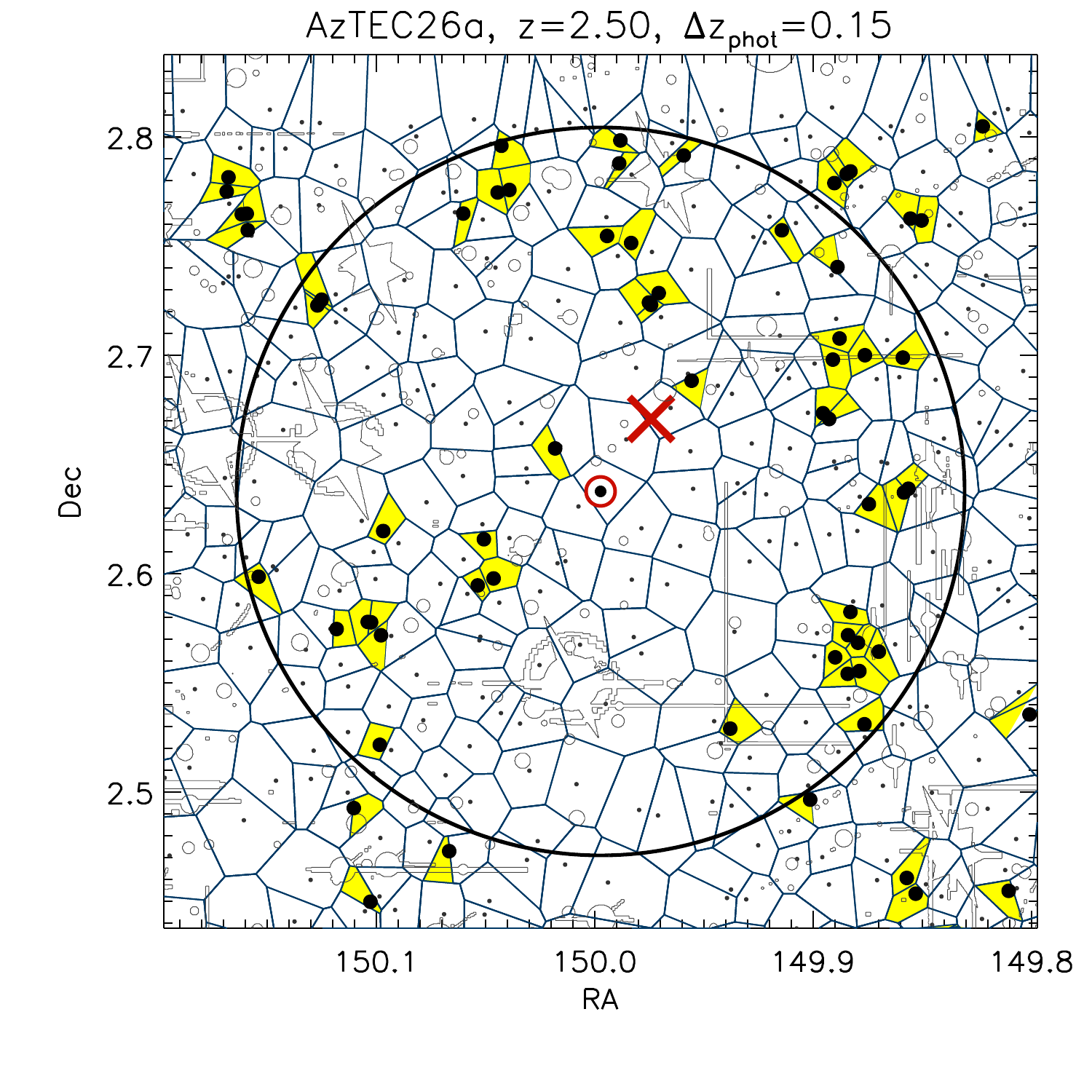}
\includegraphics[width=0.31\textwidth]{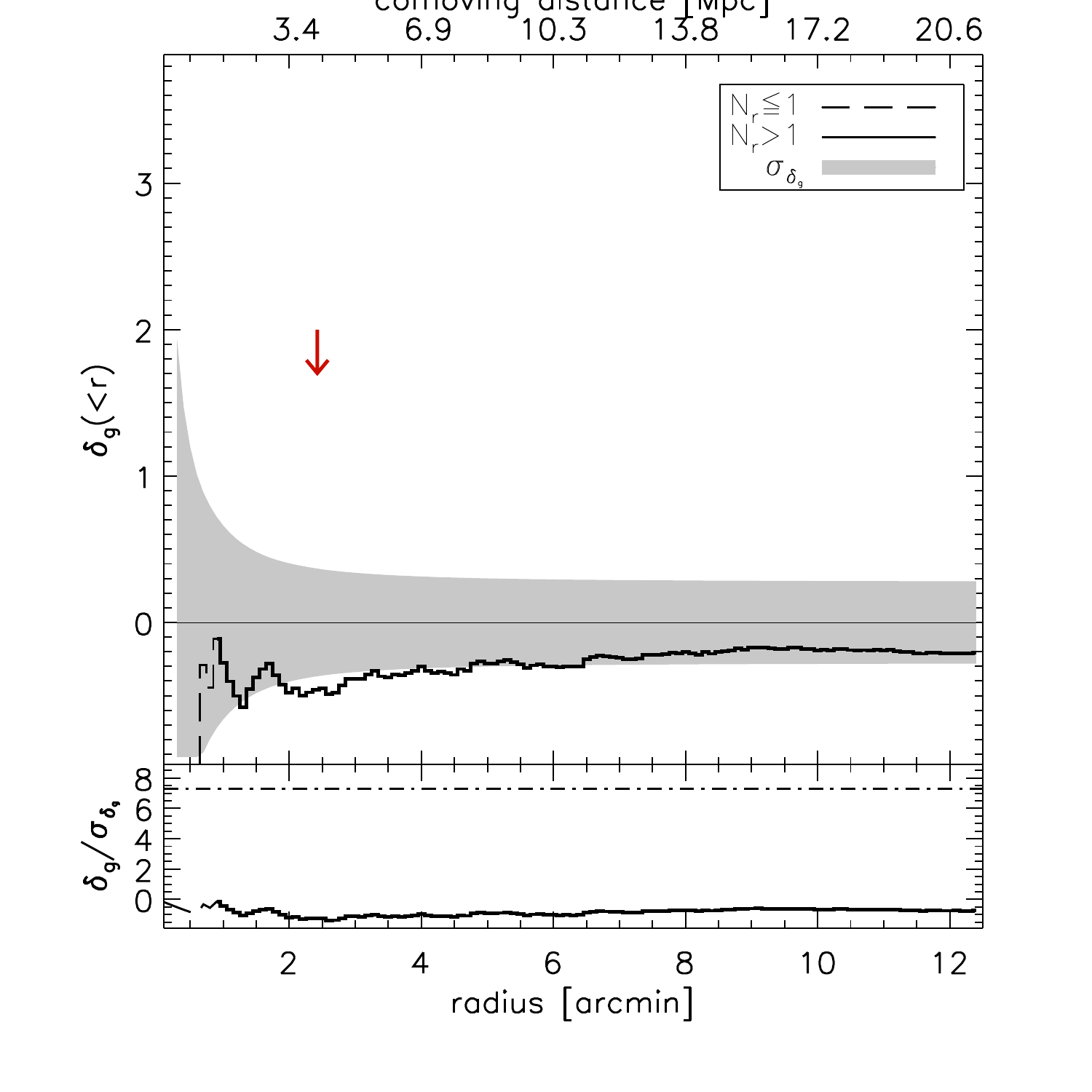}\\
\includegraphics[width=0.31\textwidth]{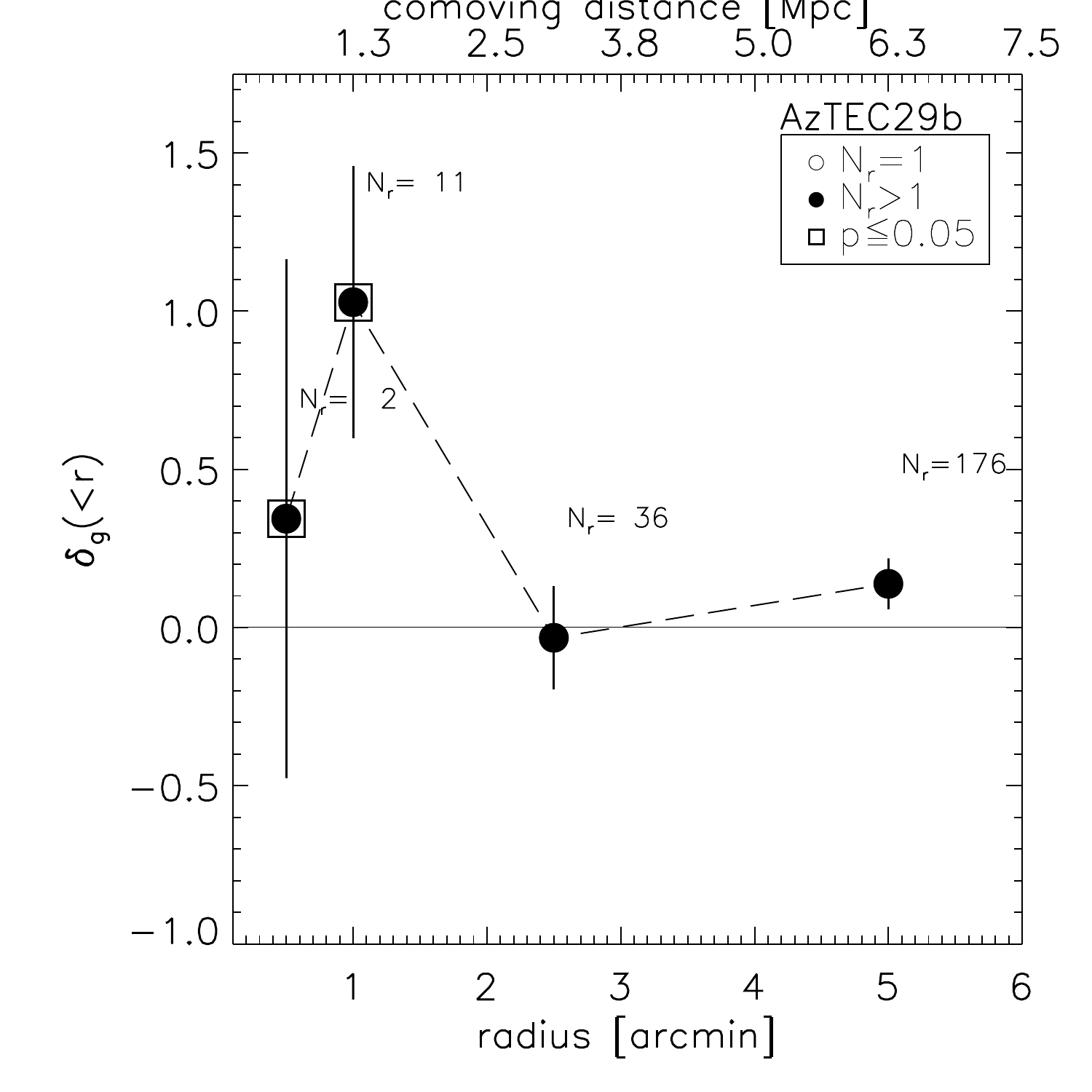}
\includegraphics[width=0.31\textwidth]{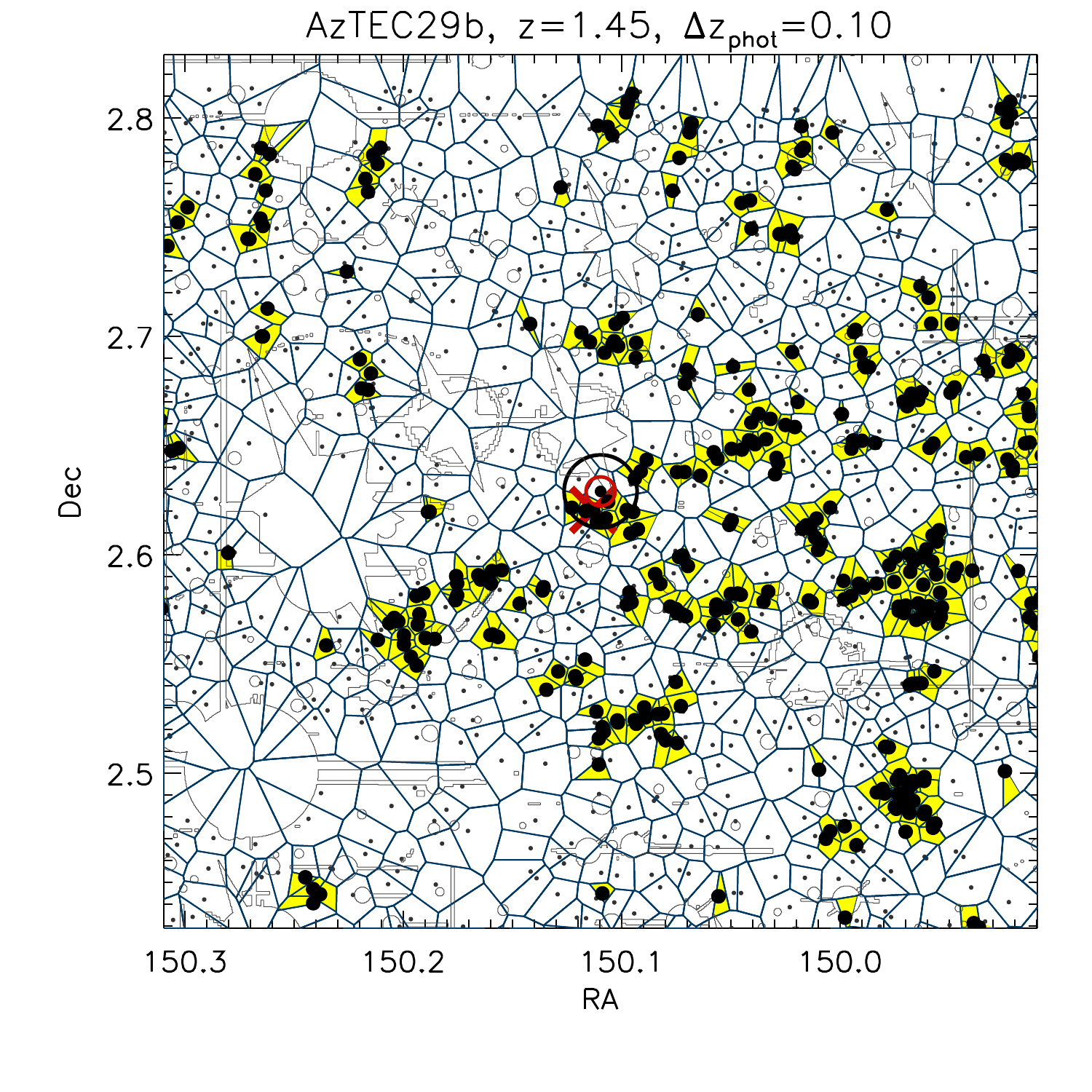}
\includegraphics[width=0.31\textwidth]{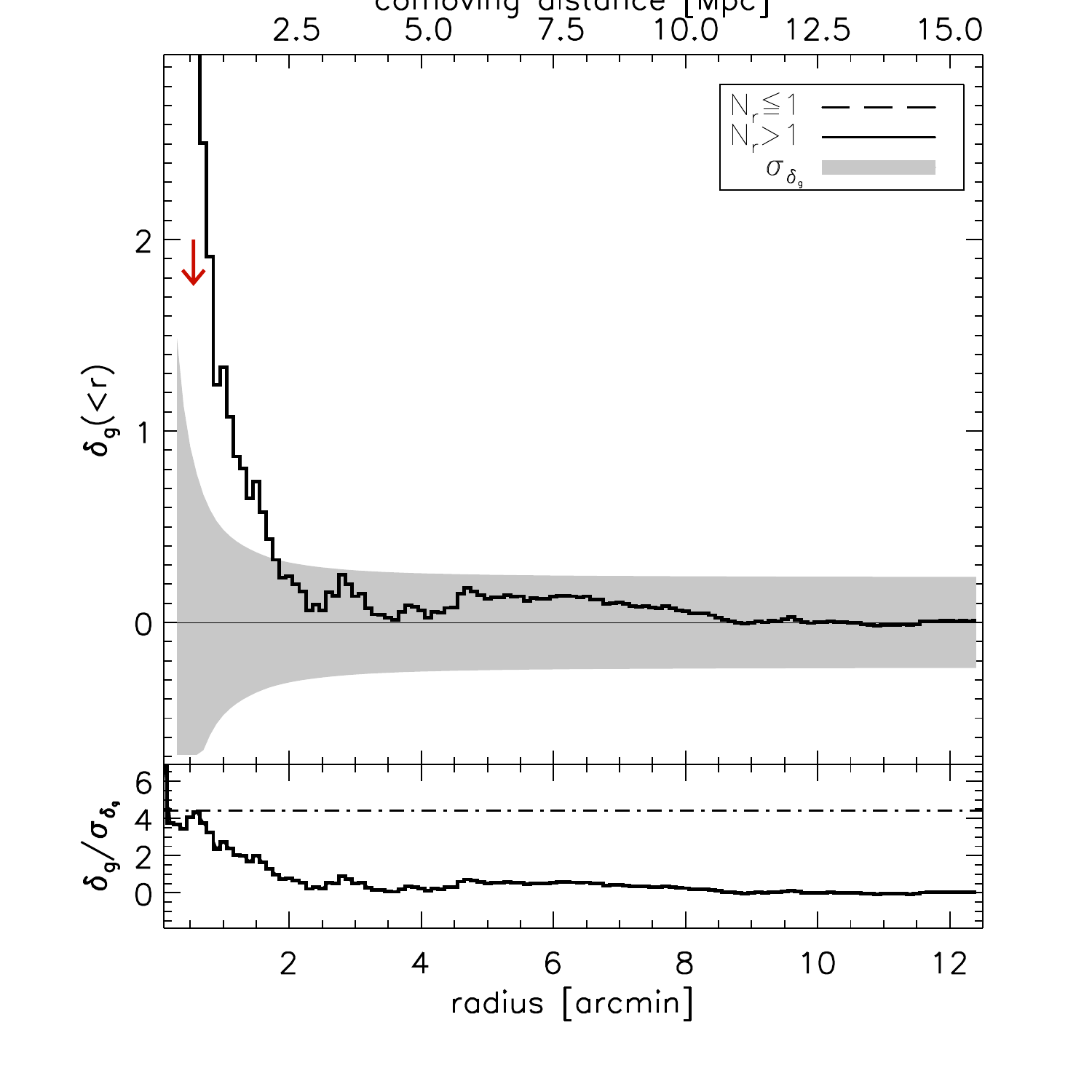}
\caption{continued.}
\end{center}
\end{figure*}

\addtocounter{figure}{-1}
\begin{figure*}
\begin{center}
\includegraphics[width=0.31\textwidth]{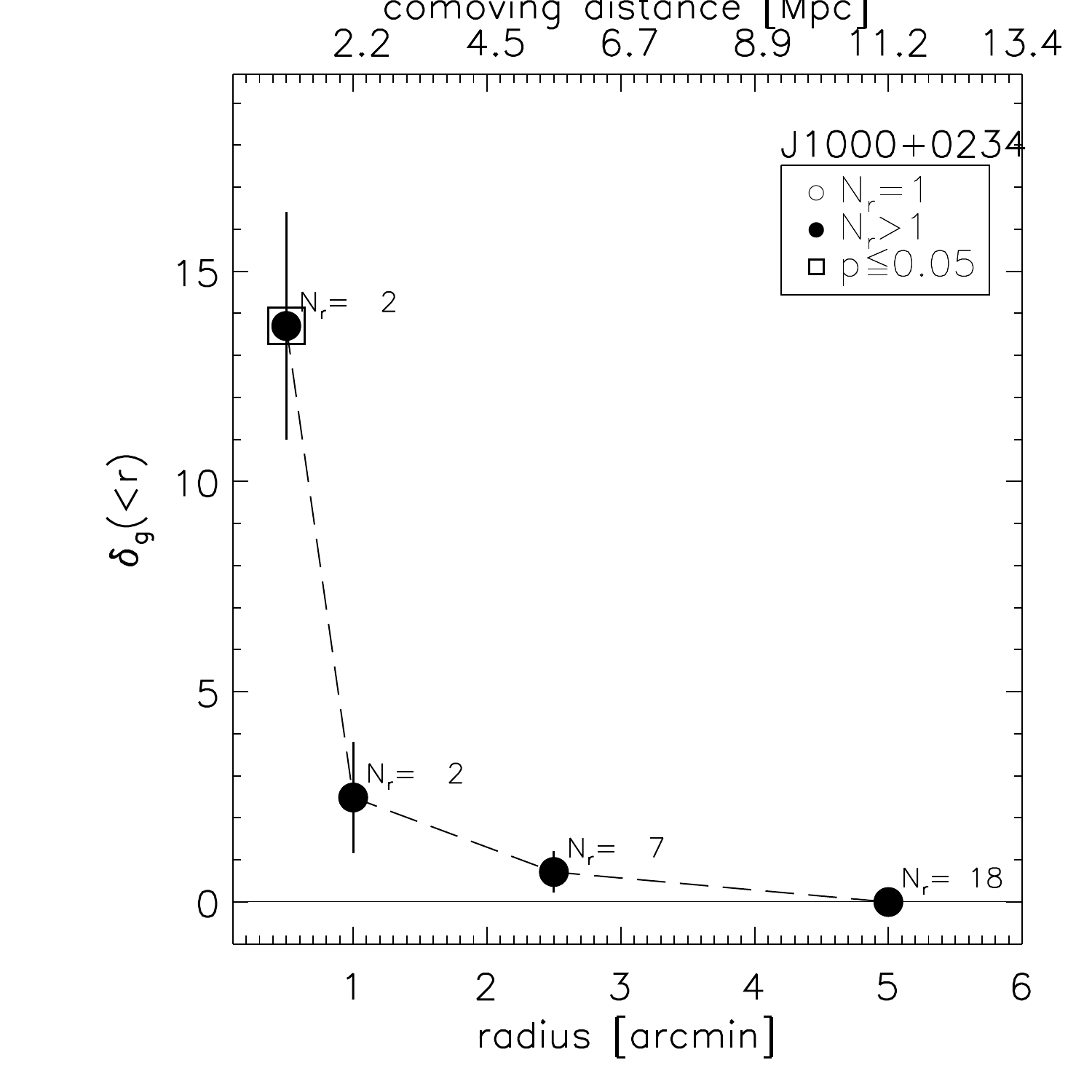}
\includegraphics[width=0.31\textwidth]{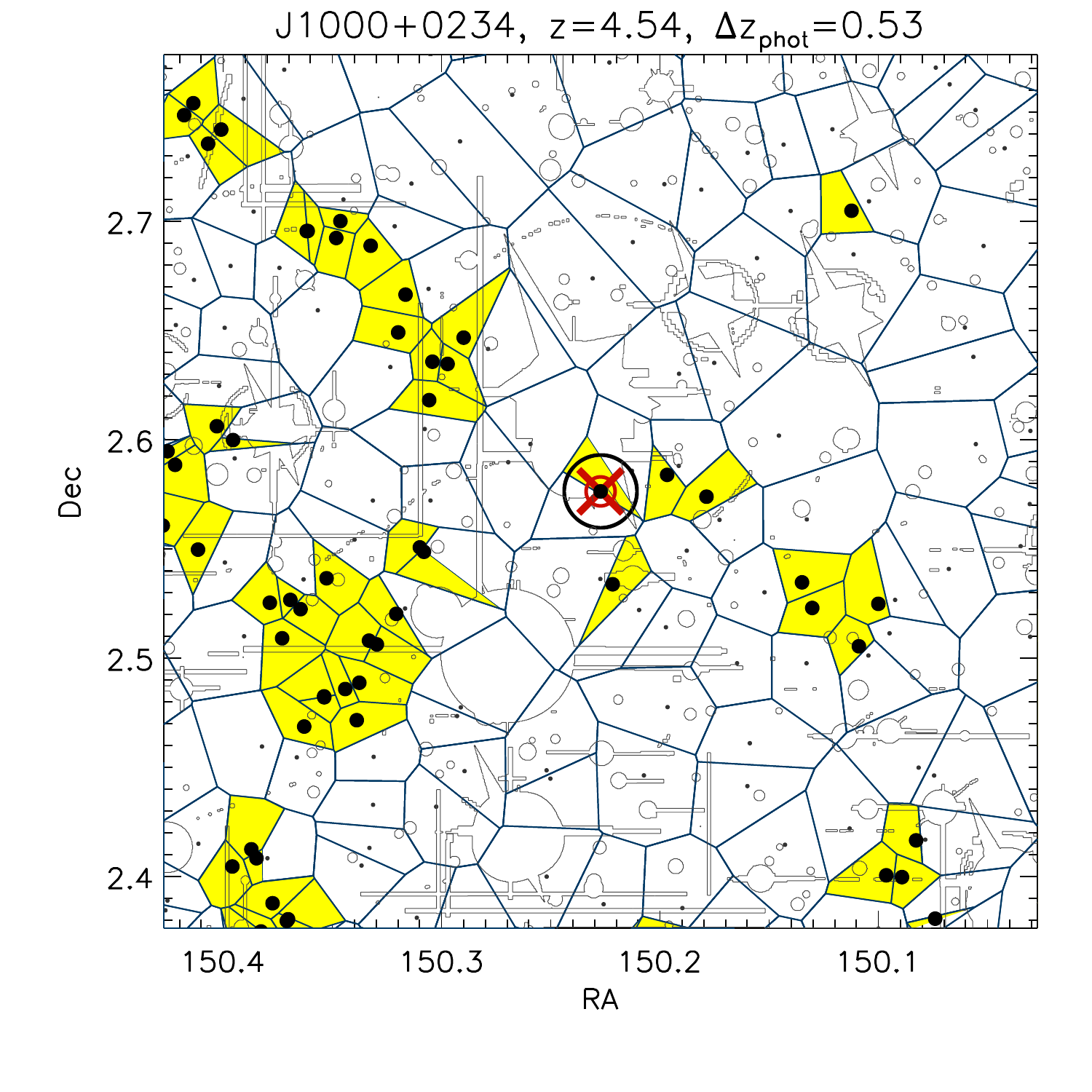}
\includegraphics[width=0.31\textwidth]{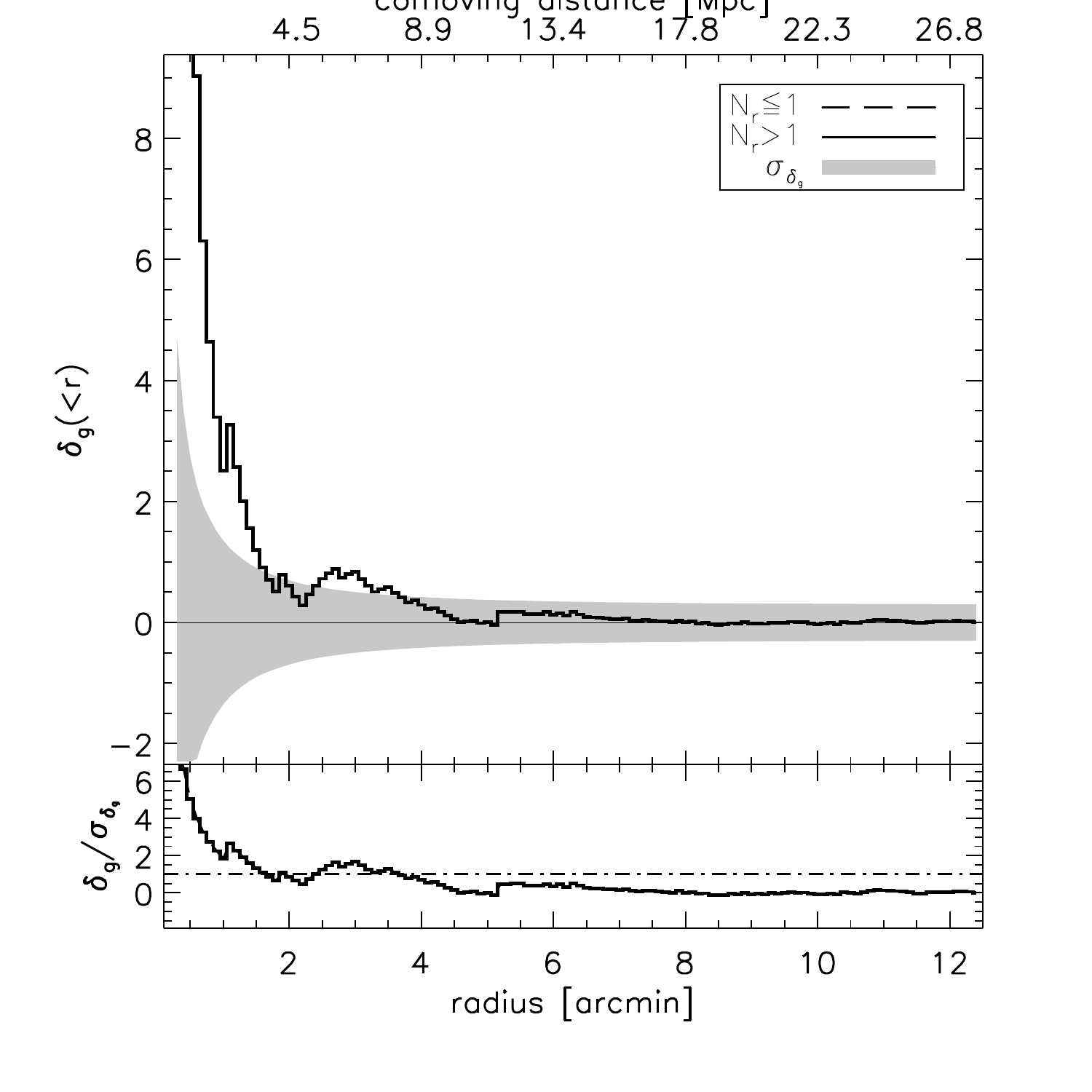}\\
\includegraphics[width=0.31\textwidth]{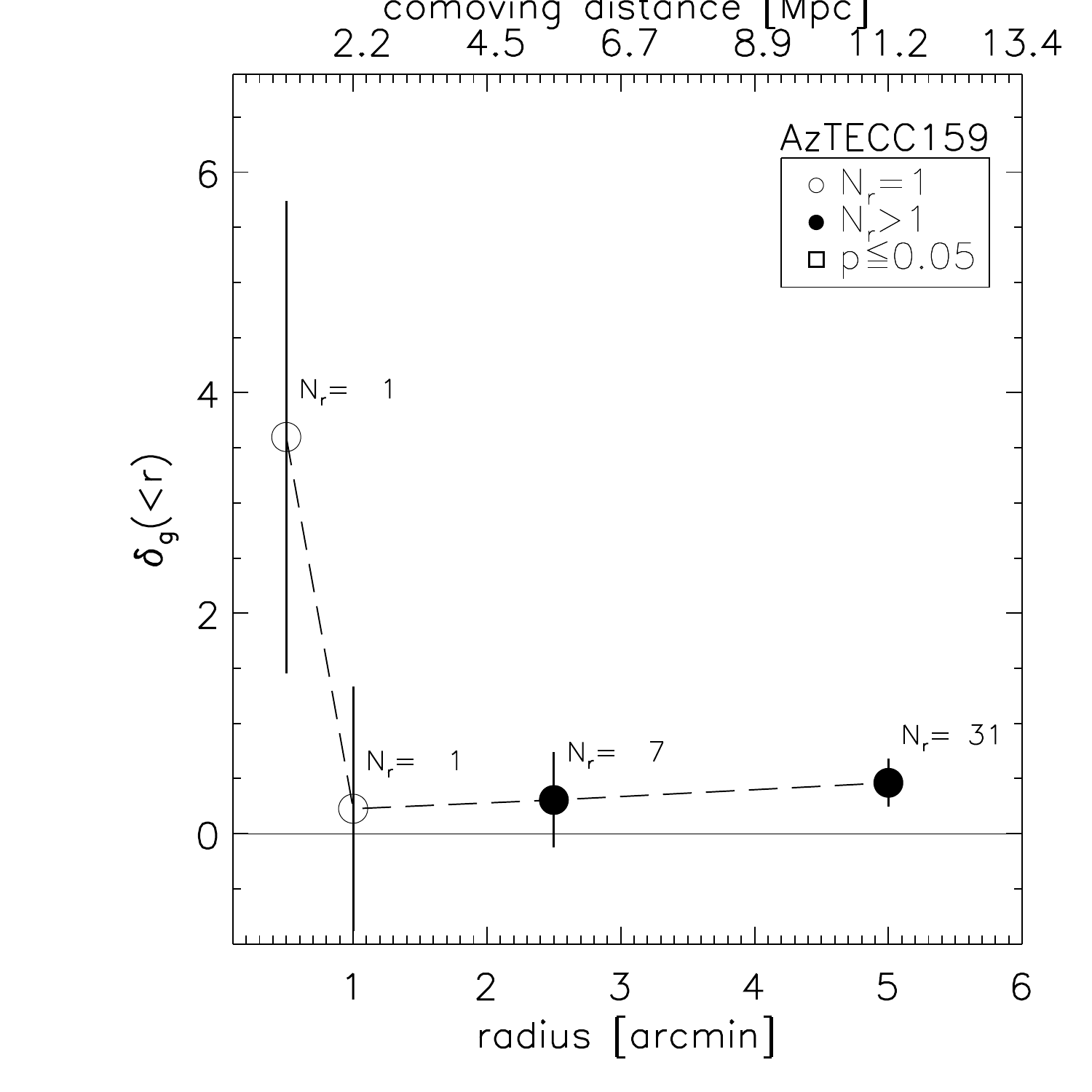}
\includegraphics[width=0.31\textwidth]{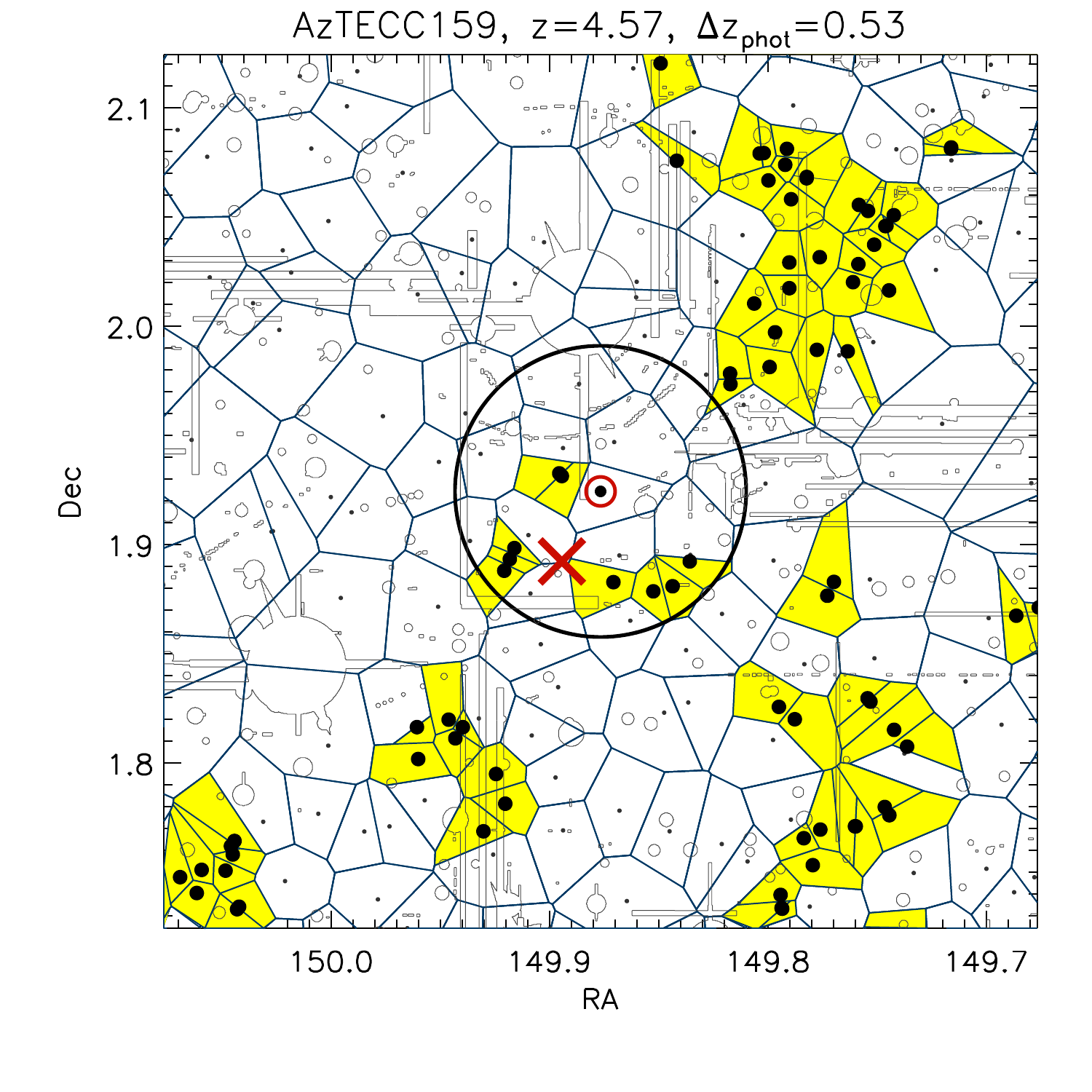}
\includegraphics[width=0.31\textwidth]{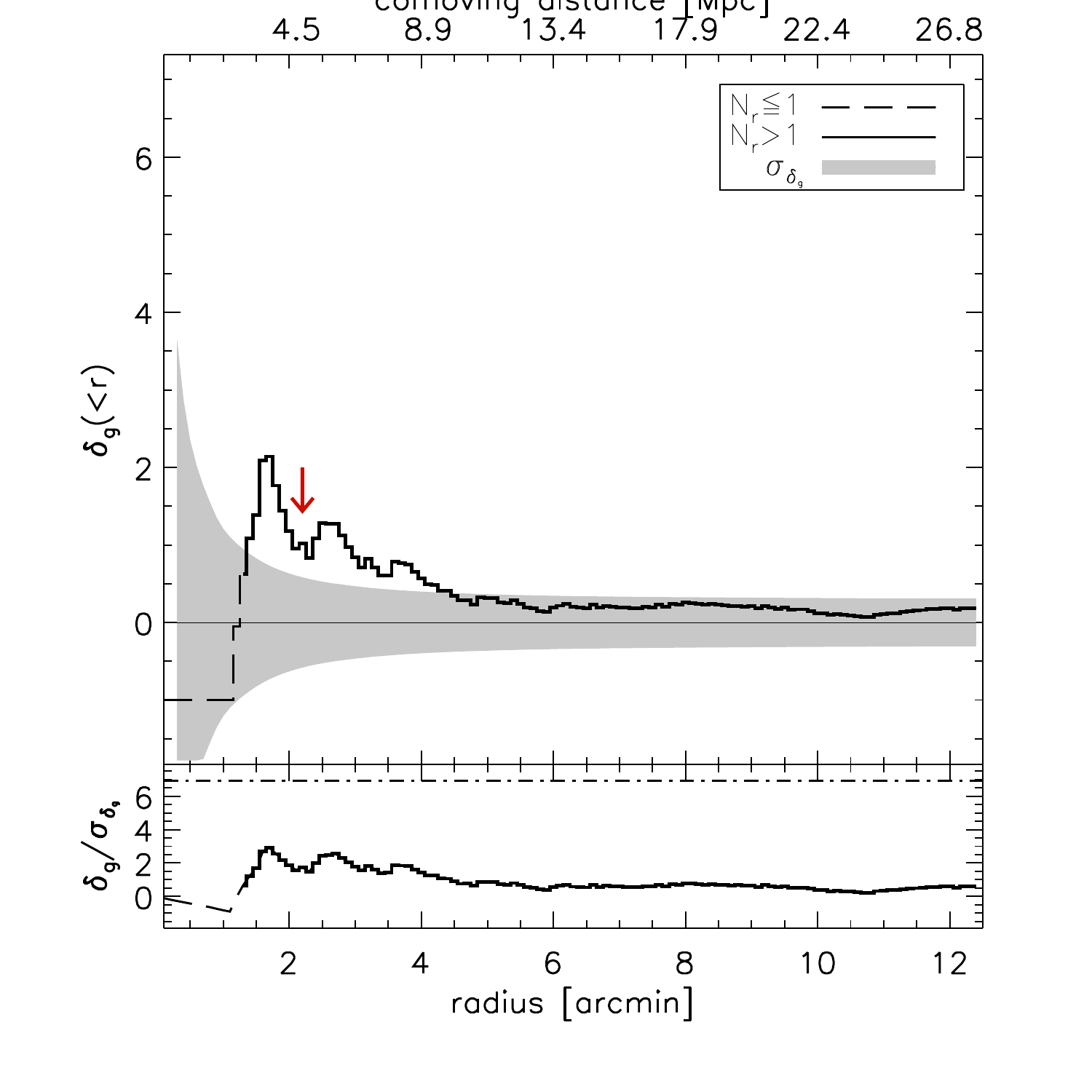}\\
\includegraphics[width=0.31\textwidth]{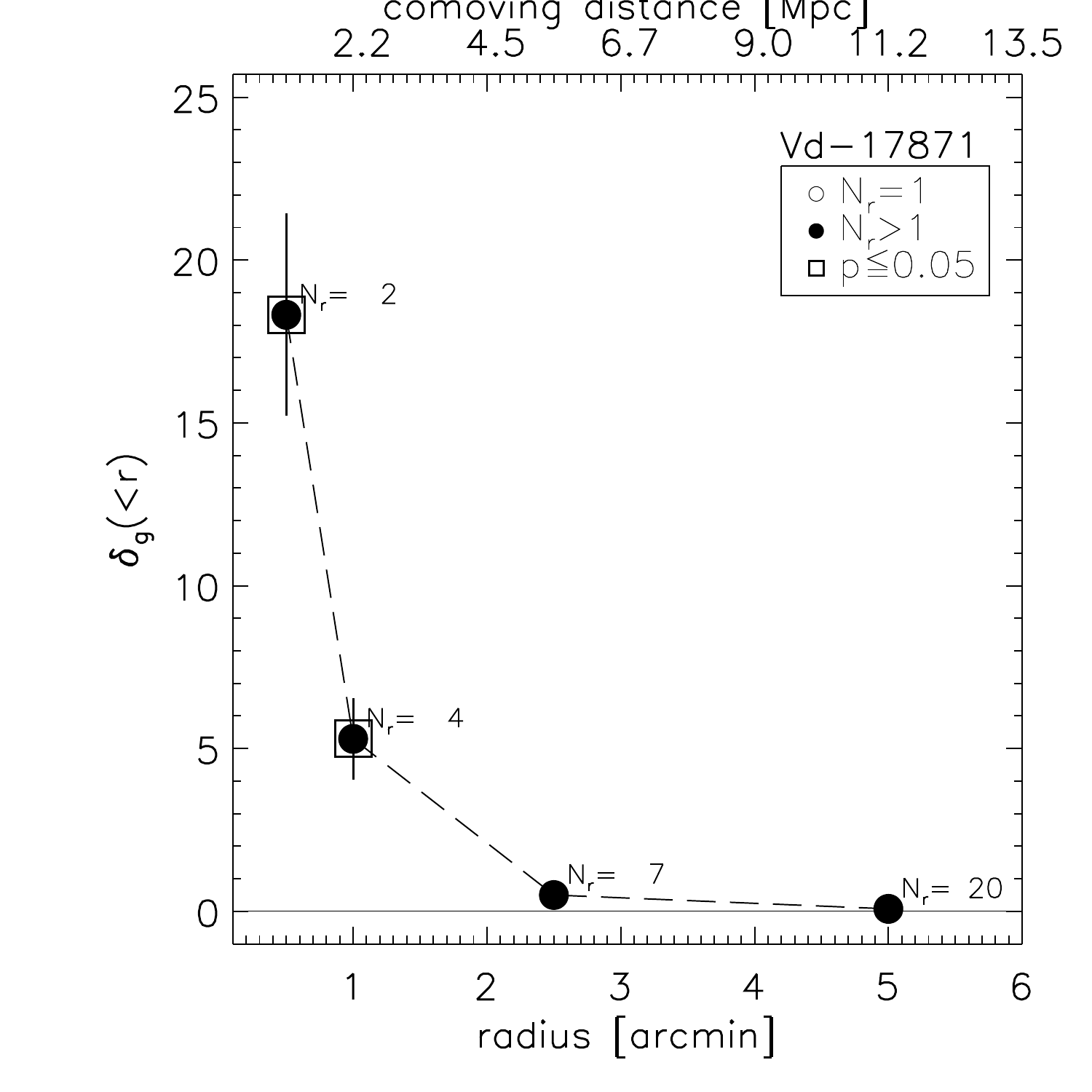}
\includegraphics[width=0.31\textwidth]{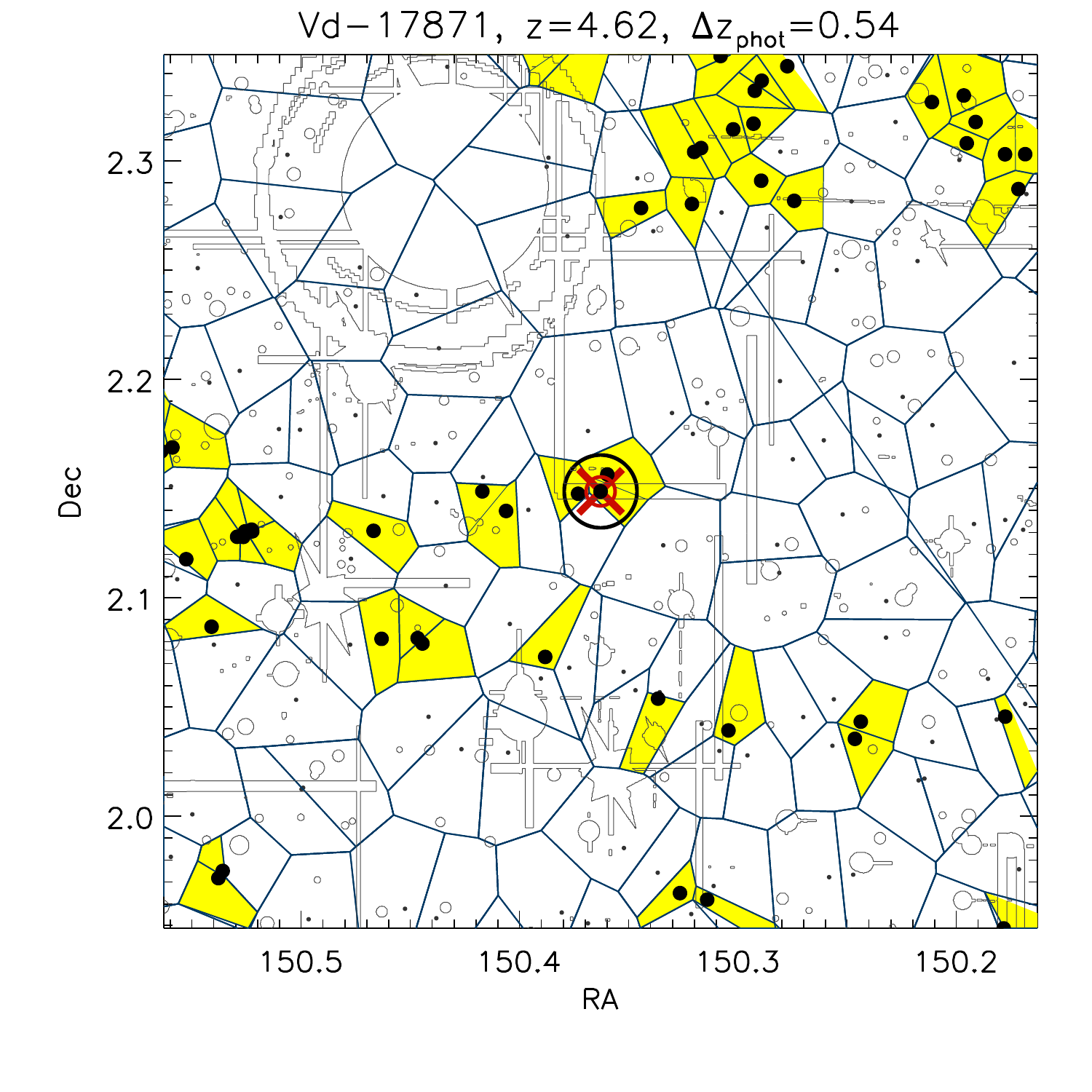}
\includegraphics[width=0.31\textwidth]{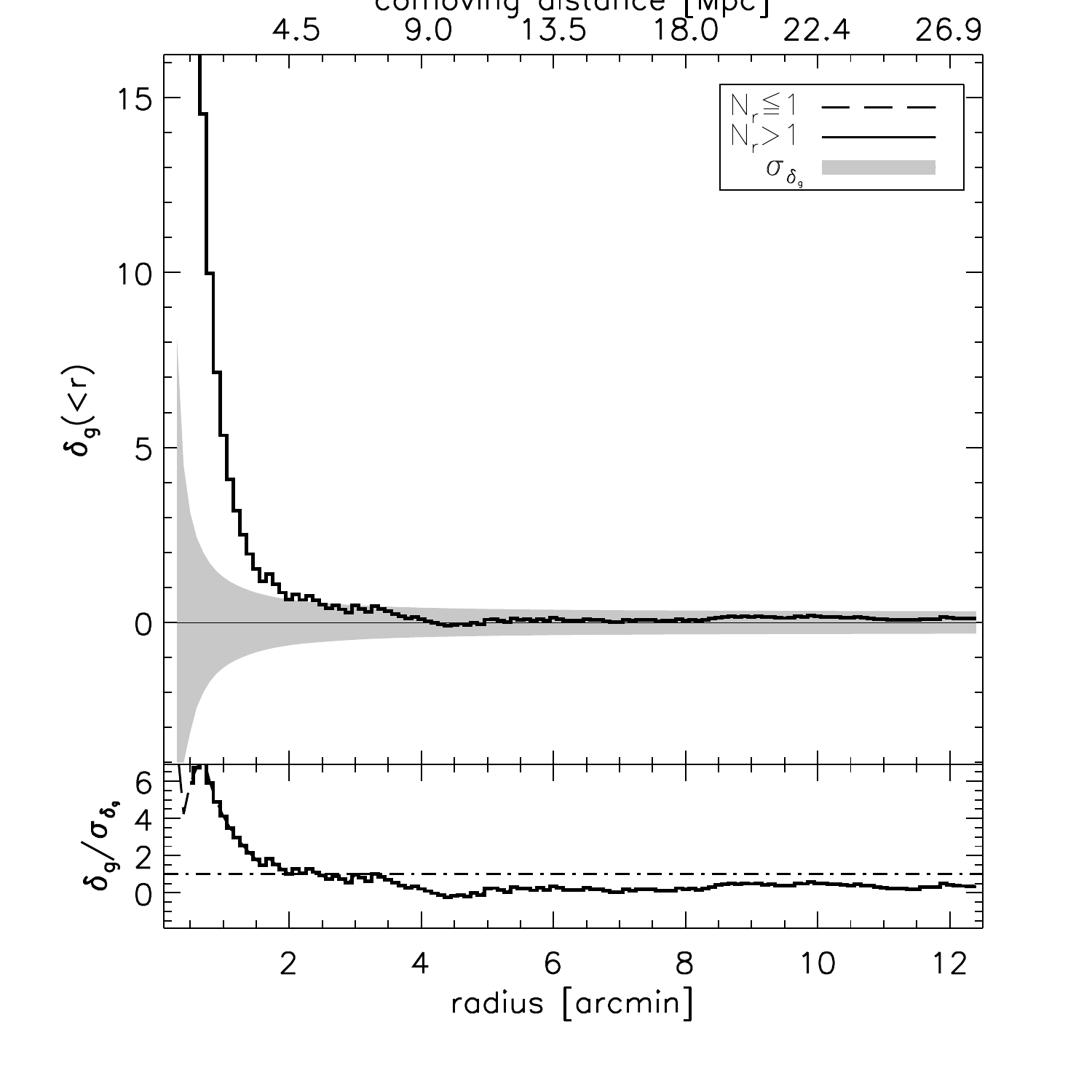}\\
\includegraphics[width=0.31\textwidth]{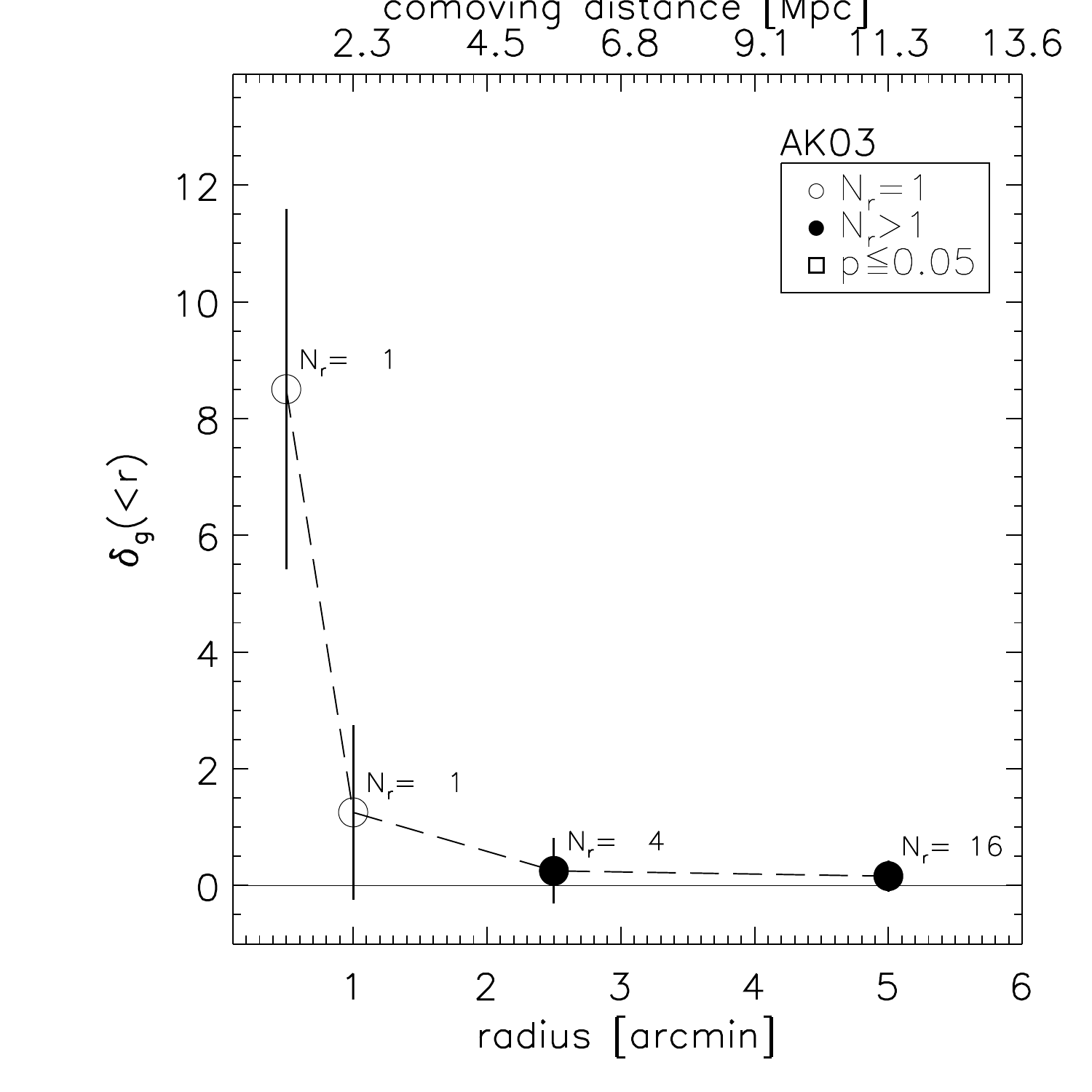}
\includegraphics[width=0.31\textwidth]{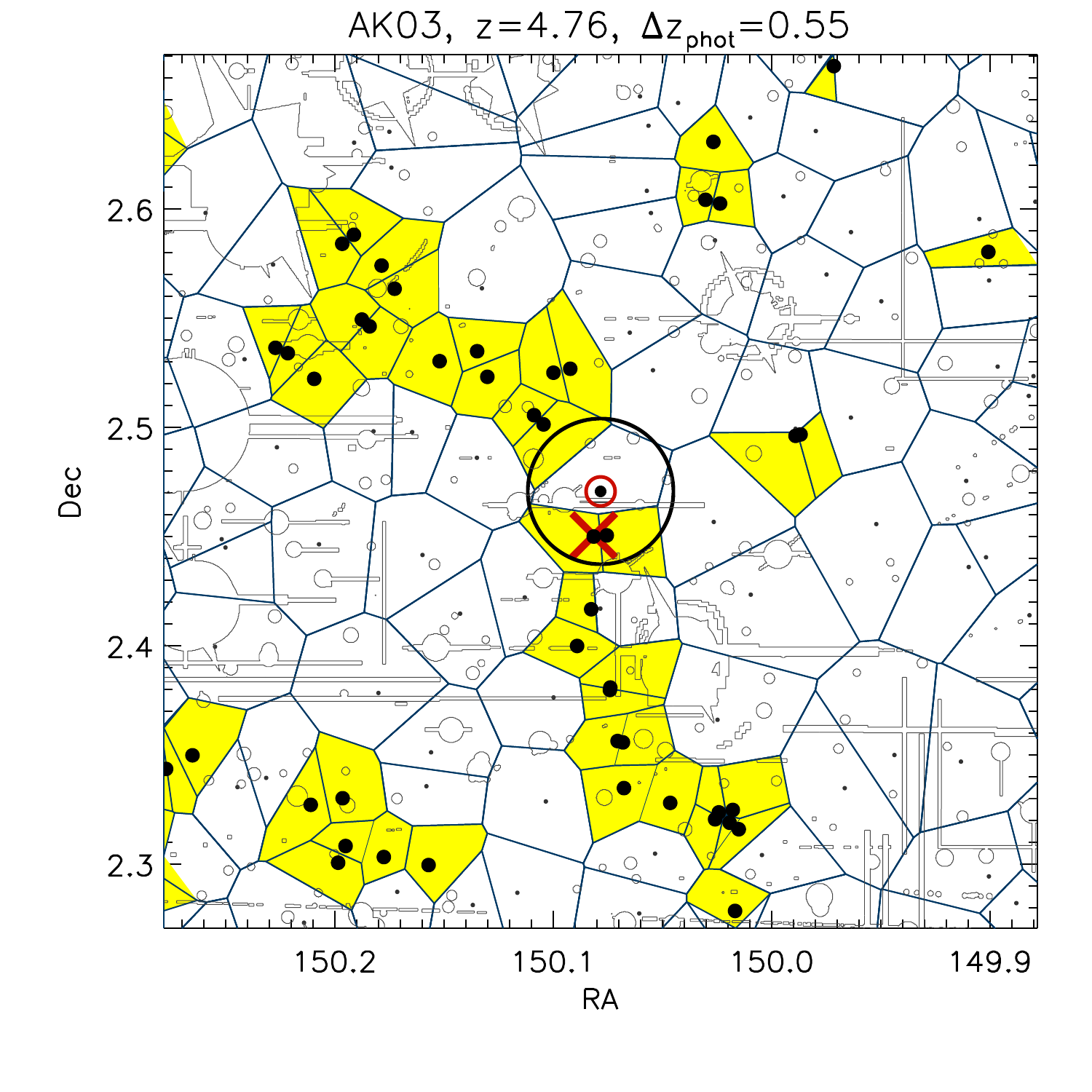}
\includegraphics[width=0.31\textwidth]{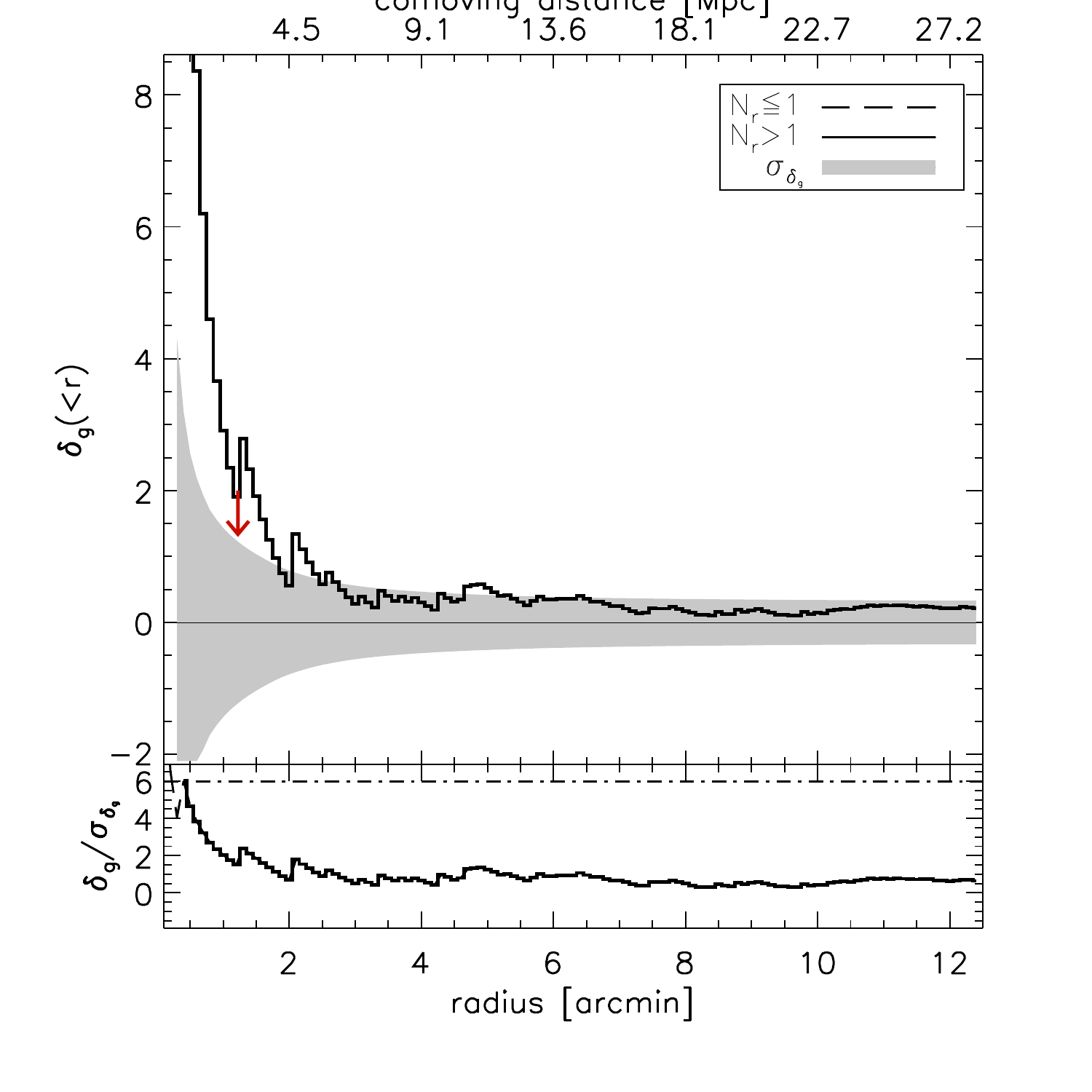}
\caption{continued.}
\end{center}
\end{figure*}

\addtocounter{figure}{-1}
\begin{figure*}
\begin{center}
\includegraphics[width=0.31\textwidth]{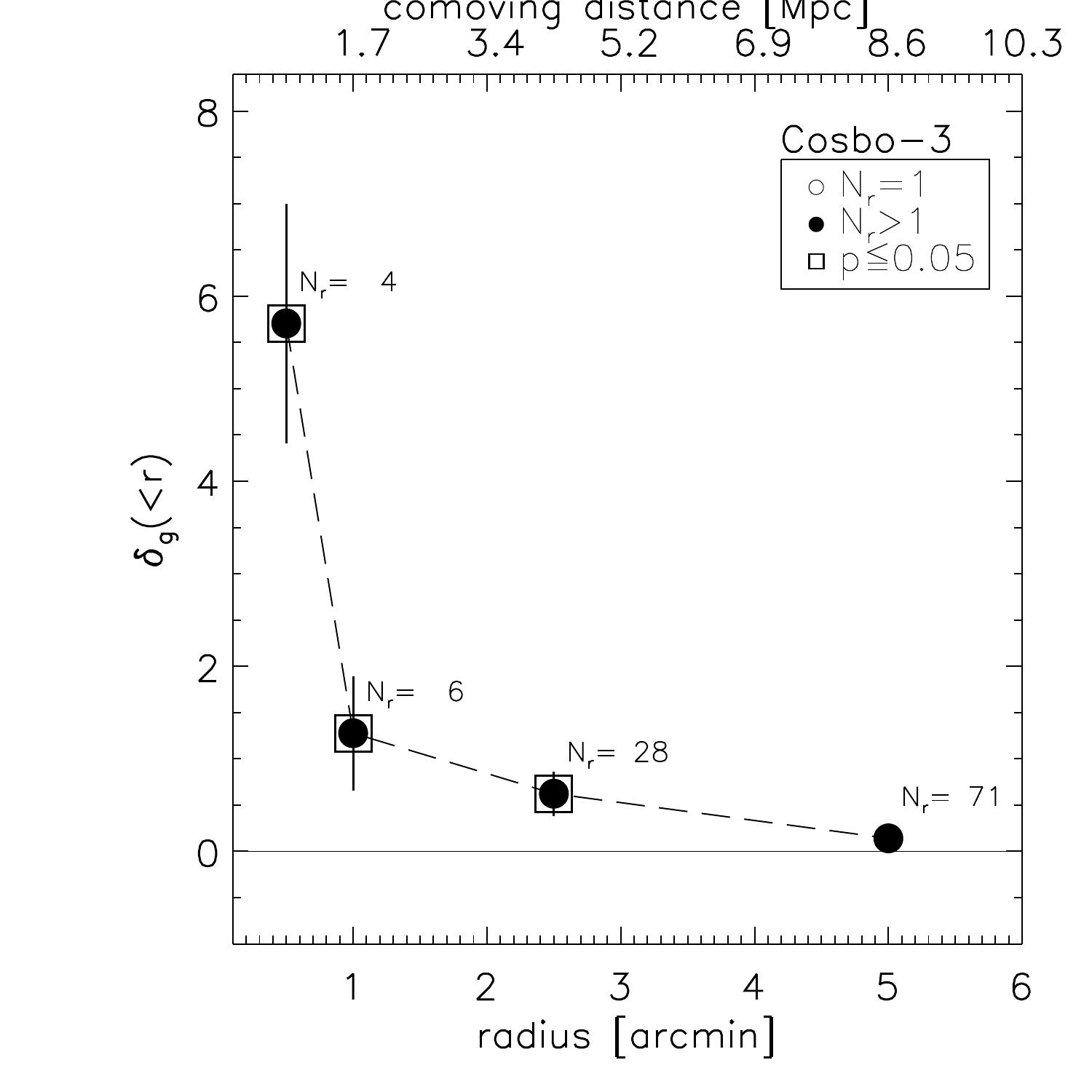}
\includegraphics[width=0.31\textwidth]{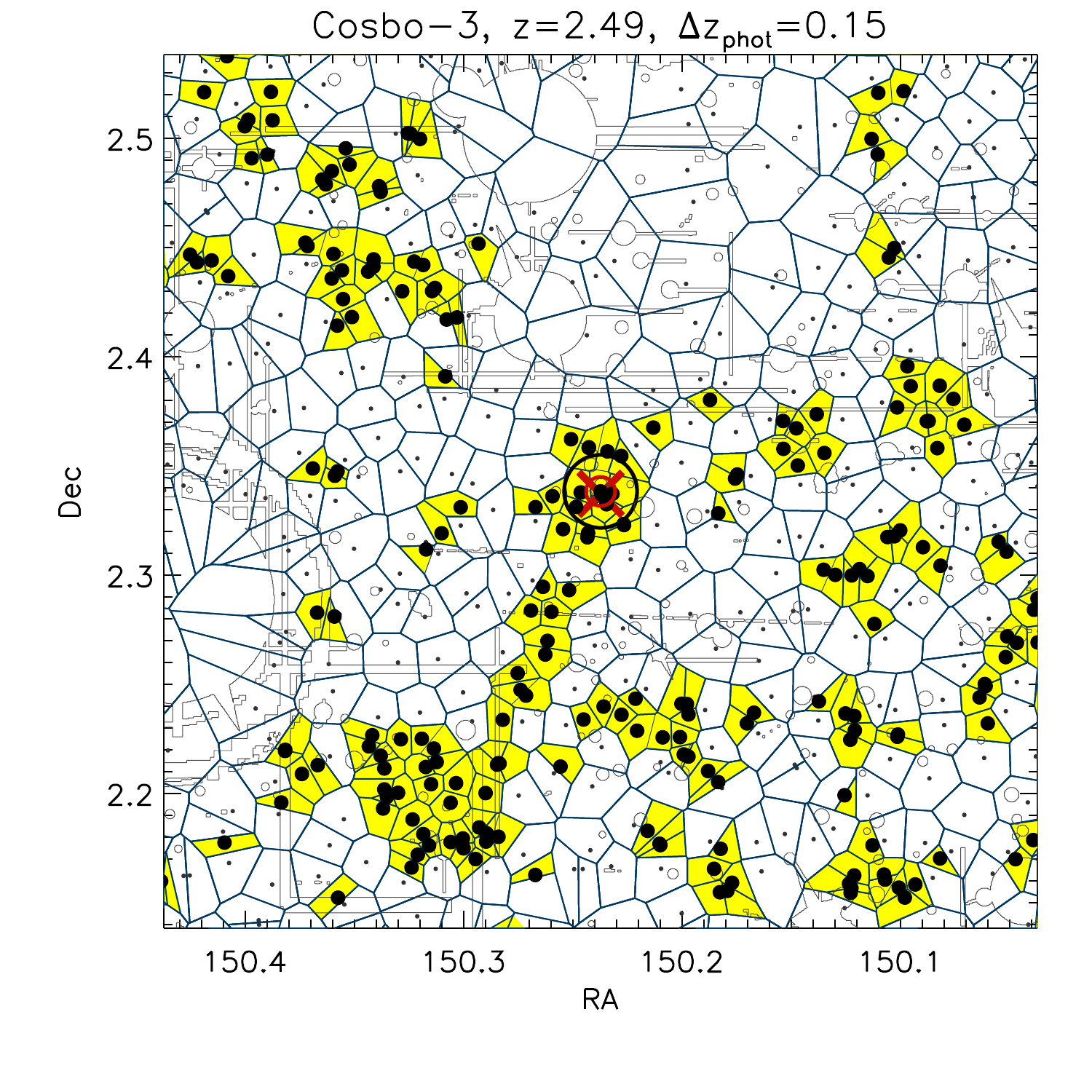}
\includegraphics[width=0.31\textwidth]{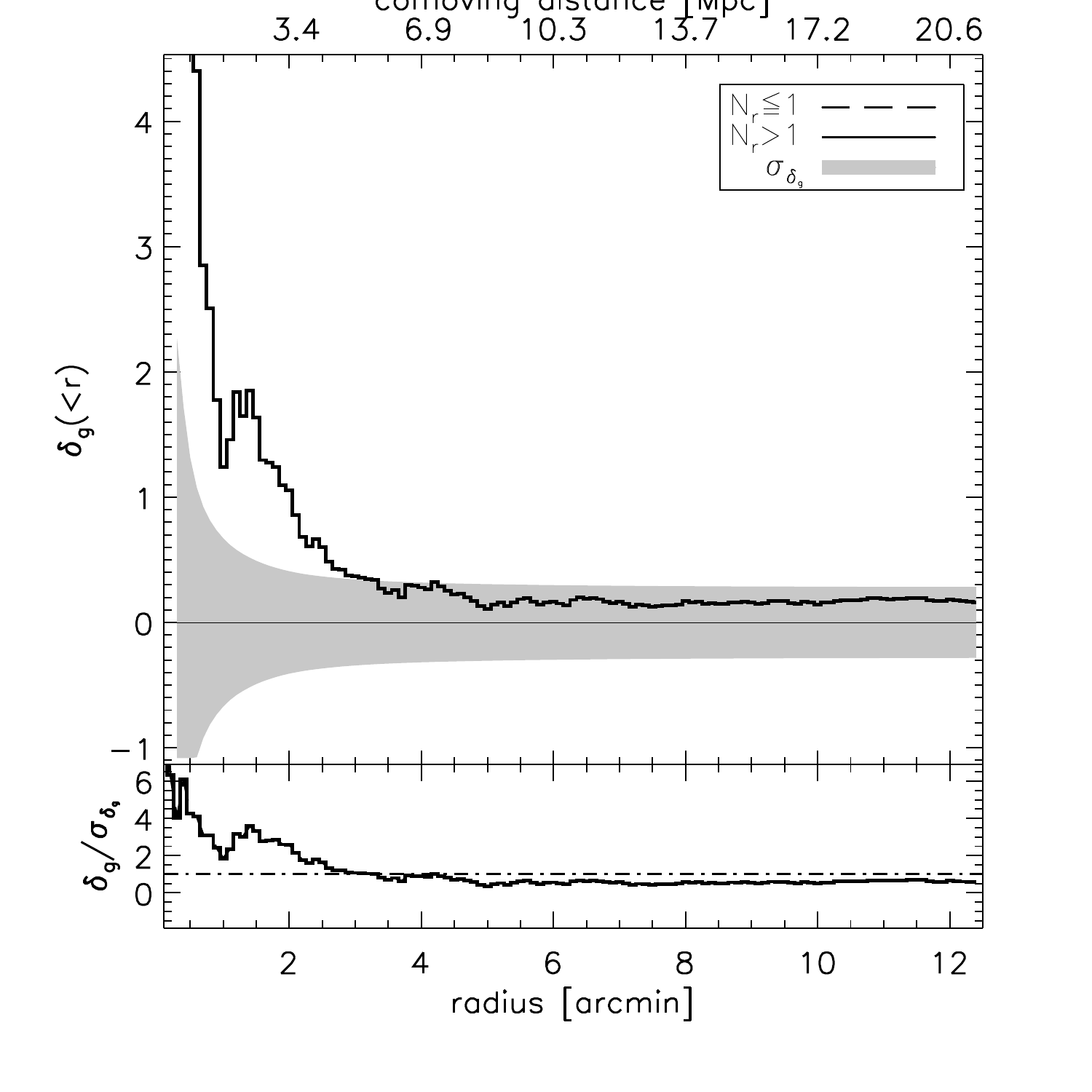}
\caption{continued.}
\end{center}
\end{figure*}

\begin{figure}
\begin{center}
\includegraphics[width=\columnwidth]{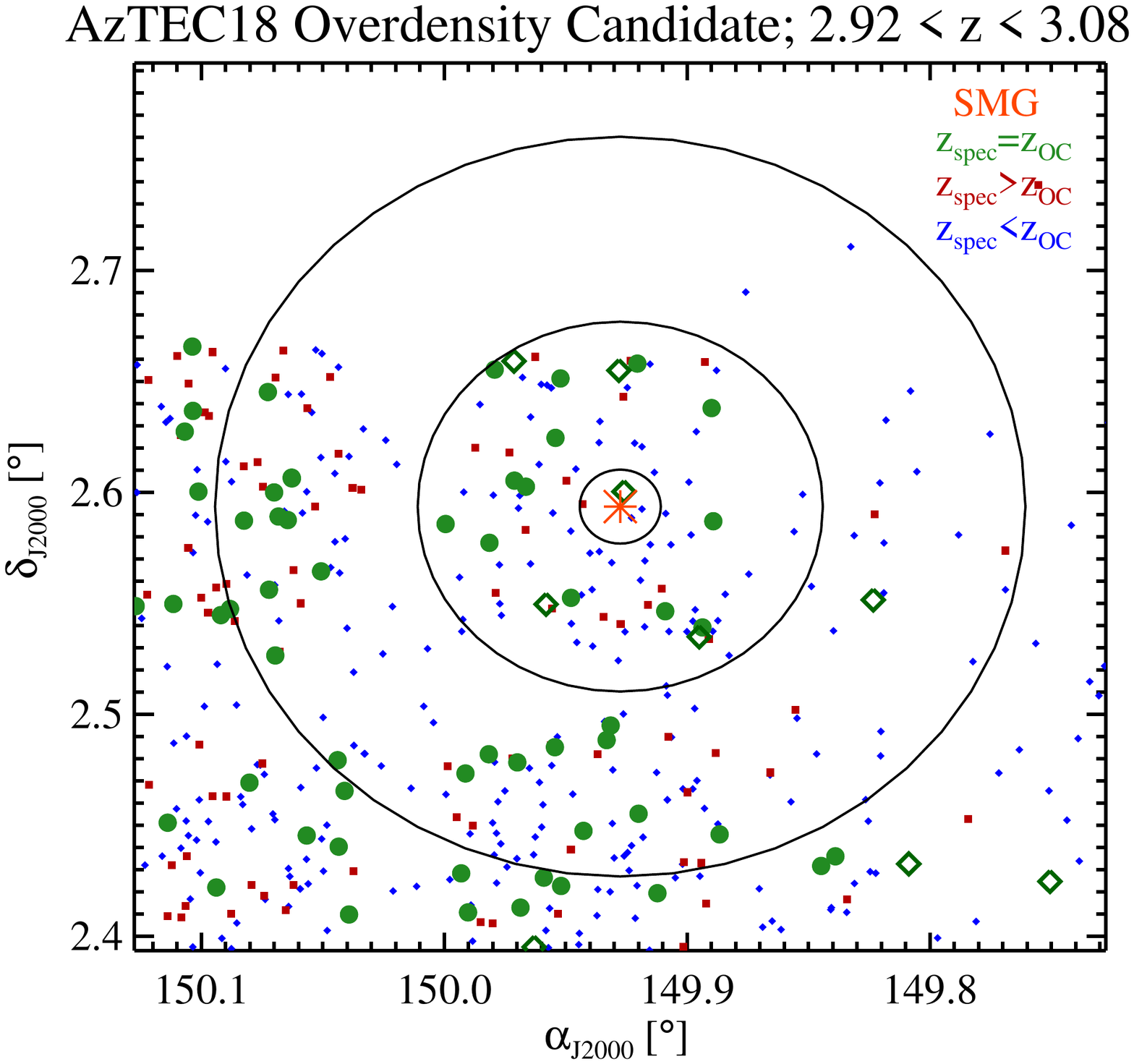}
\caption{Example of spectral, often non-uniform coverage, shown for the overdensity candidate (OC) associated with AzTEC18, confirmed to also yield a spectroscopic overdensity (see text for details). The left panel shows the galaxies with secure (filled symbols) and less secure (open symbols) spectroscopic redshifts in or outside the redshift range considered ($2.92< z_\mathrm{OC}\leq3.08$), as listed in the legend. The circles illustrate radii of $r=1\arcmin,5\arcmin,10\arcmin$. The panels on the right show the spectroscopic redshift distribution for galaxies within $5\arcmin$ (top right) and $10\arcmin$ (bottom righ) from the overdensity candidate with secure (filled histogram) and less secure (open histogram) spectroscopic redshifts. 
}
\label{fig:spec}
\end{center}
\end{figure}

\appendix

\section{Recovering known groups and simulated proto-clusters}
\label{sec:mc}

To  test our overdensity search method, described in detail in Sect.~\ref{sec:method}, at $z<1.5$ we apply  it to positions of X-ray detected groups in the COSMOS field taken from Finoguenov et al.\ (2007). We retrieve every single group with high significance. This is in accordance with results from previous studies that used the VTA with photometric redshifts to recover galaxy overdensities  (\cite{smolcic2007}; \cite{oklopcic2010}; \cite{jelic2012}). 

To test our overdensity search method at $z>1.5$ we apply  it to simulated proto-clusters at $z=2$, 3, 4, and 5. We generate mock catalogues of proto-clusters following the overdensity profiles of proto-cluster galaxies with stellar mass $M_{\star}>10^9$~\msol \ for three present-day cluster mass bins ($M_{z=0}>10^{15}$~M$_{\sun}$, $M_{z=0}=3-10\times10^{14}$~M$_{\sun}$, and $M_{z=0}=1.37-3\times10^{14}$~M$_{\sun}$) at redshifts 2, 3, 4, and 5 adopted from Chiang et al.\ (2013). Chiang et al. (2013) have derived these stacked profiles based on about 3\,000 clusters from the Millennium Simulation, whose evolution they tracked  from early cosmic times till today. The number of our generated galaxies as a function of  distance from the proto-cluster center is then set by the proto-cluster profiles for a given redshift ($z=2,3,4,5$; see Fig.~3 in \cite{chiang2013}). We then further 'scatter' the galaxies in redshift space \textit{i)} assuming a line-of-sight (LOS) velocity dispersion (in the range of 200--1\,000~\kms, which corresponds to $\Delta z\leq0.015$), and \textit{ii)} taking the COSMOS photometric redshift uncertainty into account. Both the LOS velocity and $\sigma_{\Delta z/(1+z)}$ distributions are taken to be Gaussian.  
Such generated cluster galaxies are then superposed onto background galaxies generated such that they \textit{i)} are uniformly distributed over a 1~deg$^2$ field (at small and large scales), and \textit{ii)} follow the photometric redshift distribution of galaxies in the COSMOS field ($i^+\leq25.5$). We generated 100 simulated proto-clusters at each redshift, $z=2,3,4$, and 5, and 
recover these with at least $\approx3\sigma_{\rm \delta_g}$ significance. 

This validates our approach to identify overdensities using photometric redshifts in the COSMOS field, in accordance with the results from Chiang et al.\ (2014) who searched for proto-cluster candidates in the COSMOS field at $1.6<z<3.1$ using photometric redshifts, and identified 36 candidate systems. 

\section{Overdensity search using within narrower redshift bins}
\label{sec:1.5sig}

Here we present the results of the overdensity search performed using galaxies with photometric redshifts $z_\mathrm{phot}=z_\mathrm{SMG}\pm 1.5\sigma_{\Delta z/(1+z)} (1+z_\mathrm{SMG})$. The 
small-scale and large-scale overdensity search results are listed in \t{table:smallscale1.5sigma} , and \t{table:lss1.5sigma} , respectively, while the combined results are listed in \t{table:all1.5sigma} . 

We find 6/23 (26\%) small-scale overdensities in our main-SMG sample, 4/6 (67\%) in our main-SMG subsample with spectroscopic redshifts for our SMGs, and 2/5 (40\%) in our additional-SMG sample. Within our overdensity search with potentially off-center SMGs we find 5/23 (22\%), 3/6 (50\%), and 3/5 (60\%) overdense systems in our main-SMG sample, main-SMG subsample with spectroscopic redshifts assigned to the SMGs, and additional-SMG sample, respectively. In total, we find 7/23 (30\%), and 3/5 (60\%) overdensities in our main-, and additional-SMG samples, respectively. Taking the expected spurious fraction in the S/N-limited 1.1~mm sample the SMGs were drawn from into account this amounts to a range of 19-45\% of SMGs occupying overdense systems.

Combining all overdensities identified using both redshift ranges, $\Delta z=z_\mathrm{SMG}\pm k\cdot\sigma_{\Delta z/(1+z)} (1+z_\mathrm{SMG})$, where k=$1.5$ and $3$, then yields a total of 13/23 (56\%), and 5/5 (100\%) overdensities in our main-, and additional-SMG samples, respectively. Taking the expected spurious fraction into account would imply a range of 42-68\% of SMGs occupying overdensities in our S/N-limited 1.1mm-selected SMG sample.

\clearpage 

\begin{table}[h!]
\renewcommand{\footnoterule}{}
\caption{
 Small-scale overdensity search results [using $\Delta z=z_\mathrm{SMG}\pm 1.5\cdot\sigma_{\Delta z/(1+z)} (1+z_\mathrm{SMG})$]. The systems with statistically significant overdensities are indicated in bold-faced.
}
{\small
\begin{minipage}{1\columnwidth}
%\centering
\label{table:smallscale1.5sigma}
\begin{tabular}{l l c c c c c }
\hline
name & radius & $N_\mathrm{r}$ & Poisson probability &  false detection \\ 
    & [$\arcmin$] & & $p(\geq N_\mathrm{r}, n_\mathrm{r})$&  probability, P$_\mathrm{FD}$ \\ 
\hline 
 {\bf          AzTEC1 }& 0.5 & 2 & 0.032 & 0.009 \\
    AzTEC2 & 1.0 & 3 & 0.198 & 0.727 \\
 {\bf     AzTEC3 }& 0.5 & 2 & 0.003 & 0.000 \\
    AzTEC4         & -- & -- & -- & -- \\
          AzTEC5         & -- & -- & -- & -- \\
    AzTEC7 & 1.0 & 2 & 0.161 & 0.382 \\
  {\bf    AzTEC8 }& 0.5 & 2 & 0.054 & 0.030 \\
    AzTEC9 & 1.0 & 3 & 0.195 & 0.557 \\
   AzTEC10 & 0.5 & 2 & 0.076 & 0.054 \\
   AzTEC11 & 1.0 & 2 & 0.191 & 0.732 \\
   AzTEC12 & 1.0 & 2 & 0.159 & 0.344 \\
 AzTEC14-W & 1.0 & 6 & 0.189 & 0.202 \\
   {\bf  AzTEC15 }& 0.5 & 2 & 0.054 & 0.026 \\
  {\bf  AzTEC17a} & 0.5 & 4 & 0.144 & 0.024 \\
  AzTEC17b & 1.0 & 2 & 0.105 & 0.120 \\
    {\bf AzTEC18} & 0.5 & 2 & 0.060 & 0.034 \\
  AzTEC19a & 1.0 & 3 & 0.138 & 0.077 \\
  AzTEC19b & 1.0 & 6 & 0.197 & 0.445 \\
  AzTEC21a         & -- & -- & -- & -- \\
  AzTEC21b & 1.0 & 2 & 0.169 & 0.423 \\
   AzTEC23 & 0.5 & 2 & 0.109 & 0.149 \\
  AzTEC26a & 1.0 & 2 & 0.158 & 0.345 \\
  AzTEC29b & 1.0 & 5 & 0.195 & 0.354 \\
  \hline
   Cosbo-3 & 1.0 & 2 & 0.158 & 0.340 \\
 {\bf J1000+0234 }& 0.5 & 2 & 0.023 & 0.006 \\
 AzTECC159         & -- & -- & -- & -- \\
  {\bf  Vd-17871 }& 0.5 & 2 & 0.020 & 0.003 \\
      AK03         & -- & -- & -- & -- \\
  \hline 
\end{tabular} 
\end{minipage} }
\end{table}

 \begin{table}[t!]
\renewcommand{\footnoterule}{}
\caption{
 Large-scale overdensity search results [using $\Delta z=z_\mathrm{SMG}\pm 1.5\cdot\sigma_{\Delta z/(1+z)} (1+z_\mathrm{SMG})$]. The systems with statistically significant overdensities are indicated in bold-faced.
}
{\small
\begin{minipage}{1\columnwidth}
%\centering
\label{table:lss1.5sigma}
\begin{tabular}{l l c }
\hline
SMG & SMG  \\ 
    &  distance [$\arcmin$]  \\
\hline 
{\bf AzTEC1} &       0.281\\  
AzTEC2 &         2.804\\ 
{\bf AzTEC3} &        0.027\\   
AzTEC4 &         10.265\\  
AzTEC5 &    6.554\\    
AzTEC7 &      1.588\\  
AzTEC8 &    4.923\\  
AzTEC9 &          1.261\\   
{\bf AzTEC10} &       0.931\\    
AzTEC11&      1.193\\   
AzTEC12 &       0.696\\   
AzTEC14-W &       0.451\\  
AzTEC15 &        2.992\\  
{\bf AzTEC17a} &        0.199\\    
AzTEC17b &       0.959\\ 
{\bf AzTEC18} &       0.198\\   
AzTEC19a  &     0.875\\   
AzTEC19b &      0.832\\ 
AzTEC21a &      1.377\\ 
AzTEC21b &       1.286\\
AzTEC23 &       2.538\\  
AzTEC26a &  4.793\\ 
AzTEC29b &      0.542\\
\hline
Cosbo-3  &          0.673\\
{\bf J1000+0234 } &       0.138\\  
{\bf AzTEC/C159 }&         3.020\\  
{\bf Vd-17871 } &        0.000\\    
AK03 &      3.478\\  
\hline 
\end{tabular} \\
\end{minipage} }
\end{table}

  \begin{table*}[h]
\renewcommand{\footnoterule}{}
\caption{
Overdensity search results  based on small- and large-scale analyses [using $\Delta z=z_\mathrm{SMG}\pm 1.5\cdot\sigma_{\Delta z/(1+z)} (1+z_\mathrm{SMG})$]. Only sources associated with statistically significant overdensities are listed. Source redshifts and assumed distances between the SMG and overdensity center are also given. Spectroscopically verified systems (as described in detail in \s{sec:specverif} ) are indicated in bold-faced.
}
{\small
\begin{minipage}{1\columnwidth}
%\centering
\label{table:all1.5sigma}
\begin{tabular}{l l c c c c c}
\hline
SMG & redshift & small-scale  & large-scale  & SMG & spectroscopically \\ 
    & &  overdensity & overdensity & distance [$\arcmin$]  & verified \\
\hline 
AzTEC1 &       4.3415\tablefootmark{a} &  $\surd$ & $\surd$ & 0.281 & -- \\  
AzTEC3 &       5.298\tablefootmark{a} & $\surd$ & $\surd$ & 0.027 & $\surd$\\ 
AzTEC8 &       3.179\tablefootmark{a} & $\surd$ & -- & 0.000 & --\\  
AzTEC10 &       2.79 &   -- & $\surd$ & 0.931 & -- \\
AzTEC15 &       3.17 &   $\surd$ & -- & 0.000 & -- \\   
AzTEC17a &      0.834\tablefootmark{a} & $\surd$ &$\surd$ & 0.199 & -- \\   
AzTEC18 &       3.0 & $\surd$ &$\surd$&    0.198 & $\surd$\tablefootmark{b}\\   
\hline
J1000+0234 &       4.542\tablefootmark{a} &   $\surd$ &$\surd$ &  0.138 & $\surd$\\  
AzTEC/C159 &       4.569\tablefootmark{a} &  -- & $\surd$ & 3.020 & -- \\  
Vd-17871 &       4.622\tablefootmark{a} & $\surd$ &$\surd$ &0.000 &$\surd$\tablefootmark{b}\\    
\hline 
\end{tabular} \\
\tablefoottext{a}{Spectroscopic redshift}\\
\tablefoottext{b}{Possible verification}
\end{minipage} }
\end{table*}

\end{document}